\documentclass[fleqn,usenatbib]{mnras}

\usepackage{newtxtext,newtxmath}

\usepackage[T1]{fontenc}
\usepackage{ae,aecompl}
\usepackage{times}
\usepackage{enumitem}
\usepackage{times}
\usepackage{amsmath}
\usepackage{amssymb}
\usepackage{textcomp}
\usepackage{verbatim}
\usepackage{esdiff}

\newcommand{\msun}{{\,\rm M_\odot}}

\newcommand{\kms}{\,{\rm km}\,{\rm s}^{-1}}
\newcommand{\cm}{\,{\rm cm}}
\newcommand{\erg}{\,{\rm erg}}

\newcommand{\Myr}{\,{\rm Myr}}

\newcommand{\pkpc}{\,{\rm pkpc}}
\newcommand{\Mpc}{\,{\rm Mpc}}
\newcommand{\pMpc}{\,{\rm pMpc}}

\newcommand{\mmag}{\,{\rm mag}}

\def\aap{A\&A}
\def\apj{ApJ}

\def\apjl{ApJ}
\def\mnras{MNRAS}
\def\araa{ARA\&A}
\def\aj{AJ}

\def\physrep{Phys. Rep.}
\def\nat{Nature}

\def\apjs{ApJS}

\usepackage{tikz}

\usepackage{graphicx}	
\usepackage{amsmath}	
\usepackage{amssymb}	

\makeatletter
\newcommand{\rmnum}[1]{\romannumeral #1}
\newcommand{\Rmnum}[1]{\expandafter\@slowromancap\romannumeral #1@}

\renewcommand\paragraph{\@startsection{paragraph}{4}{\z@}{3.25ex\@plus1ex\@minus.2ex}{-1em}{\normalfont\it\normalsize}}

\newcommand*\circled[1]{\tikz[baseline=(char.base)]{
            \node[shape=circle,draw,inner sep=0.3pt] (char) {#1};}}

\defcitealias{Vogelsberger2019}{Paper~\Rmnum{1}}

\title[High redshift predictions from IllustrisTNG]
{High redshift {\it JWST} predictions from IllustrisTNG: \Rmnum{2}. Galaxy line and continuum spectral indices and dust attenuation curves}

      \author[Shen et al.] {\parbox{17.5cm}{
          Xuejian Shen$^1$\thanks{email:xshen@caltech.edu},
          Mark Vogelsberger$^2$,
          Dylan Nelson$^3$, 
          Annalisa Pillepich$^4$,
          Sandro Tacchella$^5$,
          Federico Marinacci$^6$,
          Paul Torrey$^7$,
          Lars Hernquist$^5$,
          Volker Springel$^3$
        }\vspace{0.3cm}\\ 
        $^1$ TAPIR, California Institute of Technology, Pasadena, CA 91125, USA\\
        $^2$ Department of Physics, Kavli Institute for Astrophysics and Space Research, Massachusetts Institute of Technology, Cambridge, MA 02139, USA\\
        $^3$ Max-Planck-Institut f\"{u}r Astrophysik, Karl-Schwarzschild-Str. 1, D-85748 Garching, Germany\\
        $^4$ Max-Planck-Institut f\"{u}r Astronomie, K\"{o}nigstuhl 17, 69117 Heidelberg, Germany\\
        $^5$ Harvard-Smithsonian Center for Astrophysics, 60 Garden Street, Cambridge, MA, 02138, USA\\
        $^6$ Department of Physics \& Astronomy, University of Bologna, via Gobetti 93/2, I-40129 Bologna, Italy\\
        $^7$ Department of Astronomy, University of Florida, 211 Bryant Space Sciences Center, Gainesville, FL 32611, USA
    }

\date{Accepted XXX. Received YYY; in original form ZZZ}

\pubyear{2020}

\begin{document}

\label{firstpage}
\pagerange{\pageref{firstpage}--\pageref{lastpage}}
\maketitle

\begin{abstract} 
We present predictions for high redshift~($z=2-10$) galaxy populations based on the IllustrisTNG simulation suite and a full Monte Carlo dust radiative transfer post-processing. 
Specifically, we discuss the ${\rm H}_{\alpha}$ and ${\rm H}_{\beta}$ + $[\rm O \,\Rmnum{3}]$ luminosity functions up to $z=8$. The predicted ${\rm H}_{\beta}$ + $[\rm O \,\Rmnum{3}]$ luminosity functions are consistent with present observations at $z\lesssim 3$ with $\lesssim 0.1\,{\rm dex}$ differences in luminosities. However, the predicted ${\rm H}_{\alpha}$ luminosity function is $\sim 0.3\,{\rm dex}$ dimmer than the observed one at $z\simeq 2$. Furthermore, we explore continuum spectral indices, the Balmer break at $4000\text{\AA}$ (D4000) and the UV continuum slope $\beta$. The median D4000 versus sSFR relation predicted at $z=2$ is in agreement with the local calibration despite a different distribution pattern of galaxies in this plane. In addition, we reproduce the observed $A_{\rm UV}$ versus $\beta$ relation and explore its dependence on galaxy stellar mass, providing an explanation for the observed complexity of this relation. We also find a deficiency in heavily attenuated, UV red galaxies in the simulations. Finally, we provide predictions for the dust attenuation curves of galaxies at $z=2-6$ and investigate their dependence on galaxy colors and stellar masses. The attenuation curves are steeper in galaxies at higher redshifts, with bluer colors, or with lower stellar masses. We attribute these predicted trends to dust geometry. 
Overall, our results are consistent with present observations of high redshift galaxies. Future {\it JWST} observations will further test these predictions.
\end{abstract}



\section{Introduction}

The $\Lambda{\rm CDM}$ model~\citep[e.g.,][]{Planck2016}, as the concordant paradigm of structure formation, provides a framework for the assembly of dark matter haloes where galaxies form. The current theory of galaxy formation~\citep{White1978,Blumenthal1984} makes further testable predictions for observed galaxy populations. Comparing these predictions with observations at multiple epochs of cosmic evolution is important for confirming or falsifying the theory. Despite the well-studied observational constraints in the local Universe, observations of high redshift galaxy populations offer a unique and much less explored channel to test the current theory of structure formation~\citep[see reviews of ][ and references therein]{Shapley2011,Stark2016,Dayal2018}.

Much effort has been devoted to the study of high redshift galaxy populations with both space- and ground-based facilities.  For the selection of galaxies, the Lyman-break technique was extended to $z\gtrsim6$~\citep[e.g.,][]{Bouwens2003,Stanway2003,Yan2003} with the Advanced Camera for Surveys on the {\it Hubble Space Telescope}~(HST). The Wide-Field Camera 3 with near-IR filters on the HST further advanced the search for star-forming galaxies beyond $z=7$~\citep[e.g.,][]{Bouwens2010,Wilkins2010,Oesch2010,Mclure2013,Finkelstein2015}. Currently, deep field extragalatic surveys performed with the HST have enabled measurements of the galaxy rest-frame UV luminosity function to $z\sim 10$~\citep[e.g.,][]{Ishigaki2018,Oesch2018,Bouwens2019}. By combining observations from the HST, the {\it Spitzer Space Telescope} and other ground-based facilities, broadband photometric spectral energy distributions (SEDs) of thousands of galaxies up to $z=8$~\citep[e.g.,][]{Eyles2005,Gonzalez2010,Stark2013,Duncan2014,Song2016} have been measured to study the assembly of galaxy stellar populations at high redshift. Meanwhile, high resolution spectroscopic observations of galaxies have been conducted beyond $z\simeq 2-3$~\citep[e.g., Keck/MOSFIRE:][]{Steidel2014,Kriek2015,Strom2017} and have revealed the physical state of galaxies at high redshift. In addition, far-infrared and sub-millimeter observations revealed a population of highly obscured star-forming galaxies at $z\gtrsim 2$~\citep[e.g.,][]{Blain2002,Solomon2005,Wei2009,Hodge2013,Simpson2014,daCunha2015,Simpson2015} and have expanded our knowledge of galaxy formation and dust physics at high redshift.

Despite these successes, measurements of physical properties of galaxies at high redshift, such as the stellar mass and the star formation rate, are still very uncertain~\citep[e.g.,][]{Papovich2001,Marchesini2009,Behroozi2010,Wuyts2011,Speagle2014,Mobasher2015,Santini2015,Leja2019}. Detection of galaxy emission at longer wavelengths than the rest-frame UV is hindered by the limited wavelength coverage of the HST, the insufficient sensitivity of infrared (IR) instruments, and the atmospheric blocking of ground-based observations. Due to these limitations, the physical properties and evolution of dust, which mainly emits in the IR, are also largely unexplored at high redshift. Dust attenuation generates additional uncertainties in the interpretation of observations. The upcoming {\it James Webb Space Telecope}~\citep[{\it JWST};][]{Gardner2006} will offer access to galaxy optical and IR emission beyond the current redshift frontier. The Near-InfraRed Camera (NIRCam) and the Mid-InfraRed Instrument (MIRI) on {\it JWST} are designed to obtain broadband photometry over the wavelength range from $0.7$ to $25.5\micron$ with unprecedented sensitivity. {\it JWST} will extend observations of the galaxy rest-frame UV luminosity function to higher redshift ($z>10$) and fainter luminosity. It will also allow detection of optical and near-infrared (NIR) SEDs of high redshift galaxies. The Near-InfraRed Spectrograph (NIRSpec) will enable high resolution spectroscopy of distant galaxies and precise measurements of spectral indices, such as emission line luminosities, line ratios, and breaks in the continuum flux. These instrumental advances will provide new insights in both the formation and evolution of high redshift galaxy populations and dust physics.

In parallel to the observational efforts, various theoretical works have made predictions for high redshift galaxy populations. This has included empirical models~\citep[e.g.,][]{Tacchella2013,Mason2015,Tacchella2018}, semi-analytical models for galaxy formation~\citep[e.g.,][]{Cowley2018,Yung2019b,Yung2019a}, and hydrodynamical simulations. For example: the First Billion Years simulation suite~\citep[e.g.,][]{Paardekooper2013}, the BlueTides simulation~\citep[e.g.,][]{Wilkins2016,Wilkins2017}, the Renaissance simulation suite~\citep[e.g.,][]{Xu2016,Barrows2017}, the FirstLight simulation~\citep[e.g.,][]{Ceverino2017}, the FIRE-2 simulation suite~\citep[e.g.,][]{Ma2018,Ma2019}, the Sphinx simulation~\citep{Rosdahl2018} and others~\citep[e.g.][]{Dayal2013,Shimizu2014,Wu2019}. However, we note that most of these hydrodynamical simulations have only been evolved down to relatively high redshift due to computational limitations. Therefore, the reliability of the high redshift predictions from these works has not been fully justified~\footnote{The FIRE-2 simulation suite contain simulations, different from the FIRE-2 high redshift simulations~\citep{Ma2018,Ma2019}, that have been evolved to $z=0$ using the same physics model~\citep{Hopkins2018}. So its local Universe predictions have been tested to some level.}, given that these models cannot be tested at lower redshift where rich observational constraints are available. To fill this gap, we present high redshift predictions from the IllustrisTNG simulation suite. The simulations have successfully produced a realistic low redshift Universe that is consistent with a wide range of observations. Our method combines the simulations with full Monte Carlo radiative transfer calculations to model dust attenuation. Moreover, the resolution corrections and the combination procedure introduced in \citetalias{Vogelsberger2019}~\citep{Vogelsberger2019} of this series enable predictions for galaxies over a wide dynamical range from $10^{7}$ to $10^{12}\msun$. The technique fully utilizes the high resolution of TNG50 and the  statistical power of TNG300. 

\begin{table*}
\begin{tabular}{llccrrrccccc}
\hline
{\bf IllustrisTNG Simulation} & run & {volume side length} & $N_{\rm gas}$  & $N_{\rm dm}$ & $m_{\rm b}$ & $m_{\rm dm}$ & $\epsilon_{\rm dm,stars}$ &  $\epsilon_{\rm gas}^{\rm min}$ \\
 & &   $[h^{-1}{\rm Mpc}]$  & & &  $[h^{-1}{\rm M}_\odot]$ & $[h^{-1}{\rm M}_\odot]$ &  $[h^{-1}{\rm kpc}]$ & $[h^{-1}{\rm kpc}]$\\
\hline
\hline
{\bf TNG300}   &  TNG300(-1)  & 205  & $2500^3$  & $2500^3$   & $7.4\times 10^6$   & $4.0\times 10^7$   & 1.0 & 0.25\\
\hline
{\bf TNG100}   &  TNG100(-1)  & 75   & $1820^3$  & $1820^3$   & $9.4\times 10^5$   & $5.1\times 10^6$   & 0.5 & 0.125 \\
\hline
{\bf TNG50}    &  TNG50(-1)   & 35   & $2160^3$  & $2160^3$   & $5.7\times 10^4$   & $3.1\times 10^5$   & 0.2 & 0.05 \\

\hline
\end{tabular}
\caption{{\bf IllustrisTNG simulation suite.} The table shows the basic numerical parameters of the three primary IllustrisTNG simulations: simulation volume side length, number of gas cells ($N_{\rm gas}$), number of dark matter particles ($N_{\rm dm}$), baryon mass resolution ($m_{\rm b}$), dark matter mass resolution ($m_{\rm DM}$), Plummer-equivalent maximum physical softening length of dark matter and stellar particles ($\epsilon_{\rm dm,stars}$), and the minimal comoving cell softening length $\epsilon_{\rm gas}^{\rm min}$. In the following we will refer to TNG50-1 as TNG50, TNG100-1 as TNG100 and TNG300-1 as TNG300. 
\label{tab:tabsims}}
\end{table*}

In \citetalias{Vogelsberger2019} of this series, technical details of our radiative transfer post-processing procedure were introduced. Our dust attenuation model was calibrated based on the galaxy rest-frame UV luminosity functions at $z=2-10$. The predictions for the {\it JWST} apparent bands' luminosity functions and number counts were presented in \citetalias{Vogelsberger2019}. As a subsequent analysis, the goal of this paper is to make further predictions for high redshift galaxy populations that involve: the $M_{\rm halo}-M_{\rm UV}$ relation, galaxy emission line luminosity functions, the Balmer break at $4000\text{\AA}$ (D4000), the UV continuum slope $\beta$ and its role as a dust attenuation indicator, and the dust attenuation curves. The comparisons of these predictions with existing observations demonstrate that our results are a solid guide for future observations conducted with {\it JWST}. 

This paper is organized as follows: In Section~\ref{sec:sim}, we briefly describe the IllustrisTNG simulation suite and state the numerical parameters of TNG50, TNG100 and TNG300 in detail. In Section~\ref{sec:method}, we describe the method we used to derive the dust attenuated broadband photometry and SEDs of galaxies in the simulations. In Section~\ref{sec:res_com}, we briefly introduce the resolution correction and the combination procedure. The main results are presented in Section~\ref{sec:results}, where we make various predictions and comparisons with observations. Summary and conclusions are presented in Section~\ref{sec:conclusions}.

\section{Simulation}
\label{sec:sim}
The analysis in this paper is based on the IllustrisTNG simulation suite~\citep[][]{Marinacci2018, Naiman2018, Nelson2018, Pillepich2018b, Springel2018}, including the newest addition, TNG50~\citep[][]{Nelson2019, Pillepich2019}. As the follow-up project to the Illustris simulations~\citep{Vogelsberger2014a, Vogelsberger2014b, Genel2014, Nelson2015, Sijacki2015}, the IllustrisTNG simulation suite consists of three primary simulations: TNG50, TNG100 and TNG300 \citep{Nelson2019b}, covering three different periodic, uniformly sampled volumes, roughly ${\approx 50^3}, 100^3, 300^3\,{\rm Mpc}^3$. The numerical parameters of these simulations are summarized in Table~\ref{tab:tabsims}. The IllustrisTNG simulation suite employs the following cosmological parameters~\citep[][]{Planck2016}: $\Omega_{\rm m} = 0.3089$, $\Omega_{\rm b} = 0.0486$, $\Omega_{\Lambda} = 0.6911$, $H_0 = 100\,h\,\kms \Mpc^{-1} = 67.74\,\kms \Mpc^{-1}$, $\sigma_{8} = 0.8159$, and $n_{\rm s} = 0.9667$. The simulations were carried out with the moving-mesh code {\sc Arepo} \citep{Springel2010,Pakmor2016} combined with the IllustrisTNG galaxy formation model~\citep[][]{Weinberger2017,Pillepich2018a} which is an update of the Illustris galaxy formation model~\citep{Vogelsberger2013, Torrey2014}. We note that TNG50, TNG100, and TNG300 differ in their highest numerical resolution as listed in Table~\ref{tab:tabsims}. In the following we will refer to TNG50-1 as TNG50, TNG100-1 as TNG100 and TNG300-1 as TNG300.

\section{Galaxy luminosities and SEDs}
\label{sec:method}
In this section, we will introduce our method of deriving the intrinsic and dust attenuated galaxy luminosities and SEDs. We will follow the Model C procedure introduced in \citetalias{Vogelsberger2019}. We will briefly review the procedure and refer the readers to \citetalias{Vogelsberger2019} for details. We assigned intrinsic emission to stellar particles in the simulations using the stellar population synthesis method and post-process the intrinsic emission from stellar particles with full Monte Carlo radiative transfer calculations to model dust attenuation. To be specific, we adopt the Flexible Stellar Population Synthesis ({\sc Fsps}) code~\citep{Conroy2009,Conroy2010} to model the intrinsic SEDs of old stellar particles (age > 10\Myr) and the {\sc Mappings-\Rmnum{3}} SED library~\citep{Groves2008} to model those of young stellar particles (age < 10\Myr). The {\sc Mappings-\Rmnum{3}} SED library self-consistently considers the dust attenuation in the birth clouds of young stars which cannot be properly resolved in the simulations. In theory, sufficiently ionized gas cells resolved in the simulations should also have some emissivities. However, self-consistently modeling this effect is beyond the scope of this paper and we did not include it in the radiative transfer post-processing. After characterizing the source, we perform the full Monte Carlo dust radiative transfer calculations using a modified version of the publicly available {\sc Skirt} (version 8)~\footnote{\href{http://www.skirt.ugent.be/root/_landing.html}{http://www.skirt.ugent.be/root/\_landing.html}} code~\citep{Baes2011,Camps2013,Camps2015,Saftly2014}. Modifications were made to incorporate the {\sc Fsps} SED templates into {\sc Skirt}. Photon packages are then randomly released based on the source distribution characterized by the positions and SEDs of stellar particles. We have designed a specific wavelength grid for the radiative transfer calculations. The grid has $1168$ grid points covering the wavelength range from $0.05\micron$ to $5\micron$ and has been refined around emission lines. 

The emitted photon packages will further interact with the dust in the interstellar medium (ISM). We assume a \citet{Draine2007} dust mixture of amorphous silicate and graphitic grains, including varying amounts of Polycyclic Aromatic Hydrocarbons (PAHs) particles. This dust model is widely used~\citep[e.g.,][]{Jonsson2010,Baes2011,Kimm2013,RemyRuyer2014} and it can reproduce the extinction properties in the Milky Way. Since we have fixed the dust model, the predictions we make in this paper will not cover the potential evolution of the physical properties of dust at high redshift. Instead, we will focus on the influence of dust geometry on dust attenuation. To determine the distribution of dust in the ISM, we consider cold gas cells~(star-forming or with temperature $<8000\,{\rm K}$) from the simulations and calculate the metal mass distribution based on their metallicities. We assume that dust is traced by metals in the ISM and adopt a constant dust-to-metal ratio among all galaxies at a fixed redshift. The dust-to-metal ratios at different redshifts are calibrated based on the galaxy rest-frame UV luminosity functions at $z=2-10$~\citep[e.g.,][]{Ouchi2009,McLure2009,Bouwens2016,Finkelstein2016,Oesch2018}. We find that the redshift evolution of the dust-to-metal ratio can be well characterized by a $0.9\times (z/2)^{-1.92}$ power-law at $z=2-10$ (see \citetalias{Vogelsberger2019} for the calibration procedure and a detailed discussion of the redshift dependence of the dust-to-metal ratio). We then turn the metal mass distribution into a dust mass distribution with the dust-to-metal ratio and map the dust distribution onto an adaptively refined grid. The grid is refined with an octree algorithm to maintain the fraction of dust mass within each grid cell to be smaller than $2\times10^{-6}$. The maximum refinement level is also adjusted to match the numerical resolution of the simulations. 

Ultimately, after photons fully interacted with dust in the galaxy and escaped, they are collected by a mock detector $1\pMpc$~\footnote{Physical $\Mpc$.} away from the simulated galaxy along the positive z-direction of the simulation coordinates. The integrated galaxy flux is then recorded, which provides us with the dust attenuated SED of the galaxy in the rest frame. Galaxy SEDs without the resolved dust attenuation are derived in the same way with no resolved dust distribution imported. We note that we use the term ``without the resolved dust'' because the unresolved dust component in the {\sc Mappings-\Rmnum{3}} SED library is always present. Compared with the resolved dust attenuation, the impact of the unresolved dust attenuation on galaxy continuum emission is limited. In order to get broadband photometry, the galaxy SEDs are convolved with the transmission curves using the {\sc Sedpy}~\footnote{\href{https://github.com/bd-j/sedpy}{https://github.com/bd-j/sedpy}} code. For the calculation of apparent magnitudes, the rest-frame flux is redshifted, corrected for the IGM absorption~\citep{Madau1995,Madau1996} and converted to the observed spectra.

Since some predictions in this paper are concerning nebular emission line luminosities, we need to discuss the nebular emission model adopted in the {\sc Mappings-\Rmnum{3}} SED library in \citet{Groves2008} in more detail. The SED library adopts one dimensional photoionization and radiative transfer calculations to model the radiation from a newly formed massive stellar cluster. The radiation propagates through the surrounding spherically symmetrical H\Rmnum{2} region and photodissociation region (PDR). The covering fraction of the PDR decreases as a function of time as it is cleared away by the strong winds from the massive stellar cluster. The {\sc Mappings-\Rmnum{3}} SED of a specific young stellar particle (age < 10\Myr) is controlled by five parameters that we have chosen in the following way: {(\rmnum{1})} {\it Star Formation Rate:} The star formation rate (SFR) is required since the {\sc Mappings-\Rmnum{3}} SED templates are normalized to have $\rm{SFR}=1\msun/\rm yr$. We assume $\rm{SFR}=m_{\rm i}/10\Myr$ to maintain the conservation of stellar mass, where $m_{\rm i}$ is the initial mass of the young stellar particle. {(\rmnum{2})} {\it Metallicity of the birth cloud:} The metallicity of the birth cloud is assumed to be the initial metallicity of the stellar particle inherited from its parent gas cell(s). This represents the metallicity of the birth environment of the stellar particle. {(\rmnum{3})} {\it Pressure of the surrounding ISM:} Tracing the pressure of the stellar particle's birth environment is technically difficult in post-processing. However, as demonstrated in \citet{Groves2008}, the impact of the ISM pressure on the UV to NIR continuum spectra is negligible. We therefore simply employ a fixed typical value of $\log{(P_{0}/k_{\rm B})/\cm^{-3}\rm{K}}=5$~\citep{Groves2008,Rodriguez2018}. The pressure does have an influence on the emission line luminosities. However, we have found that varying the pressure by an order of magnitude from the fiducial value will only lead to $\lesssim 0.1\,{\rm dex}$ differences in the luminosities of the emission lines we study in this paper. {(\rmnum{4})} {\it The compactness of the birth cloud:} This describes the density of the {H\small\Rmnum{2}} region and can be calculated as $\log{C}=3/5\log{M_{\rm cl}/\msun}+2/5\log{P_{0}/k_{\rm B}/(\rm{cm}^{-3}\rm K)}$~\citep{Groves2008}, where $M_{\rm cl}$ is the mass of the stellar cluster and $P_{0}$ is the pressure of the {H\small\Rmnum{2}} region. Again, correctly deriving the properties of the stellar cluster is difficult in post-processing and may require artificial resampling. However, the compactness parameter mainly affects the temperature of dust and its emission in the FIR. It has almost no impact on the UV to NIR photometric properties~\citep{Groves2008}. We therefore also fix this to a typical value of $\log{C}=5$~\citep{Groves2008,Rodriguez2018}. For the emission line luminosities, varying the compactness by an order of magnitude from the fiducial value leads to $\lesssim 0.1\,{\rm dex}$ differences in the ${\rm H}_{\alpha}$ and ${\rm H}_{\beta}$ luminosities and $\lesssim 0.25\,{\rm dex}$ difference in the ${[\rm O\,\Rmnum{3}]}$ luminosity. {(\rmnum{5})} {\it Photodissociation region covering fraction:} This defines the time-averaged fraction of the stellar cluster solid angle that is covered by the PDR. The {\sc Mappings-\Rmnum{3}} SED library consists of SEDs with fully covered or uncovered {H\small\Rmnum{2}} regions. The final output SED is a linear combination of these two extreme cases controlled by the parameter $f_{\rm PDR}$. We adopt here the fiducial value $0.2$ for the $f_{\rm PDR}$~\citep[][]{Groves2008,Jonsson2010,Rodriguez2018}. These parameters then fully specify the {\sc Mappings-\Rmnum{3}} SED including the nebular continuum and line emission as well as the unresolved dust attenuation in the birth clouds of young stars. This model allows us to make robust predictions for the emission line luminosities within the uncertainties quoted. As stated in \citetalias{Vogelsberger2019}, at $z\geq6$, the young stellar populations were modelled by the {\sc Fsps} SED instead of the {\sc Mappings-\Rmnum{3}} SED. The nebular emission at these redshifts is therefore also modelled by the {\sc Fsps} code with default parameter settings. As illustrated in the calibration procedure of \citetalias{Vogelsberger2019}, this switch removes the unresolved dust component at $z\geq6$ where its impact should be limited~\footnote{The calibrated dust-to-metal ratio of the resolved dust component drops at higher redshift, so the unresolved dust attenuation should presumably also become weaker at higher redshift.}, and leads to a better agreement in the rest-frame UV luminosity function with observations at $z\geq6$.

As discussed in \citetalias{Vogelsberger2019}, we define a galaxy as being either a central or satellite galaxy identified by the SUBFIND algorithm~\citep{Springel2001,Dolag2009}. In our analysis, we only consider galaxies with stellar mass larger than $100$ times the baryonic mass resolution of the simulation, $100 \times m_{\rm b}$, within twice the stellar half mass radius. Galaxies below this mass threshold will not be considered in the following analysis since we assume that their structures are not reliably modelled. Furthermore, galaxy luminosities, magnitudes, stellar masses and star formation rates are all calculated within a fixed physical aperture of $30\pkpc$~\citep[e.g.,][]{Pillepich2018b} and are based on gravitationally bound particles and cells only.

\begin{figure*}
\includegraphics[width=0.246\textwidth]{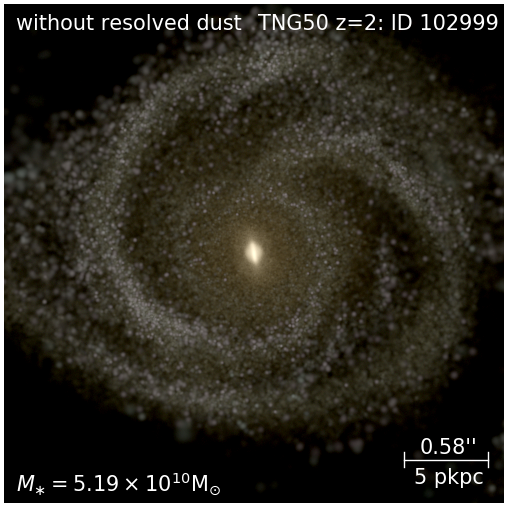}
\includegraphics[width=0.246\textwidth]{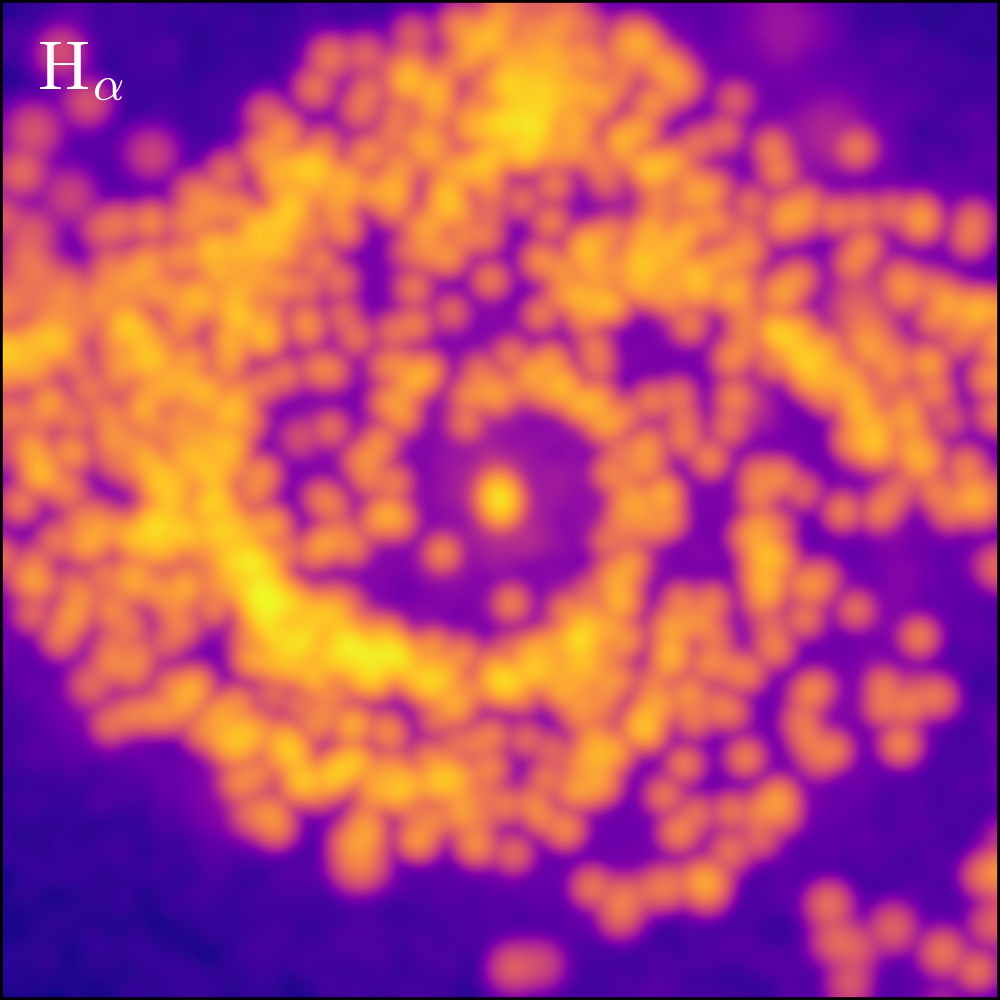}
\includegraphics[width=0.246\textwidth]{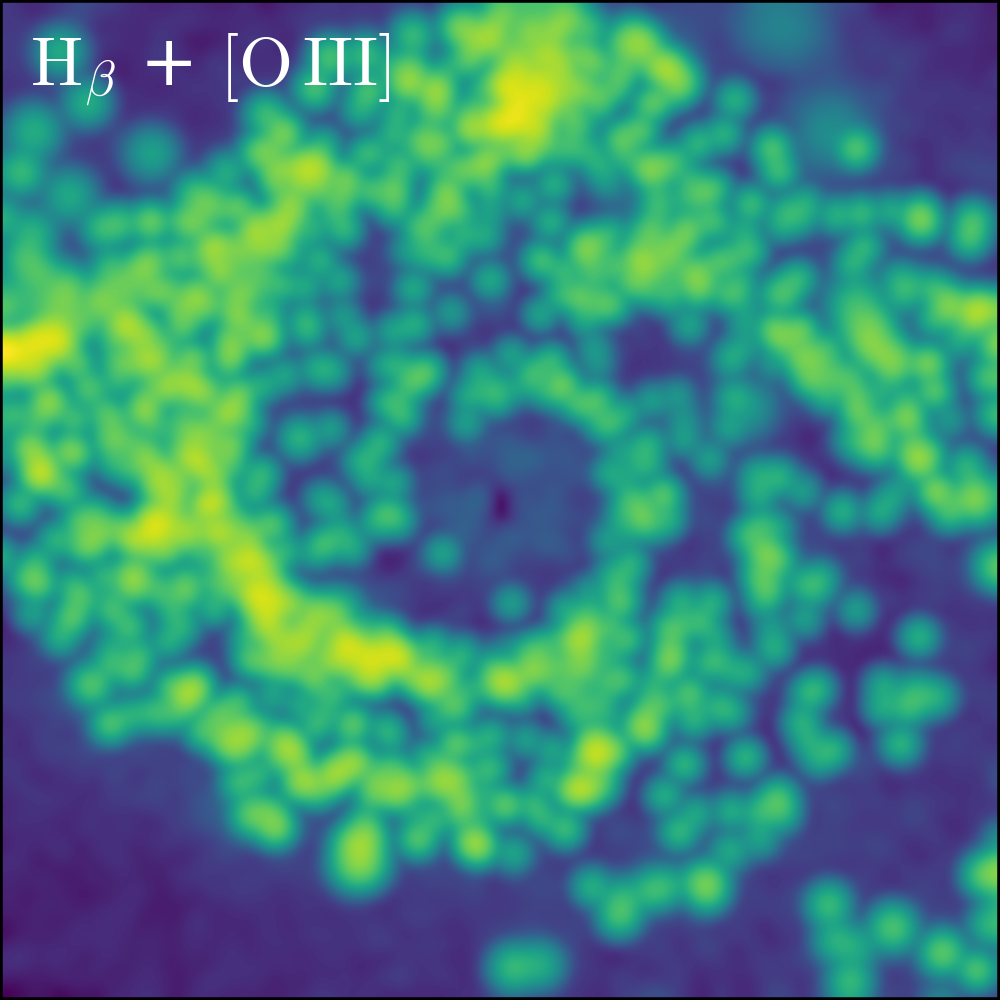}
\includegraphics[width=0.246\textwidth]{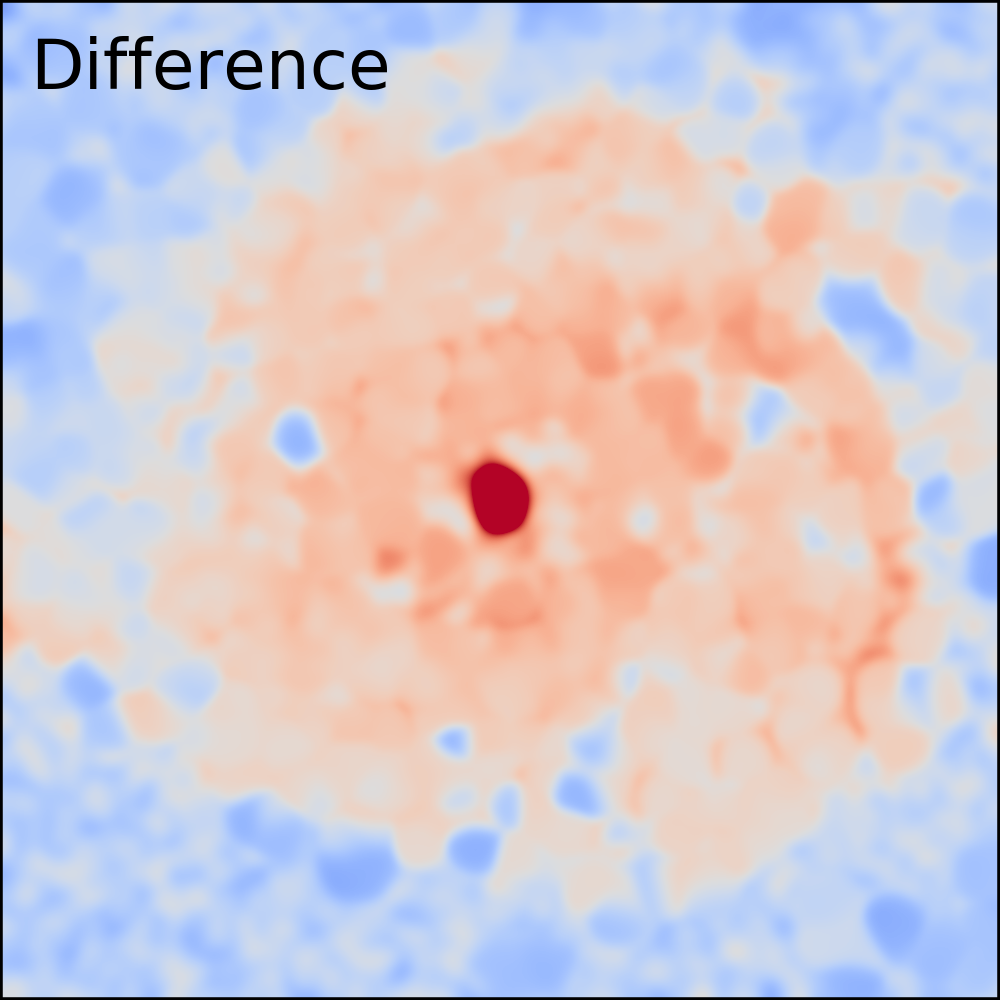}\\
\includegraphics[width=0.246\textwidth]{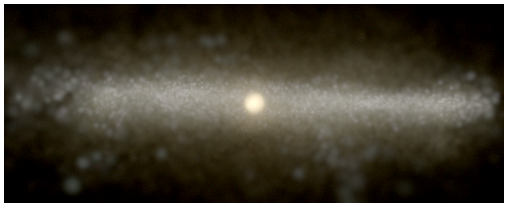}
\includegraphics[width=0.246\textwidth]{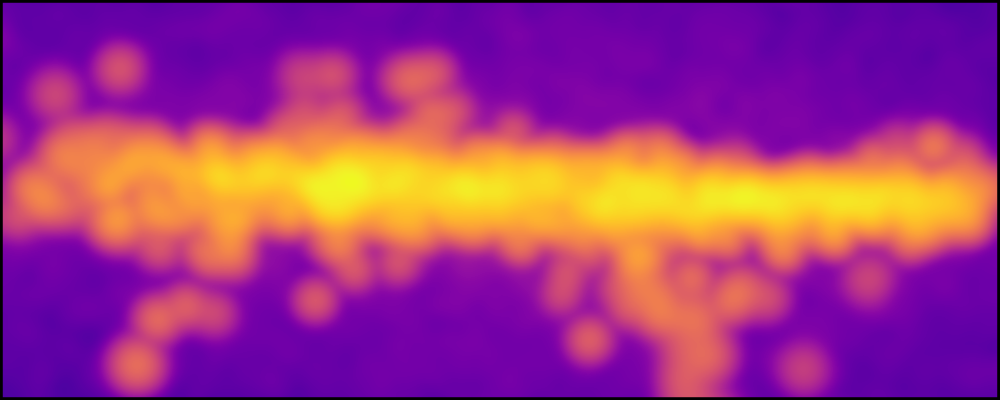}
\includegraphics[width=0.246\textwidth]{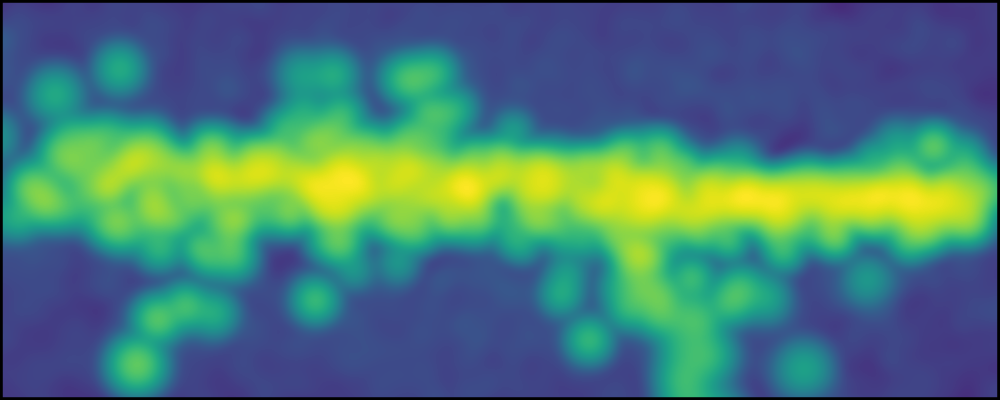}
\includegraphics[width=0.246\textwidth]{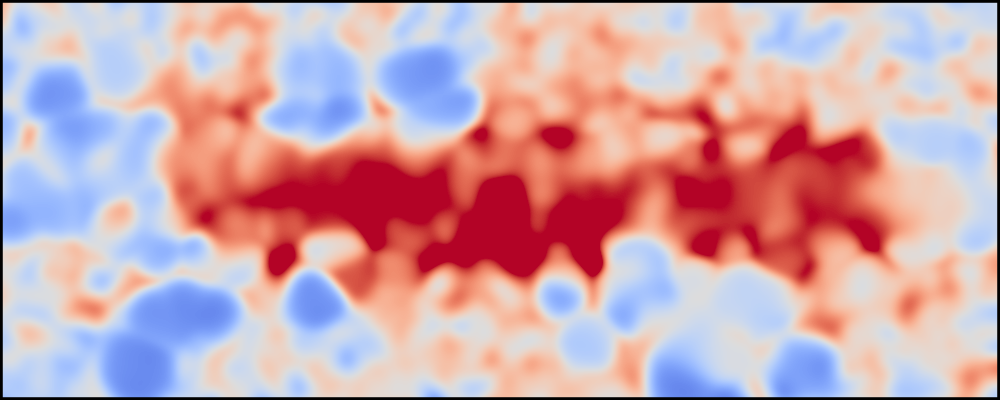}\\

\includegraphics[width=0.246\textwidth]{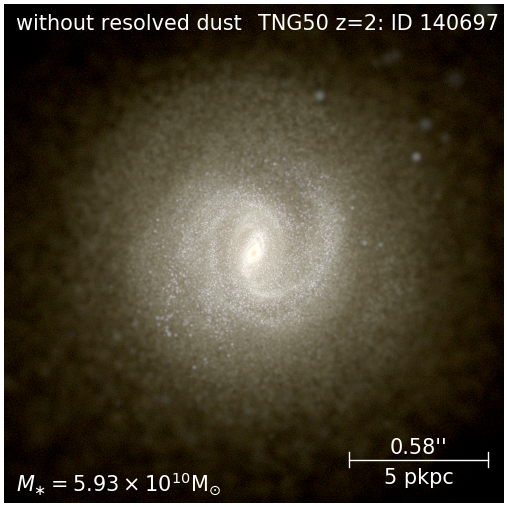}
\includegraphics[width=0.246\textwidth]{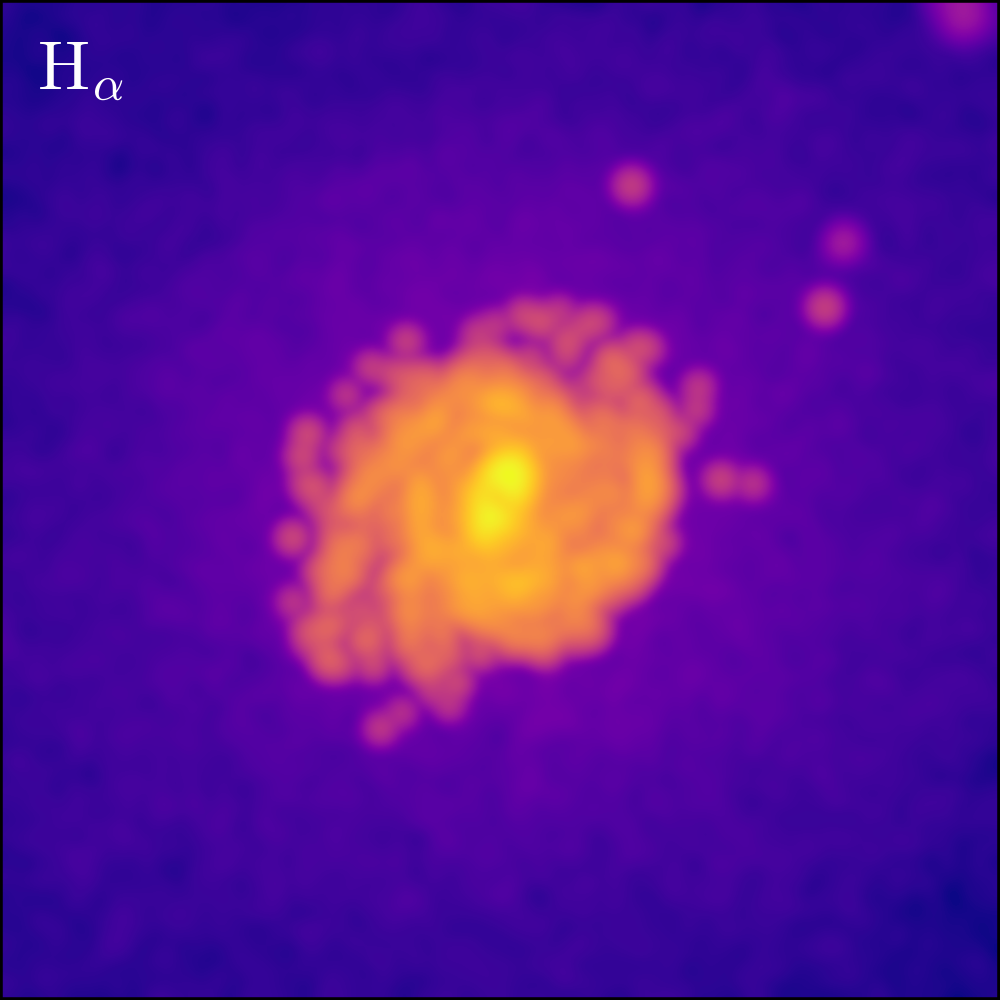}
\includegraphics[width=0.246\textwidth]{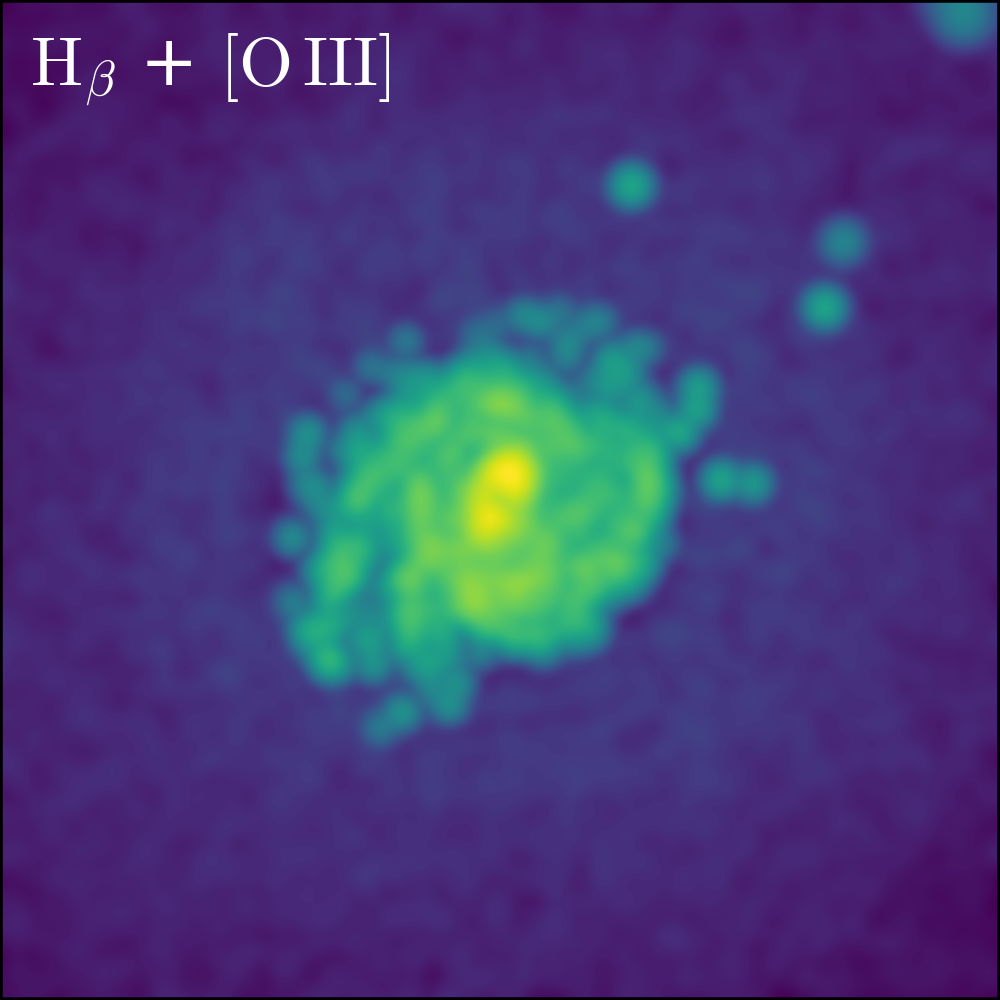}
\includegraphics[width=0.246\textwidth]{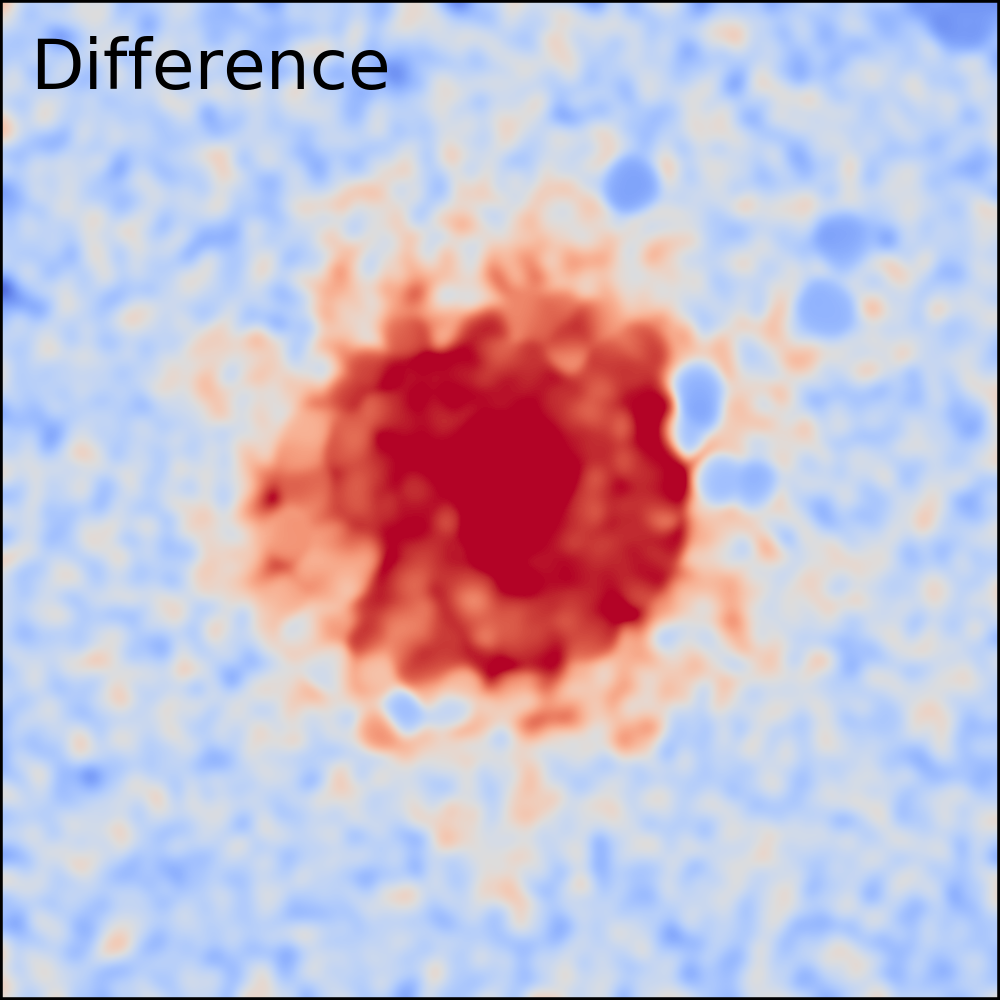}\\
\includegraphics[width=0.246\textwidth]{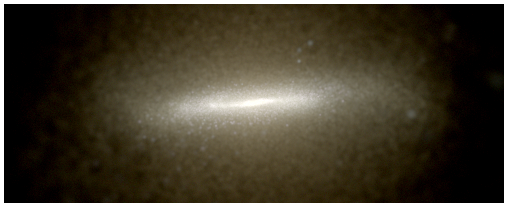}
\includegraphics[width=0.246\textwidth]{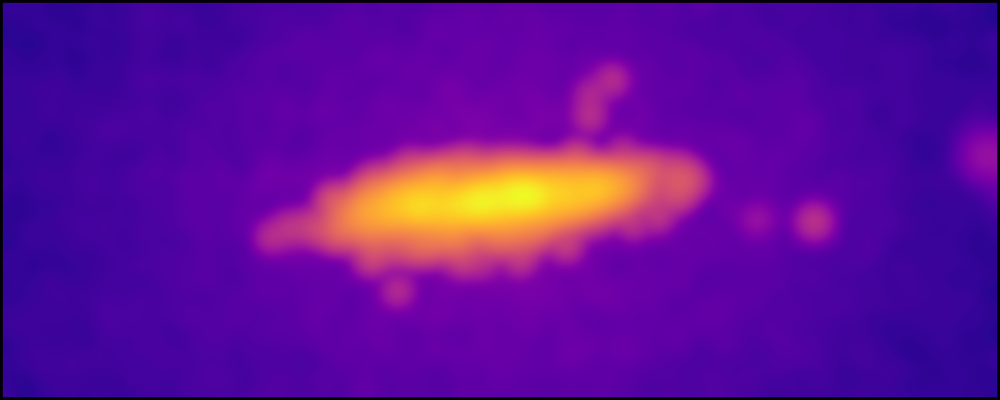}
\includegraphics[width=0.246\textwidth]{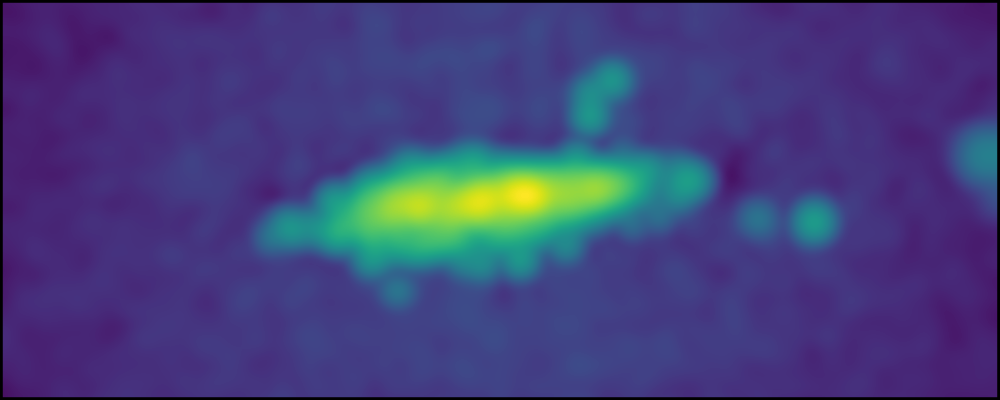}
\includegraphics[width=0.246\textwidth]{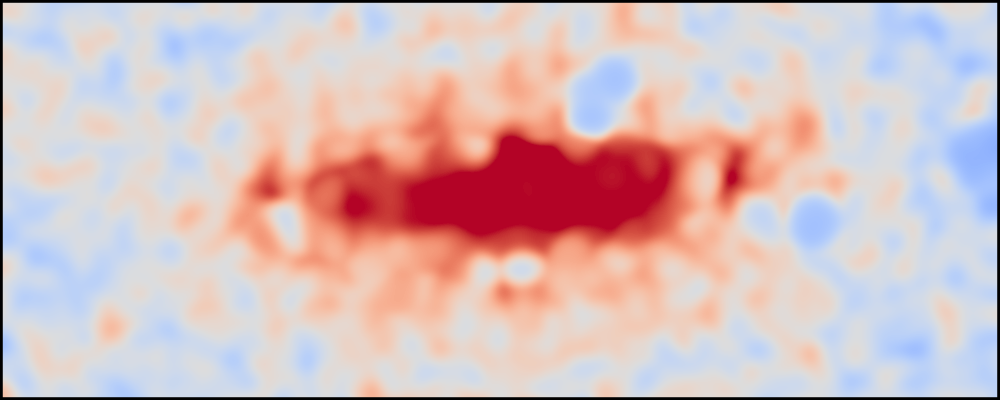}\\

\includegraphics[width=0.246\textwidth]{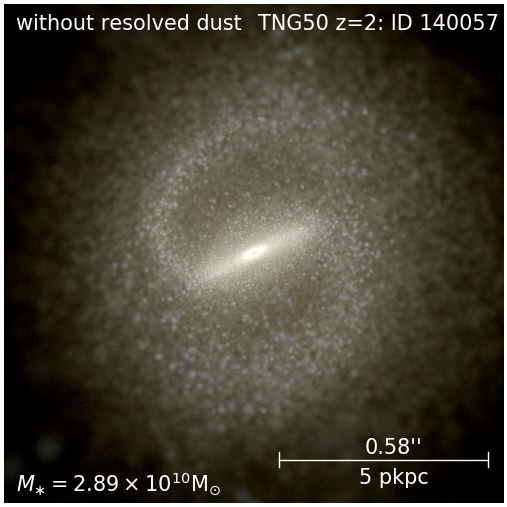}
\includegraphics[width=0.246\textwidth]{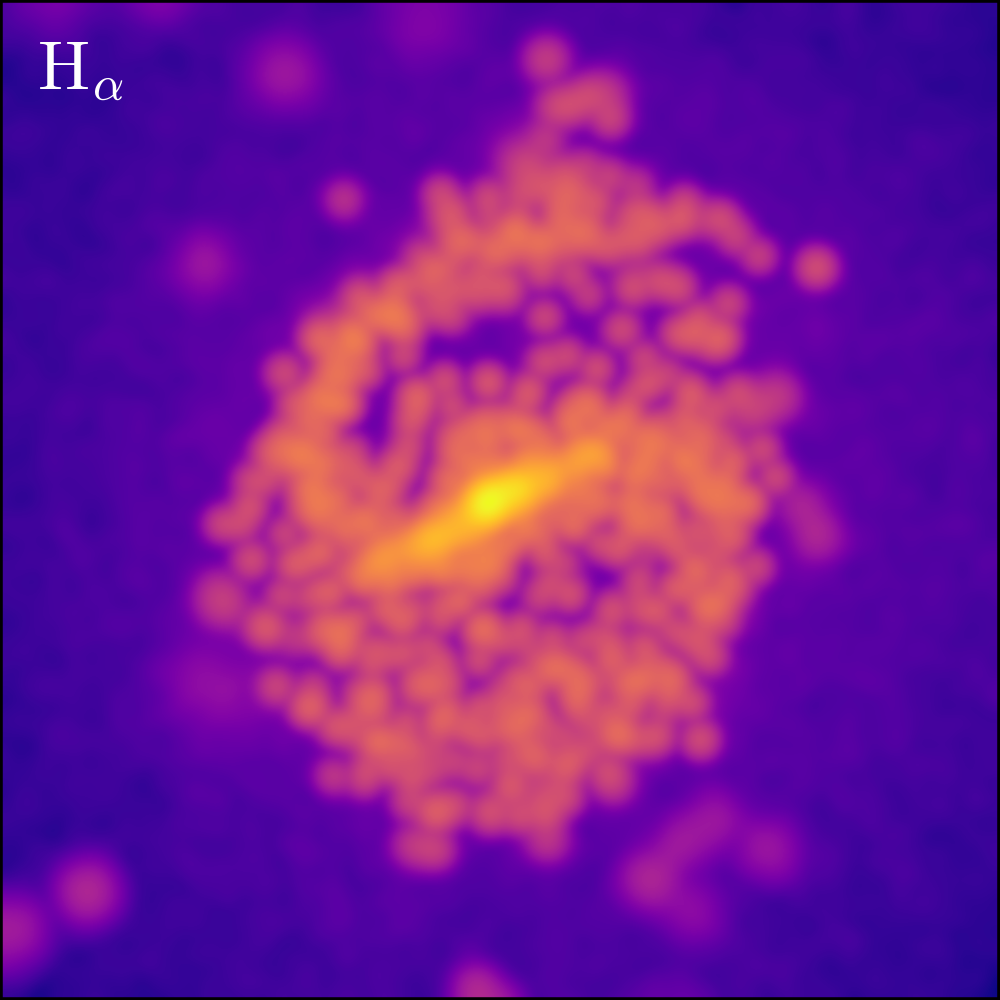}
\includegraphics[width=0.246\textwidth]{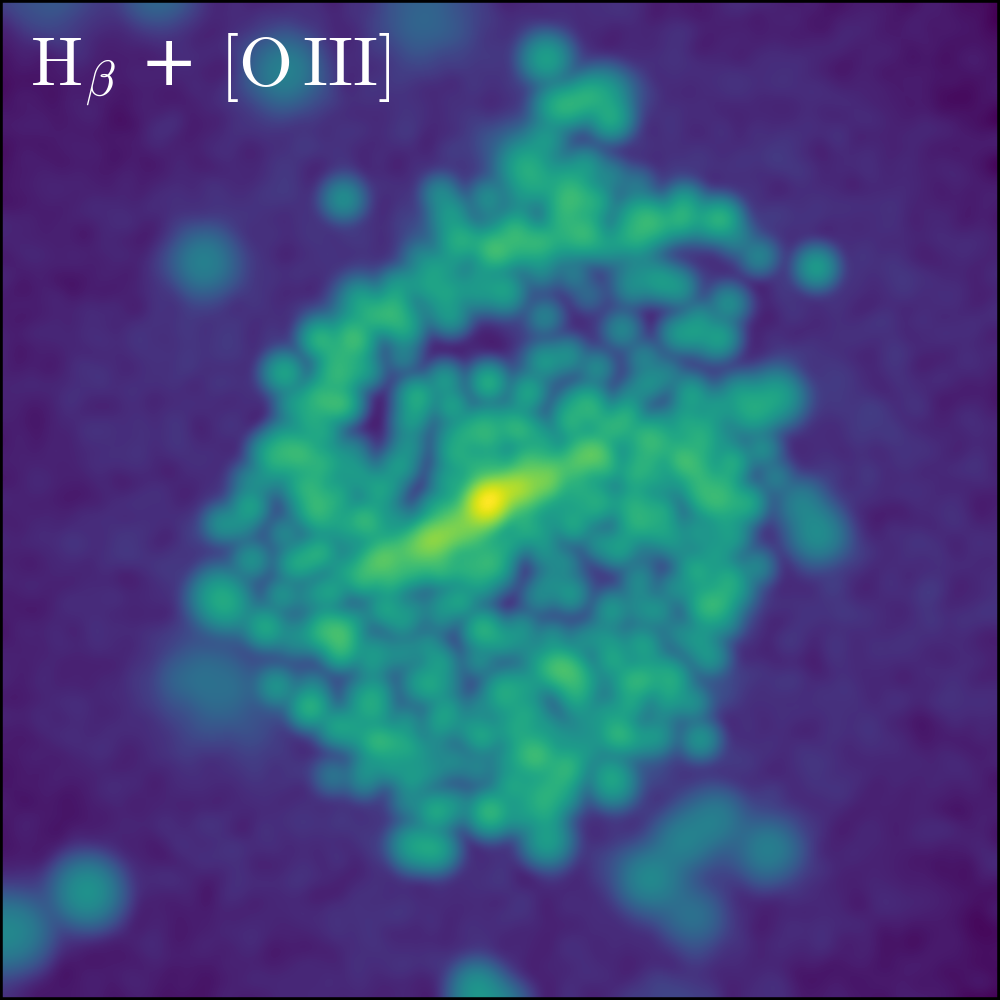}
\includegraphics[width=0.246\textwidth]{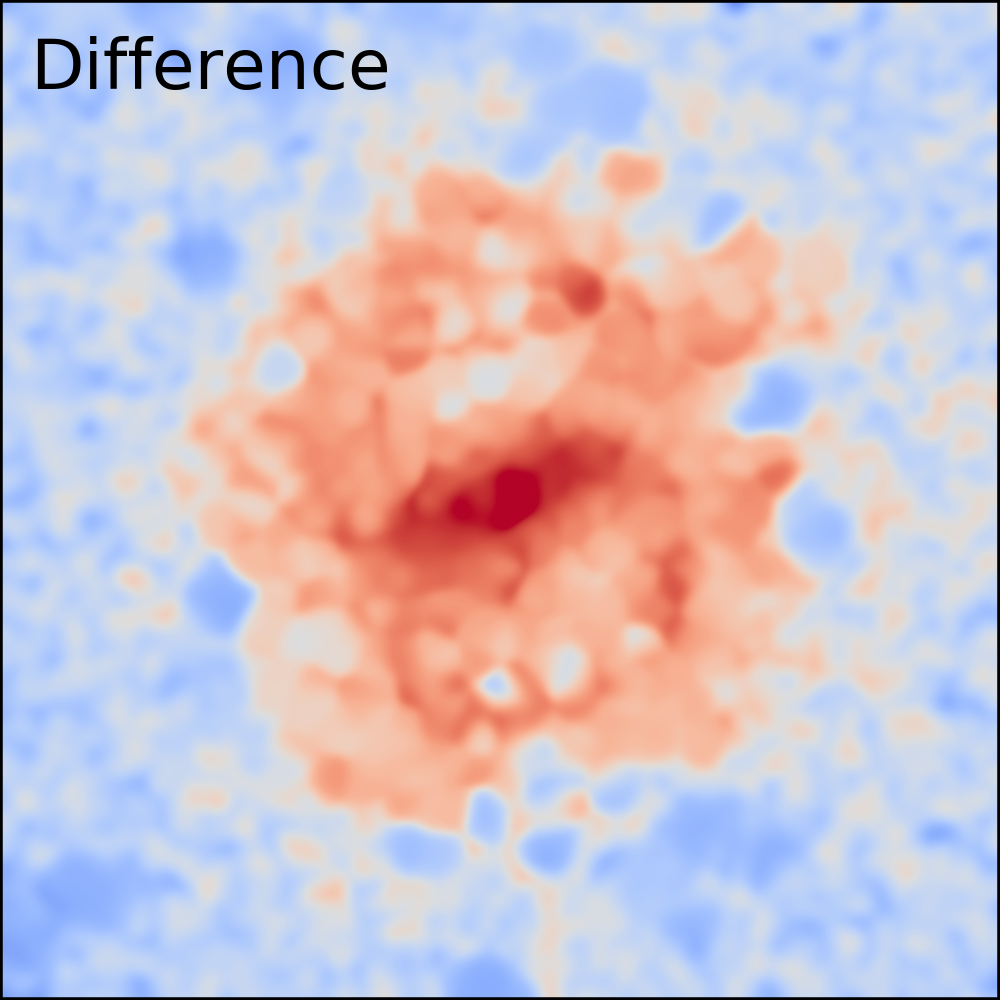}\\
\includegraphics[width=0.246\textwidth]{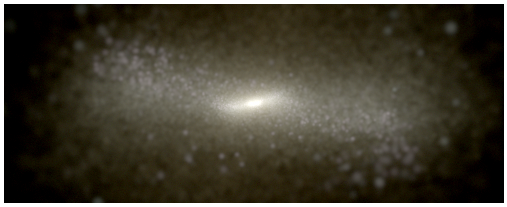}
\includegraphics[width=0.246\textwidth]{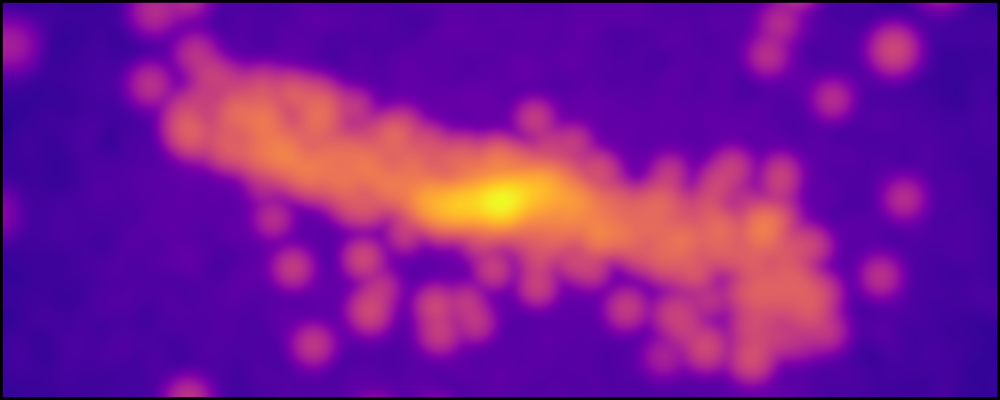}
\includegraphics[width=0.246\textwidth]{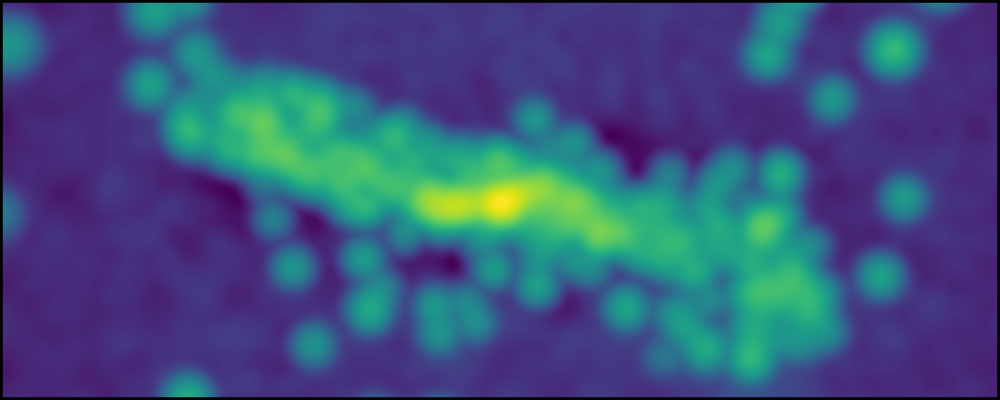}
\includegraphics[width=0.246\textwidth]{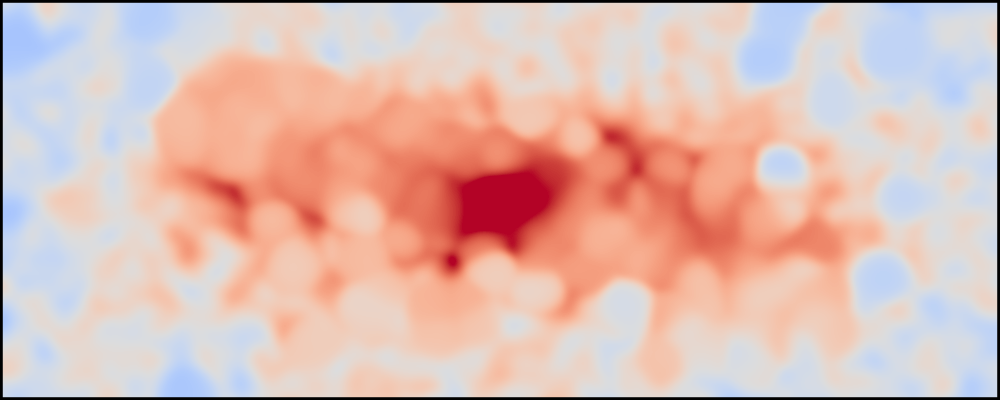}\\

\caption{ {\bf Face-on and edge-on images of selected TNG50 galaxies at $\mathbf{z=2}$.} {\it Left column:} {\it JWST} NIRCam F115W, F150W, F200W bands synthetic images~(without point spread and noise); {\it Middle left (right) column:} emission line strength maps of ${\rm H}_{\alpha}$ (${\rm H}_{\beta} + {[\rm O\,\Rmnum{3}]}$). The maps have been smoothed with a Gaussian kernel to alleviate artificial clumpiness. The width of the kernel is roughly the angular resolution of the {\it JWST} NIRCam, $\sim 0.03-0.06\,{\rm arcsec}$; {\it Right column:} Maps of relative differences in the emission line strength of ${\rm H}_{\alpha}$ and ${\rm H}_{\beta} + {[\rm O\,\Rmnum{3}]}$, $(\log{L_{{\rm H}_{\beta} + {[\rm O\,\Rmnum{3}]}}}-\log{L_{{\rm H}_\alpha}}) - (\log{L_{{\rm H}_{\beta} + {[\rm O\,\Rmnum{3}]}}}-\log{L_{{\rm H}_\alpha}})_{\rm mean}$. Red regions indicate stronger ${\rm H}_{\alpha}$ emission and blue regions indicate stronger ${\rm H}_{\beta} + {[\rm O\,\Rmnum{3}]}$ emission~(note that ${\rm H}_{\alpha}$ is generally stronger than ${\rm H}_{\beta} + {[\rm O\,\Rmnum{3}]}$ in these galaxies, so we have already subtracted the mean difference before making this comparison). We find that ${\rm H}_{\alpha}$ is relatively stronger in the star-forming regions, e.g. spiral arms and disks, while ${\rm H}_{\beta} + {[\rm O\,\Rmnum{3}]}$ is relatively stronger at the outskirts of the galaxies. Such phenomena are mainly driven by the difference in the metallicity of star-forming gas in these regions. As we will show, the difference in the spatial origins of the emission lines will have a significant impact on the dust attenuation of the lines. All the images shown here do not include the resolved dust attenuation.}
\label{fig:image1}
\end{figure*}

\begin{figure*}
\includegraphics[width=0.49\textwidth]{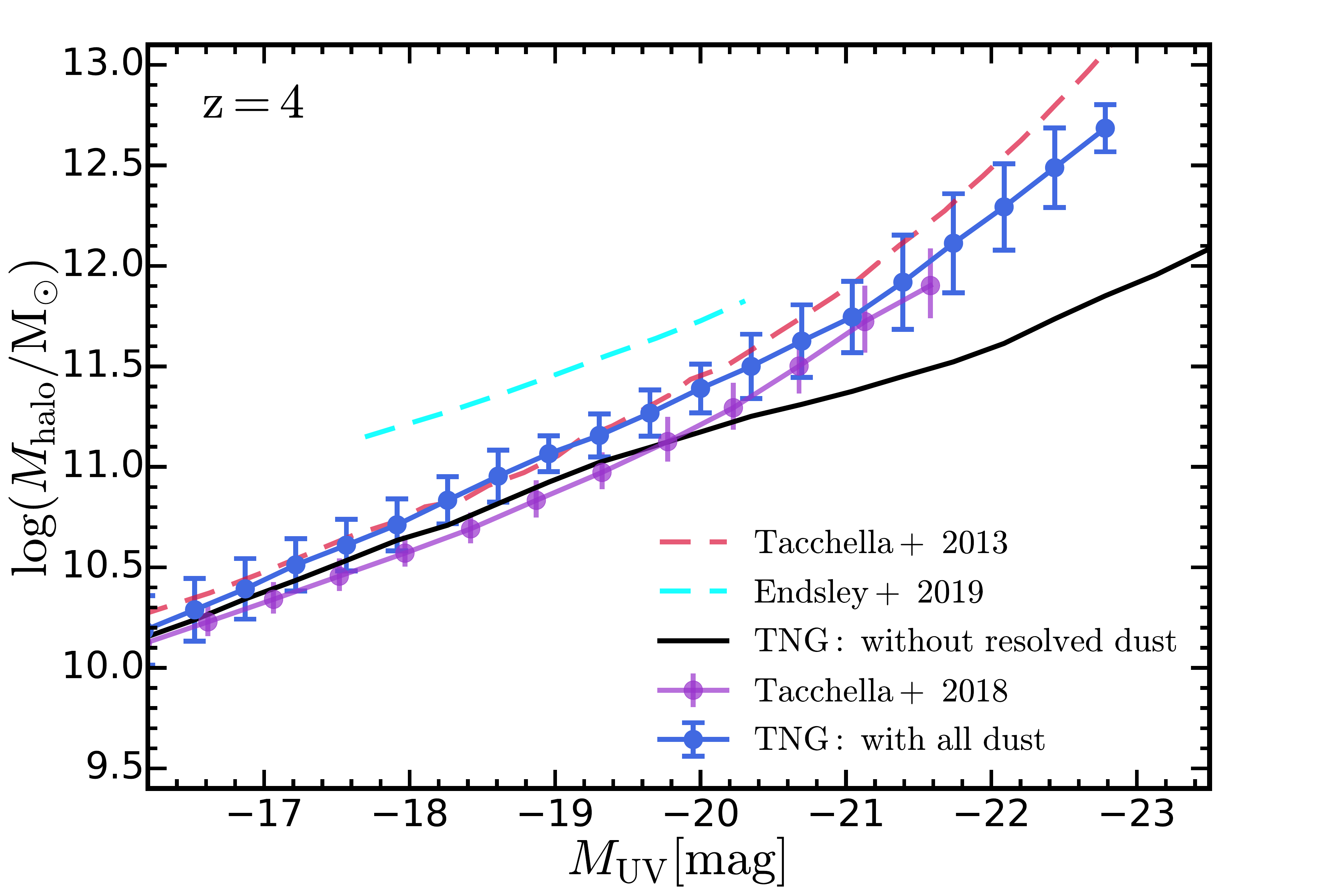}
\includegraphics[width=0.49\textwidth]{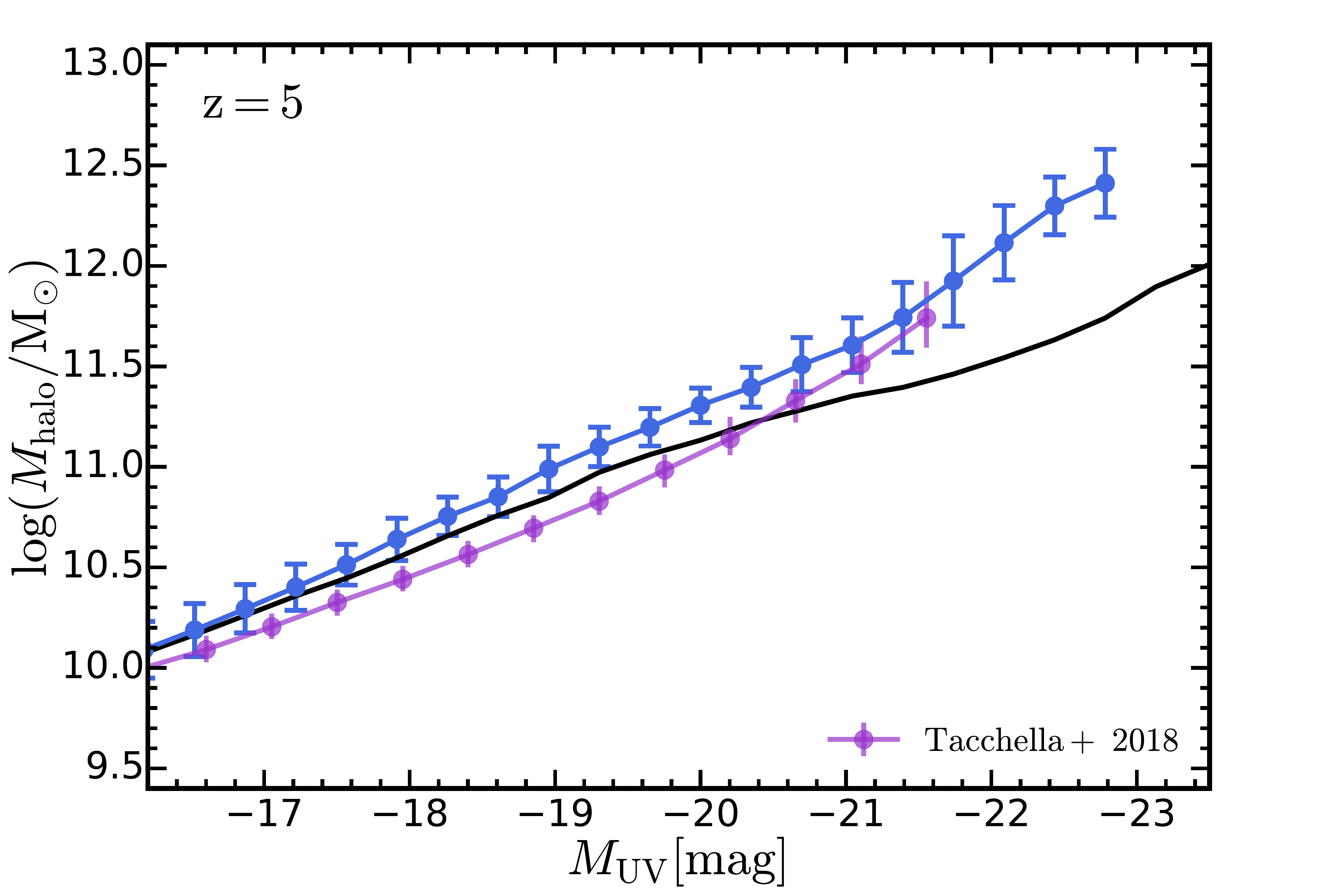}
\includegraphics[width=0.49\textwidth]{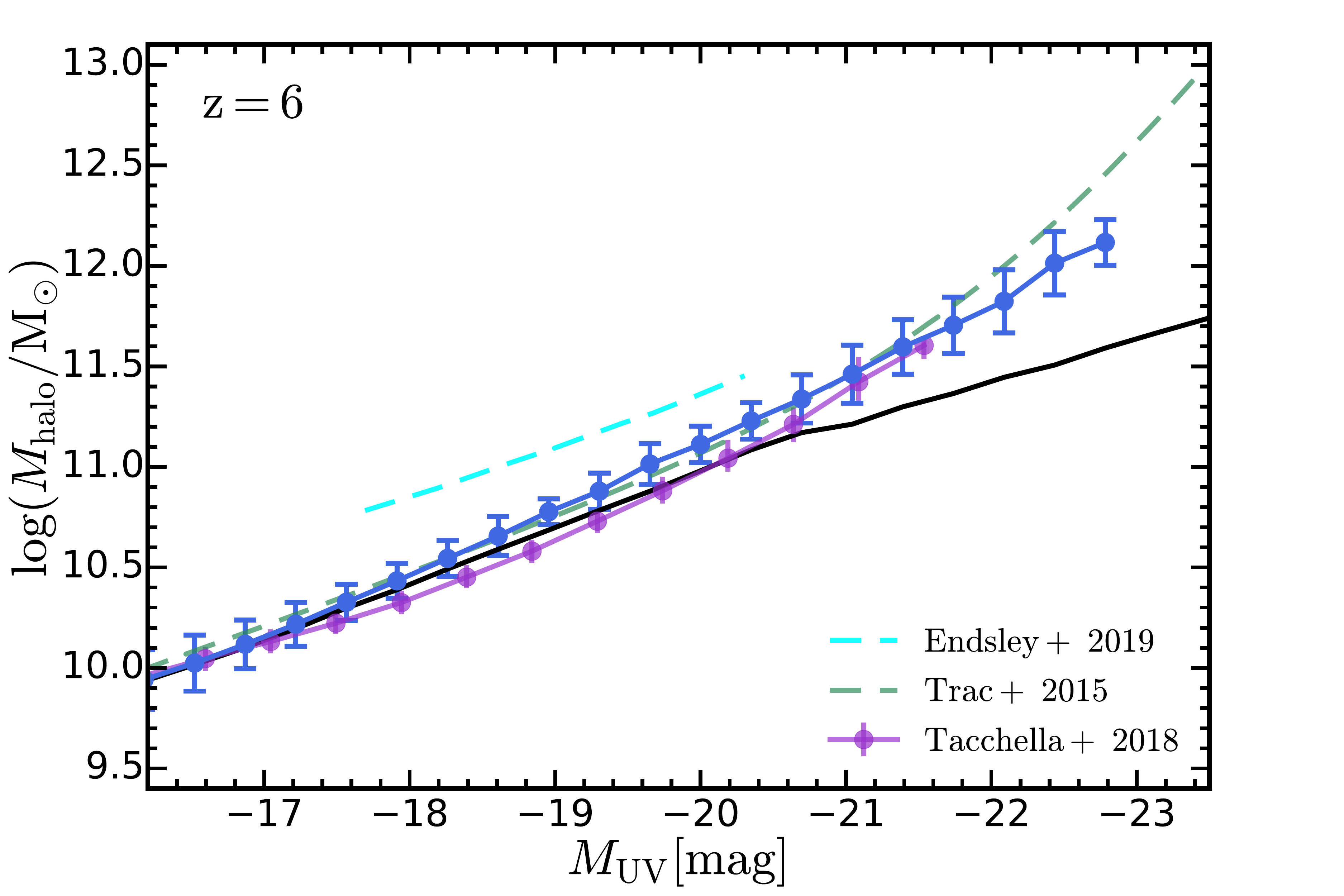}
\includegraphics[width=0.49\textwidth]{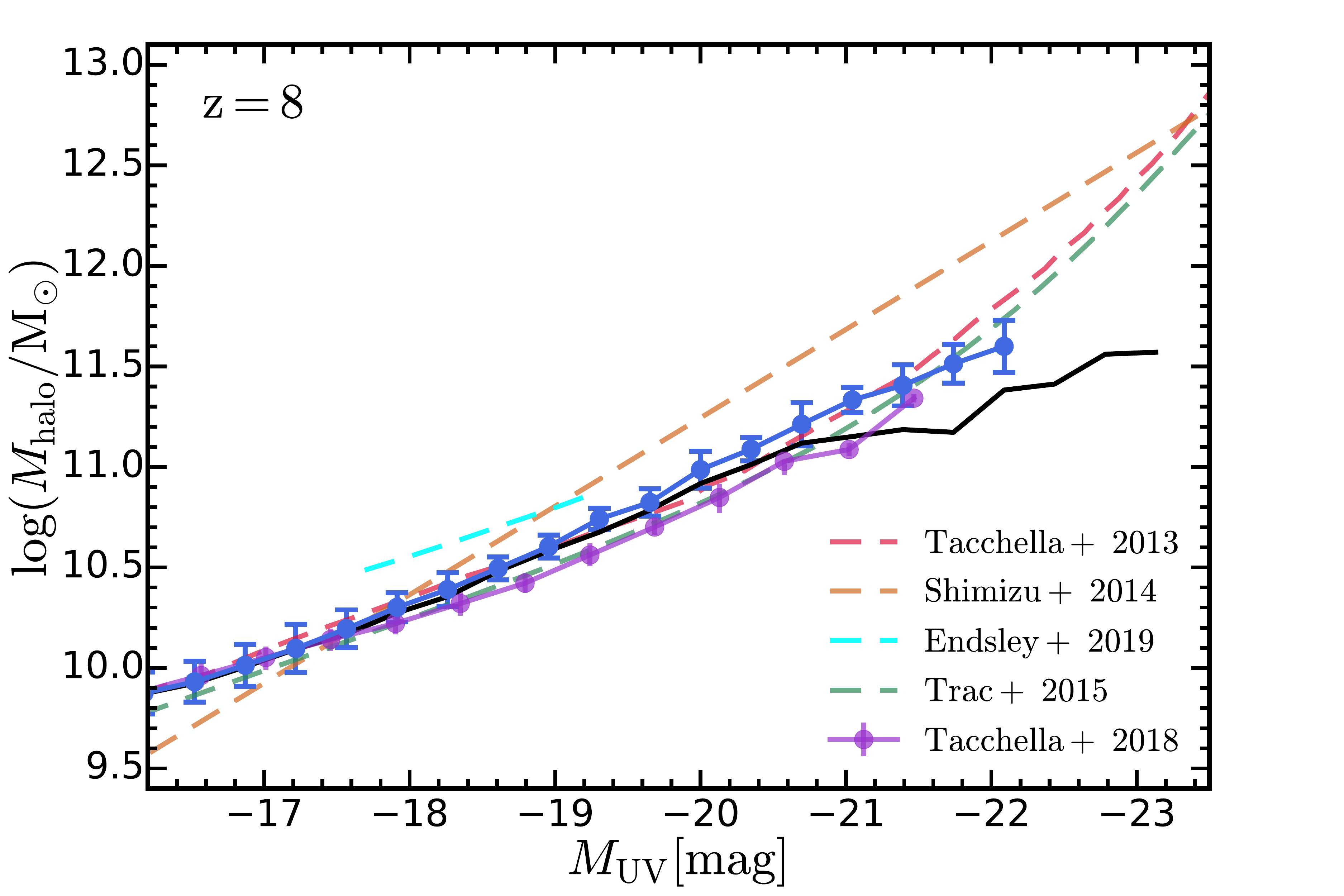}
\caption{{\bf Galaxy host halo mass versus rest-frame UV magnitude relations at $z=4,5,6,8$ from the IllustrisTNG simulations.} Blue circles and lines show the median $\log{(M_{\rm halo}/{\rm M}_{\odot})}$ within each dust attenuated UV magnitude bin. Error bars represent the $1\sigma$ dispersions in $\log{(M_{\rm halo}/{\rm M}_{\odot})}$ of the samples. Black lines indicate the relations when no resolved dust attenuation is taken into account. We compare our results with the relations obtained in empirical models: \citet{Tacchella2013}, red dashed lines; \citet{Trac2015}, green dashed lines; \citet{Tacchella2018}, purple lines and circles; \citet{Endsley2019}, cyan dashed lines. We also compare them with the prediction from the hydrodynamical simulation of~\citet{Shimizu2014} shown in the orange dashed line. Our predictions are qualitatively consistent with these studies in that the relations monotonically increase and exhibit a steepening induced by the dust attenuation at the bright end at all redshifts.}
\label{fig:Mhalo_Muv}
\end{figure*}

\section{Resolution correction and combination procedure}
\label{sec:res_com}
As discussed in \citetalias{Vogelsberger2019}, our high redshift predictions depend on the combination of TNG50, TNG100 and TNG300 simulations, which have different simulation volumes and numerical resolutions. To combine them together, a resolution correction procedure is needed for TNG100 and TNG300 simulation results to match the numerical resolution of TNG50, which provides the highest numerical resolution of the IllustrisTNG simulation suite. To be more specific, to resolution correct an arbitrary physical quantity X, for each simulation, we first divide galaxies into $20$ logarithmically uniformly spaced halo mass bins ranging from $10^{9}\msun$ to $10^{14}\msun$. Here, we assume that the halo mass is a robust quantity against numerical resolution. This assumption is supported by the fact that the halo mass functions from different simulations overlap perfectly in overlapping dynamical ranges. In each halo mass bin, we then calculate for each simulation the median X of galaxies. By comparing the median X of TNG100 (TNG300) with the median X of TNG50, we can derive the correction factor~\footnote{The correction factor is an additive factor if the quantity X is the logarithm of a physical quantity, e.g. magnitudes. In other cases, it is a multiplicative factor.} of the physical quantity X for TNG100 (TNG300) to the resolution level of TNG50 at a given halo mass. The simulation volume of TNG50 is too small to provide a large number of galaxies with large halo masses. We therefore do not apply the correction once the number of galaxies in a halo mass bin drops below $10$ for TNG50. This approach is supported by the fact that TNG100 and TNG300 have nearly identical galaxy luminosity (stellar mass) functions at the bright (massive) end. We can therefore safely assume a vanishing correction in this regime. We apply this whole procedure of deriving correction factors at all redshifts of interest. The resulting correction factors as a function of halo mass and redshift were stored such that we can easily correct arbitrary physical quantities at any redshift. 

This resolution correction procedure has been applied to all physical quantities studied in this paper. We use these resolution corrected quantities to derive statistical distributions (e.g., luminosity functions) and scaling relations (e.g., the $M_{\rm halo}-M_{\rm UV}$ relation). It is therefore by construction that all of the distributions and relations derived from the three simulations agree well in their shared dynamical ranges after the resolution correction. We then perform the combination procedure to smoothly connect them to get distributions or relations over a wide dynamical range. Usually, the high resolution simulation is superior at predicting distributions or scaling relations at the low mass (faint) end and suffers from Poisson noise at the massive (bright) end, where results from a simulation with lower resolution but larger simulated volume are required. In doing this, in the shared dynamical range of two simulations, for an arbitrary physical quantity X (already resolution corrected) depending on another physical quantity Y, we calculate the combined X in a bin of Y as:
\begin{equation}
X_{\rm combined}(Y)  =  \frac{X_{\rm 50}(Y) \, N_{\rm 50}^2 + 
      X_{\rm 100}(Y) \, N_{\rm 100}^2  + 
      X_{\rm 300}(Y) \, N_{\rm 300}^2}{N_{\rm 50}^2  +  N_{\rm 100}^2 + N_{\rm 300}^2},
\label{eq:combine}
\end{equation}
where $N_{\rm \#}$ refers to the number of galaxies of TNG\# in the bin. The weighting $N_{\rm \#}^{2}$ is chosen empirically to maintain the dominance of the simulation that has the best statistic in the bin.

\section{Results}
\label{sec:results}
In this section, we explore high redshift predictions from the IllustricTNG simulation suite and our post-processing procedure. These high redshift predictions involve: the scaling relation between galaxy host halo mass and rest-frame UV luminosity, the $M_{\rm halo}-M_{\rm UV}$ relation; the ${\rm H}_{\alpha}$ and ${\rm H}_{\beta}$ + $[\rm O \,\Rmnum{3}]$ luminosity functions; the Balmer break at $4000\text{\AA}$ (D4000) as a star formation indicator; the UV continuum slope $\beta$ and its role as a dust attenuation indicator; the dust attenuation curves of galaxies and the origin of its variety. In Figure~\ref{fig:image1}, we provide a visual presentation of star-forming galaxies at $z=2$ selected from TNG50, including the JWST bands synthetic images, the emission line strength maps of ${\rm H}_{\alpha}$ and ${\rm H}_{\beta}$ + $[\rm O \,\Rmnum{3}]$ and the relative differences between the strength of ${\rm H}_{\alpha}$ and ${\rm H}_{\beta}$ + $[\rm O \,\Rmnum{3}]$. The $3$ selected galaxies are all star-forming galaxies of stellar masses $\sim 10^{10-11} \msun$ with apparent disks and spiral arms. The line emission preferentially originates in the star-forming regions, e.g. disks and spiral arms, compared to the sparser distribution of broadband light. We also find that ${\rm H}_{\alpha}$ is relatively stronger in the star-forming regions while ${\rm H}_{\beta} + {[\rm O\,\Rmnum{3}]}$ is relatively stronger at the outskirts of the galaxies. This phenomenon will be discussed in more detail in Section~\ref{sec:diff_att}.

\subsection{$M_{\rm halo}-M_{\rm UV}$ relation}
\label{sec:halo-UV}
Understanding the formation and evolution of high redshift galaxies is one of the most important topics that {\it JWST} will help explore~\citep[e.g.,][]{Gardner2006,Cowley2018,Williams2018,Yung2019b}. The evolution of galaxies during this epoch is related to the evolution of their host dark matter haloes, which naturally leads to statistical correlations between various galaxy and halo properties, known as ``galaxy-halo connections''~\citep[e.g., see the review of][and references therein]{Wechsler2018}. 
These empirical relations help interpreting physical information of dark matter haloes from observed galaxy properties. Moreover, predictions of galaxy properties can be made from the evolutionary history of dark matter haloes~\citep[e.g.,][]{Conroy2007,Behroozi2013,Moster2013,Tacchella2013,Tacchella2018} without explicitly modeling the baryonic processes of star formation. Due to accessibility in observations, the rest-frame UV luminosity of galaxies serves as an important observable in building such scaling relations. It has been demonstrated to be a good proxy for the stellar mass assembly at high redshift~\citep[e.g.,][]{Stark2013,Song2016} where direct measurement of galaxy stellar mass is limited by the sensitivity and resolution of IR instruments. In \citetalias{Vogelsberger2019} we demonstrated that the IllustrisTNG simulation suite and our radiative transfer post-processing procedure produce galaxy rest-frame UV luminosity functions that are consistent with observations at $z=2-10$. We have also studied the galaxy rest-frame UV luminosity versus stellar mass relation at $z\geq4$ and found agreement with observations. In this paper, we explore the scaling relation between galaxy host halo mass and rest-frame UV luminosity (magnitude), the ${M_{\rm halo}-M_{\rm UV}}$ relation. Future galaxy clustering measurements with {\it JWST} are expected to place new constraints on this relation~\citep[e.g.,][]{Endsley2019}.

In Figure~\ref{fig:Mhalo_Muv}, we present the predicted relation between galaxy host halo mass and rest-frame UV magnitude at $z=4,5,6,8$. We note that the halo mass is here defined as the total mass of all the particles that are gravitationally bound to the host halo of the galaxy. Unless specified otherwise, the rest-frame UV magnitude, $M_{\rm UV}$, in all subsequent analysis in this paper is the dust attenuated magnitude. To derive the ${M_{\rm halo}-M_{\rm UV}}$ relation, we perform a binning in the UV magnitude with 23 bins linearly spaced from $-16 \mmag$ to $-24 \mmag$. For each bin, we calculate the median $\log{(M_{\rm halo}/{\rm M}_{\odot})}$ and the $1\sigma$ dispersion $\sigma_{\log{M_{\rm halo}}}$ for galaxies in TNG50, TNG100 and TNG300, respectively. We combine the median $\log{(M_{\rm halo}/{\rm M}_{\odot})}$ or the $\sigma_{\log{M_{\rm halo}}}$ using Equation~\ref{eq:combine} in the magnitude range shared by two or three simulations. At the faint end, where TNG100 (TNG300) does not provide sufficient resolution and therefore deviates from TNG50 (TNG100), we only use TNG50 (TNG50 and TNG100) to construct the combined ${M_{\rm halo}-M_{\rm UV}}$ relation. In the combination procedure, we do not consider bins with fewer than $10$ galaxies in TNG300. The combined median relations are presented as blue solid lines in Figure~\ref{fig:Mhalo_Muv}. Error bars represent $1\sigma$ dispersions in $\log{(M_{\rm halo}/{\rm M}_{\odot})}$. We note that all the subsequent analysis in this paper undergo similar binning and combination procedures and all the presented relations/distributions are resolution corrected and combined unless specified otherwise. Figure~\ref{fig:Mhalo_Muv} reveals a strong correlation between galaxy host halo mass and rest-frame UV luminosity. Dust attenuation is clearly important and it shifts the relation towards the high halo mass direction at the bright end. We compare our results with the relations predicted by empirical models~\citep{Tacchella2013,Tacchella2018,Trac2015,Endsley2019} and the simulation of~\cite{Shimizu2014}. The predictions are qualitatively consistent with these previous studies in that the relations monotonically increase and exhibit a steepening induced by the dust attenuation at the bright end at all redshifts. Our results almost overlap with the relations found in \citet{Tacchella2013,Tacchella2018,Trac2015} with $\lesssim 0.3\,{\rm dex}$ differences in halo masses at all luminosities and redshifts. However, we predict a less steep relation at $z=8$ compared with \citet{Shimizu2014}. At all redshifts, we predict a much brighter UV luminosity for a given halo mass compared with the results of \citet{Endsley2019}. This is likely because we include all the subhaloes (both central and satellite galaxies) identified in the simulations in our analysis while they only included haloes (galaxy clusters) in their analysis.

Compared with the emipirical model in \citet{Tacchella2018}, our predicted relation exhibits similar scatter in halo mass at the bright end while having significantly larger scatter at the faint end. In \citet{Tacchella2018}, the star formation history of a galaxy is uniquely determined by the accretion history of its host halo.  Haloes with the same mass may have different assembly histories and thus different amounts of stellar component. This results in scatter in scaling relations involving halo mass and galaxy properties. However, this empirical model assumed that the star formation rate is proportional to the gas accretion rate onto the halo, and that the star formation efficiency solely depends on the halo mass. In simulations, given the same mass accretion history of a host halo, baryonic matter can be accreted and cool down for star formation in different modes and the star formation efficiency can be affected by properties other than the halo mass; e.g. the halo concentration and the environment in which the halo resides. This results in additional scatter in the scaling relation.

\begin{figure}
    \centering
    \includegraphics[width=0.49\textwidth]{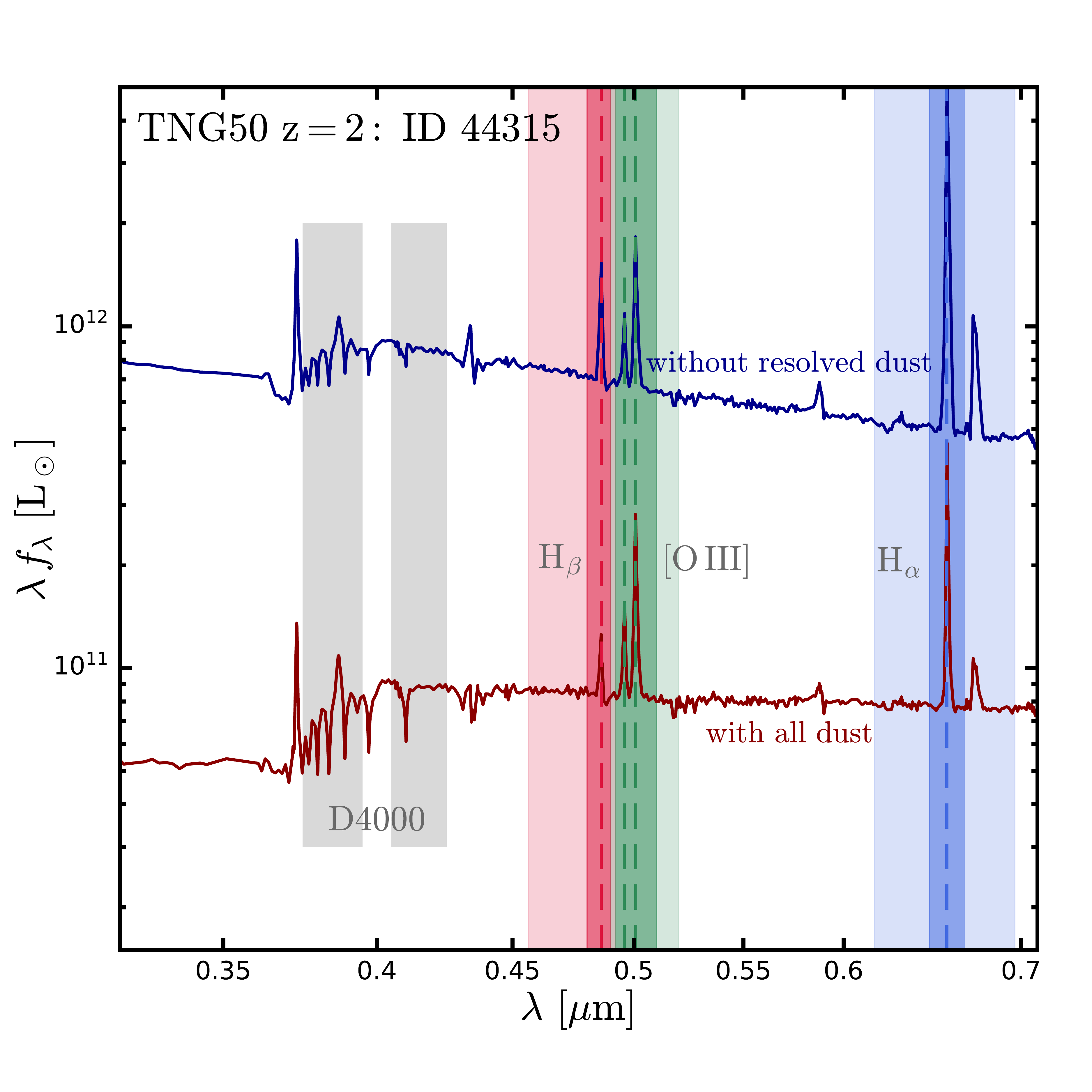}
    \caption{{\bf Constructed filters and galaxy SEDs.} We show filters constructed to derive the emission line luminosities and the Balmer break at $4000\text{\AA}$ (D4000). Synthetic SEDs of a star-forming galaxy in TNG50 are presented for reference. The SED with all dust attenuation (both resolved and unresolved) is shown in the red line while the one without the resolved dust attenuation is shown in the blue line. The centers of the emission lines studied in this paper are marked by vertical dashed lines. Shaded regions indicate the wavelength ranges covered by the filters. Lighter (darker) shaded regions correspond to broad (narrow) band filters.}
    \label{fig:filters}
\end{figure}

\subsection{Galaxy emission line luminosity functions}

\begin{figure*}
    \centering
    \includegraphics[width=0.49\textwidth]{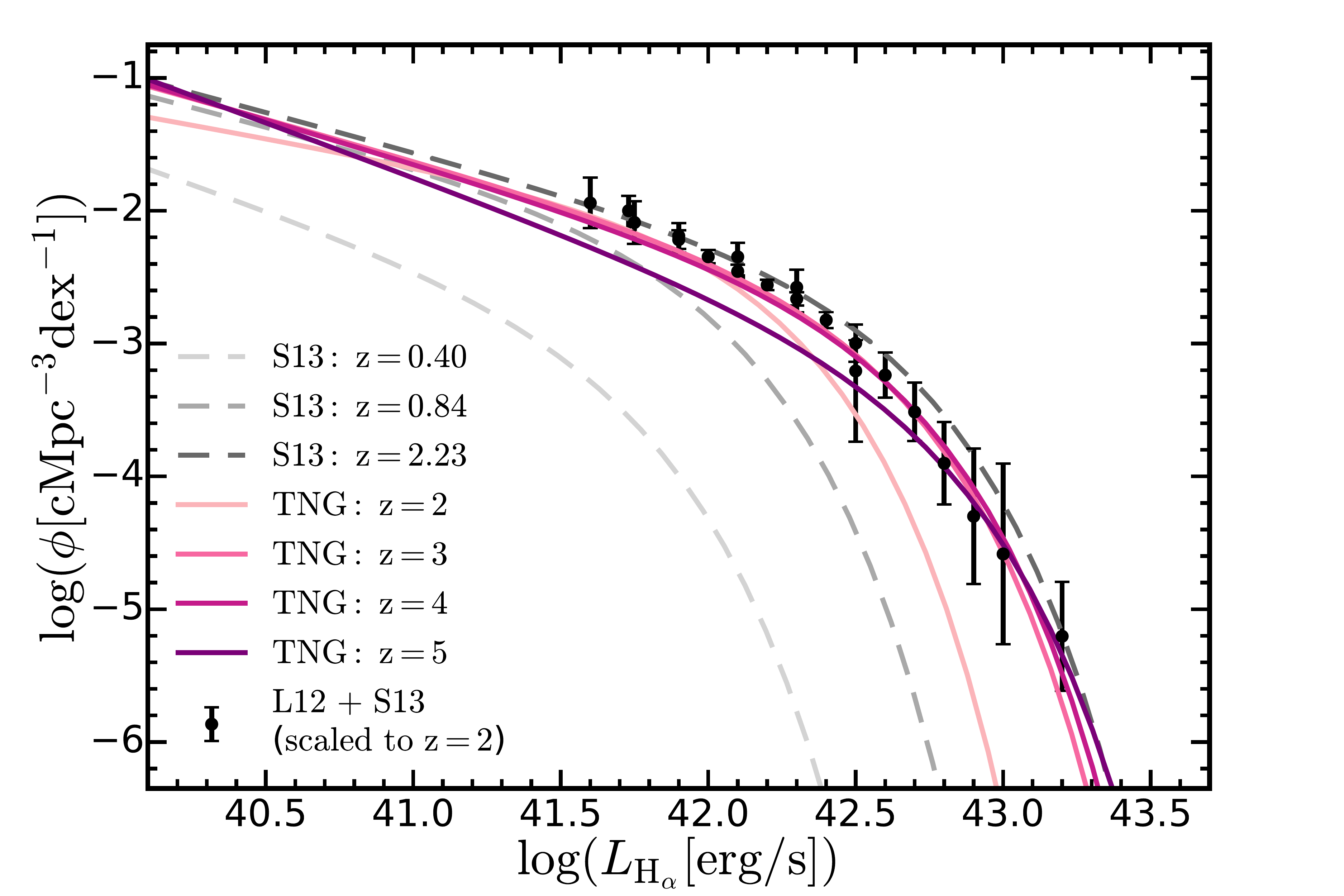}
    \includegraphics[width=0.49\textwidth]{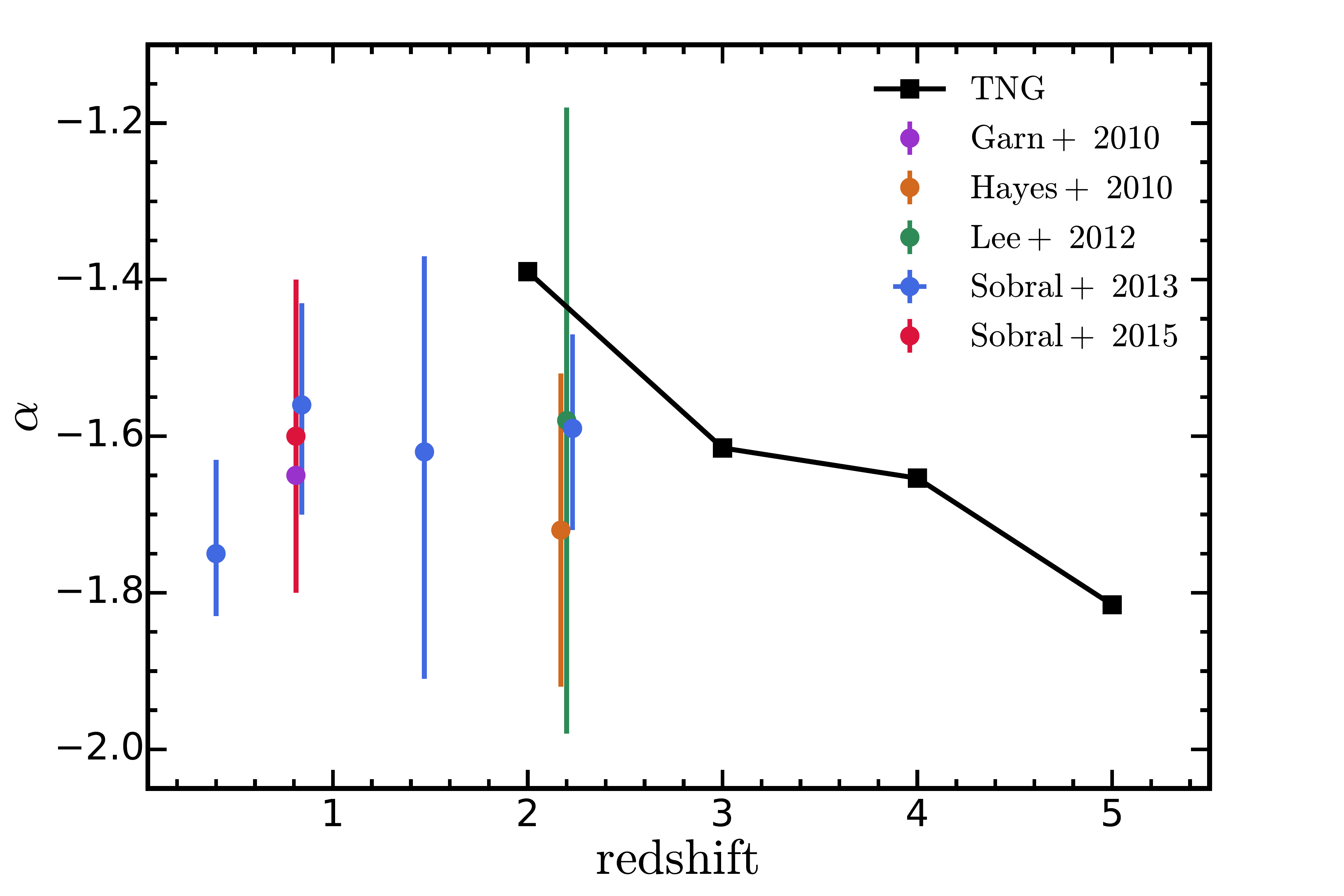}
    \includegraphics[width=0.49\textwidth]{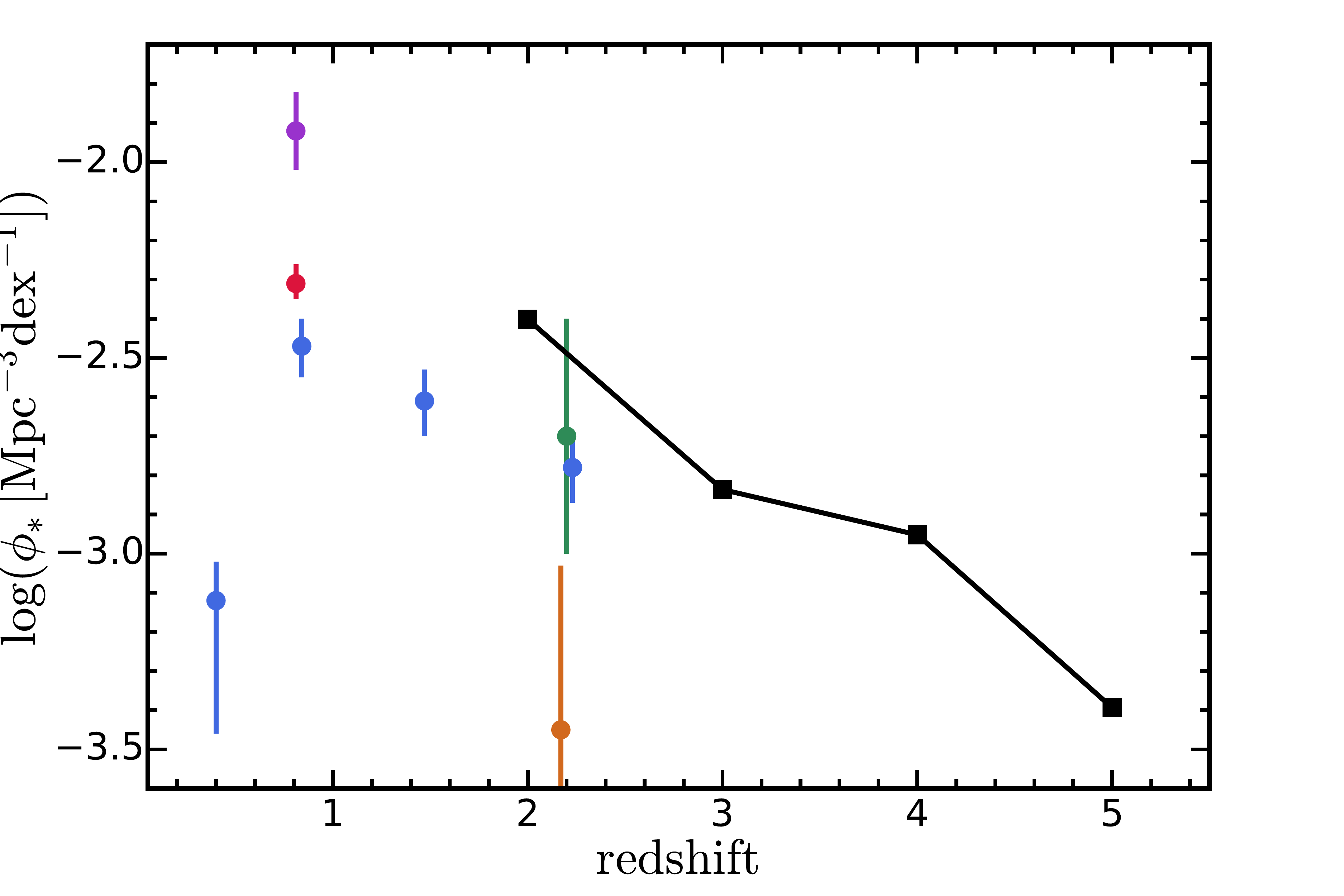}
    \includegraphics[width=0.49\textwidth]{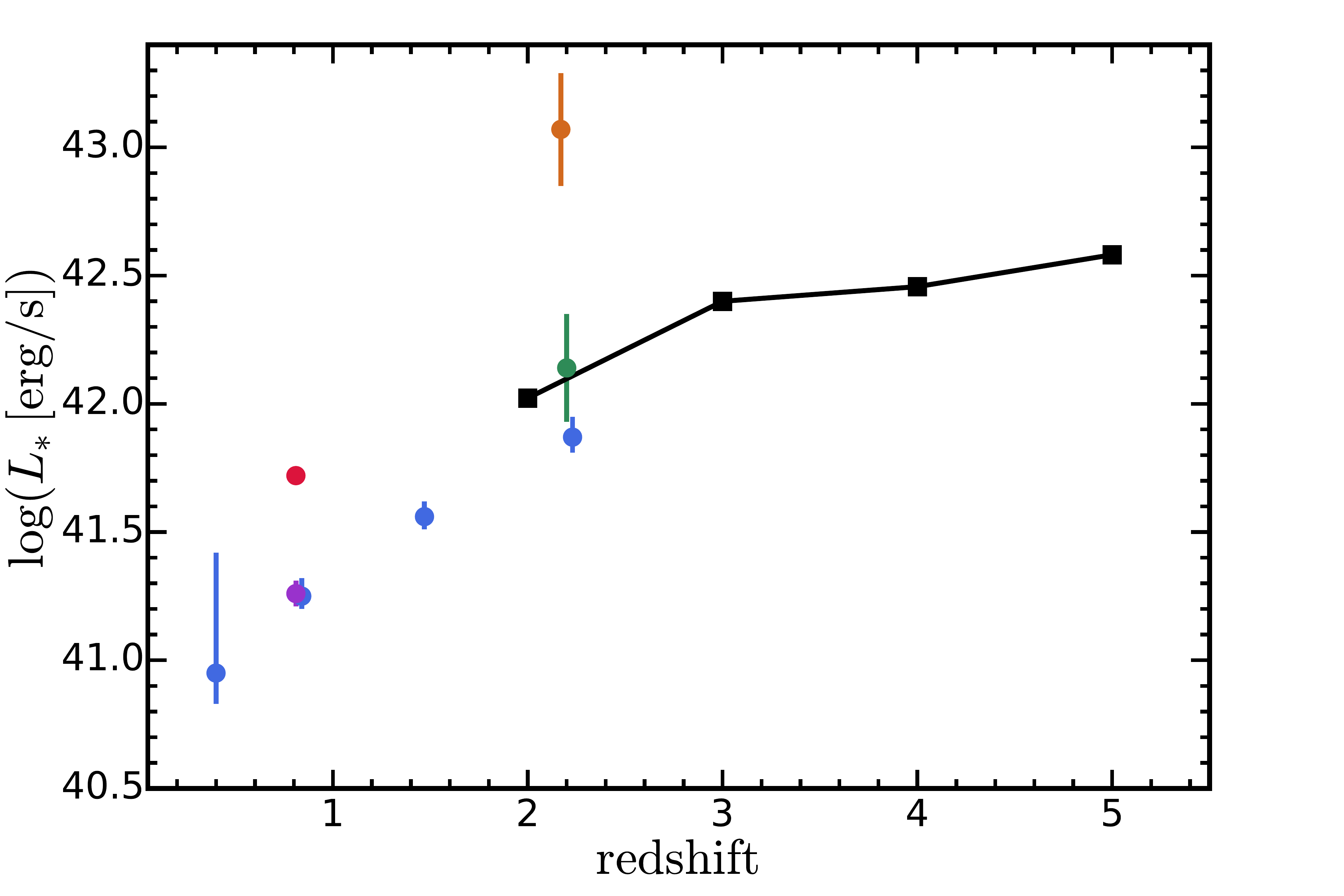}
    \caption{{\bf Evolution of the ${\rm H}_{\alpha}$ luminosity function from the IllustrisTNG simulations.} {\it Top left:} Evolution of the ${\rm H}_{\alpha}$ luminosity functions at $z=2-5$ predicted based on the IllustrisTNG simulations. Lines with darker colors indicate luminosity functions at higher redshifts. We compare the predictions with the best-fit luminosity functions found in observations at $z\leq2.23$~\citep{Sobral2013}. We also compare the results with the observational binned estimations at $z\simeq 2$ taken from \citet{Lee2012} and \citet{Sobral2013} (denoted as L12 and S13). We scale these binned estimations that are not performed exactly at $z=2$ to $z=2$ using the redshift-dependent luminosity function fits in \citet{Sobral2013}. The ${\rm H}_{\alpha}$ luminosity function exhibits different evolutionary patterns at low and high redshift separated at $z\simeq 2-3$. {\it Top right and bottom row:} Best-fit Schechter function parameters of the ${\rm H}_{\alpha}$ luminosity function. We compare our predictions with observations \citep{Hayes2010,Garn2010,Lee2012,Sobral2013,Sobral2015}. The predicted evolution of these parameters at $z\geq 2$ is consistent with the evolutionary trend found in observations at $z\lesssim 2$. The rapid increase in the break luminosity $L_{\ast}$ at $z\lesssim 2-3$ is followed by a decrease in the number density normalization at $z\gtrsim 2-3$.} 
    \label{fig:halpha}
\end{figure*}

\subsubsection{${\rm H}_{\alpha}$ luminosity function}

Determining the cosmic star formation history is important for understanding galaxy formation and evolution. Numerous efforts~\citep[e.g.,][]{Madau2014,Duncan2014,Salmon2015} have been made to measure the star formation rate density (SFRD) at high redshifts ($z\gtrsim 4$). Most of these works measured the SFRs of galaxies based on broadband photometry, which suffers from caveats with respect to both completeness~\citep[e.g.,][]{Inami2017} and contamination with strong nebular emission~\citep[e.g.,][]{Schaerer2009,Stark2013,Wilkins2013}. Nebular emission line luminosities will become a better proxy to measure the SFRs of galaxies at high redshift with the help of {\it JWST}. Massive, short-lived stars are direct tracers of recent star formation. The large amount of ionizing photons produced by these stars ionize surrounding gas. Hydrogen recombination cascades then produce nebular line emission, including the Balmer series lines like ${\rm H}_{\alpha}$ and ${\rm H}_{\beta}$ as well as forbidden lines like $[\rm O\,\Rmnum{2}]$ and $[\rm O \,\Rmnum{3}]$. Among them, ${\rm H}_{\alpha}$ is the most widely used and most sensitive star formation indicator. Compared with the UV, ${\rm H}_{\alpha}$ is located at a longer wavelength and is naturally considered less affected by dust attenuation. For example, a typical value of the dust attenuation of ${\rm H}_{\alpha}$ is often assumed to be $A_{{\rm H}_{\alpha}}\simeq1 \mmag$~\citep{Garn2010,Sobral2012,Sobral2013,Ibar2013,Stott2013} at $z\lesssim 2$ while the attenuation in the UV at similar redshifts can easily reach $\gtrsim 2\mmag$ in galaxies with stellar masses larger than $10^{10}\msun$~\citep[e.g.,][]{Heinis2014,Pannella2015,Reddy2018}.

In \citetalias{Vogelsberger2019}, we have presented the ${\rm H}_{\alpha}$ luminosity functions at $z=2,4$ and compared those with observations where we found a marginal consistency. We have also studied the scaling relation between SFR and ${\rm H}_{\alpha}$ luminosity at $z=2$ and found excellent agreement with the local calibrations~\citep{Kennicutt1998,Murphy2011}. Currently, ground based observations can only probe the ${\rm H}_{\alpha}$ luminosity function at $z\lesssim 3$. Since {\it JWST} will probe the ${\rm H}_{\alpha}$ luminosity functions at $z\gtrsim 3$ for the first time, it is timely to make more predictions for the ${\rm H}_{\alpha}$ luminosity function at high redshift with our model as a guide for future observations. 

We follow the method discussed in \citetalias{Vogelsberger2019} to derive the ${\rm H}_{\alpha}$ luminosity of a galaxy. We first construct broad band and narrow band top-hat filters around the emission line. The detailed information of these filters are listed in Table~\ref{tab:line_band}. We then convolve the SED of the galaxy with these two filters and obtain the filter averaged fluxes~(luminosity per unit wavelength), $f_{\rm BB}$ (broad band) and $f_{\rm NB}$ (narrow band). We then derive the luminosity and the equivalent width (EW) of the line as~\citep[e.g.,][]{Lee2012,Sobral2013,Sobral2015,Matthee2017}:
\begin{equation}
L_{\rm line}=\Delta\lambda_{\rm NB}\,\,\dfrac{f_{\rm NB}-f_{\rm BB}}{1-\dfrac{\Delta\lambda_{\rm NB}}{\Delta\lambda_{\rm BB}}}, \quad
{\rm EW}=\Delta\lambda_{\rm NB}\,\,\dfrac{f_{\rm NB}-f_{\rm BB}}{f_{\rm BB}-f_{\rm NB}\dfrac{\Delta\lambda_{\rm NB}}{\Delta\lambda_{\rm BB}}},
\end{equation}
where $\Delta \lambda$ is the width of the top-hat filter. We perform this calculation for every galaxy in the simulations. Then we perform the same resolution correction, binning and combination procedure as in Section~\ref{sec:halo-UV} in deriving the luminosity function. We also fit the predicted ${\rm H}_{\alpha}$ luminosity function with a Schechter function:
\begin{equation}
    \phi(\log{L})= \ln{10}\,\,\phi_{\ast} \Big(\dfrac{L}{L_{\ast}}\Big)^{1+\alpha}\exp{\Big(-\dfrac{L}{L_{\ast}}\Big)},
\end{equation}
where $\alpha$ is the faint-end slope, $L_{\ast}$ is the break luminosity and $\phi_{\ast}$ is the number density normalization. We will refer to this best-fit Schechter function as the ``predicted ${\rm H}_{\alpha}$ luminosity function'' in the following analysis. In the top left panel of Figure~\ref{fig:halpha}, we present the predicted ${\rm H}_{\alpha}$ luminosity functions at $z=2-5$ compared with the best-fit luminosity functions found in observations at $z\leq 2.23$~\citep{Sobral2013}. We also compare the results with the observational binned estimations at $z\simeq 2$ taken from \citet{Lee2012} and \citet{Sobral2013} (denoted as L12 and S13). We scale these binned estimations that are not performed exactly at $z=2$ to $z=2$ using the redshift-dependent luminosity function fits in \citet{Sobral2013}. We note that these studies have performed their own dust attenuation corrections, so we reintroduce the dust attenuation here based on the estimated $A_{{\rm H}_{\alpha}}$s in these papers. Our prediction at $z=2$ is $\sim 0.3\,{\rm dex}$ dimmer than these binned estimations at the bright end as we have already found in \citetalias{Vogelsberger2019}. We note that the unresolved dust attenuation implemented through the {\sc MAPPINGS-\Rmnum{3}} SED library has been calibrated for the local Universe~\citep{Groves2008} and uncertainties arise when applying this to high redshift galaxies. In addition, contamination from active galatic nuclei (AGN) also potentially contribute to the luminous bins in observations. For example, in \citet{Sobral2013}, $\sim 15\%$ of the sources observed at $z=1.47$ and $2.23$ were potentially AGN. In \citet{Matthee2017}, the X-ray fraction of the sources reached $\sim 20-50\%$ at $L_{{\rm H}_{\alpha}}\sim 10^{43} \erg/{\rm s}$ at $z\sim 2$. Both factors could explain the discrepancy at the bright end. The $\rm H_{\alpha}$ luminosity function exhibits a strong evolution in its break luminosity from $z=0$ to $z\simeq 2-3$. However, at $z\gtrsim 2-3$, it has only a mild evolution despite a slight decrease in the number density normalization. The evolution of the number density normalization at $z\gtrsim 2-3$ is likely following the hierarchical build-up of structure in the Universe. The sharp decline in the break luminosity at $z\lesssim 2-3$ is likely driven by the quenching of galaxies, which shuts down star formation and halts the nebular line emission associated with young stars. In the other three panels of Figure~\ref{fig:halpha}, we present the evolution of the best-fit Schechter function parameters compared with observations~\citep{Hayes2010,Garn2010,Lee2012,Sobral2013,Sobral2015}. Our best-fit parameters at $z=2-5$ are consistent with the evolutionary trend found at $z\lesssim 2$: the faint-end slope is stable at low redshift and becomes steeper towards higher redshift; the number density normalization is relatively stable at low redshift followed by an apparent decrease at $z\gtrsim 2-3$; the break luminosity increases rapidly at low redshift followed by mild evolution at $z\gtrsim 2-3$. We note that dust attenuation is not the primary driver for this evolutionary pattern, since such an evolutionary pattern is preserved when the resolved dust attenuation is not taken into account. However, potential underestimation of the line luminosity exists due to our inability to self-consistently model the unresolved dust component and control its strength. This would mainly affect the evolution of the break luminosity which may increase faster than predicted at $z\gtrsim 2-3$.

\begin{table}
    \centering
    \begin{tabular}{p{0.07\textwidth}|p{0.13\textwidth}|p{0.085\textwidth}|p{0.085\textwidth}}
         \hline
         Target name & Center Wavelength & Broad band & Narrow band\\
         \hline
         \hline
         ${\rm H}_{\alpha}$ & 6563\AA & 6163-6963\AA & 6463-6663\AA \\
         ${\rm H}_{\beta}$  & 4861\AA & 4561-4921\AA & 4801-4900\AA \\
         $[\rm O \,\Rmnum{3}]$ & 4959 \& 5008\AA &  4900-5200\AA & 4921-5100\AA\\
         $\rm D4000+$ & & 4050-4250\AA & \\
         $\rm D4000-$ & & 3750-3950\AA & \\
         \hline
    \end{tabular}
    \caption{{\bf Details of the filters.} The center wavelengths and the wavelength coverages of the constructed broad and narrow band filters in measuring the emission line luminosities and the Balmer break at $4000\text{\AA}$~(D4000). The filters are all designed to have a top-hat shape. For D4000, we only need two broad band filters to measure the mean fluxes.}
    \label{tab:line_band}
\end{table}

\begin{figure*}
    \centering
    \includegraphics[width=0.48\textwidth]{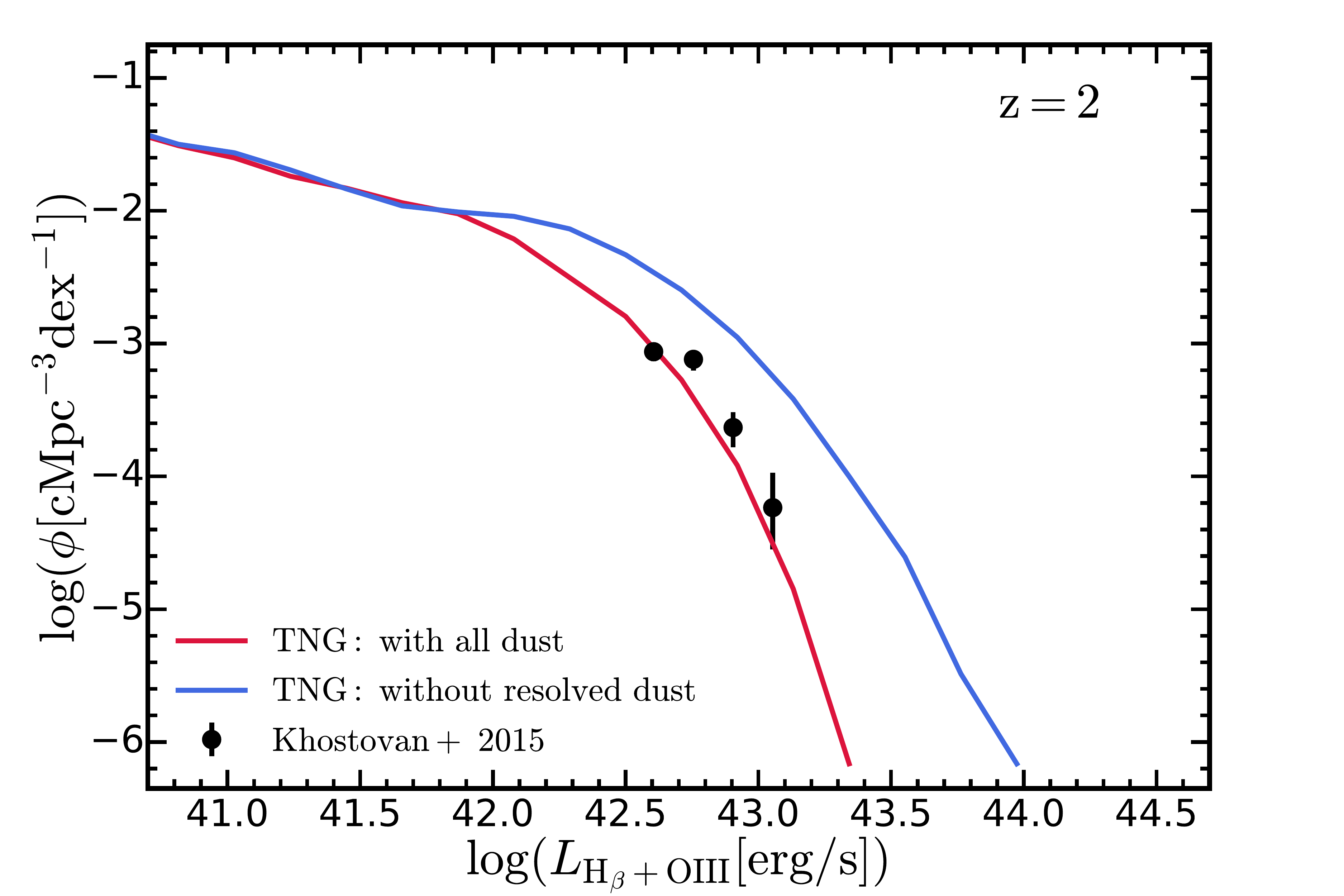}
    \includegraphics[width=0.48\textwidth]{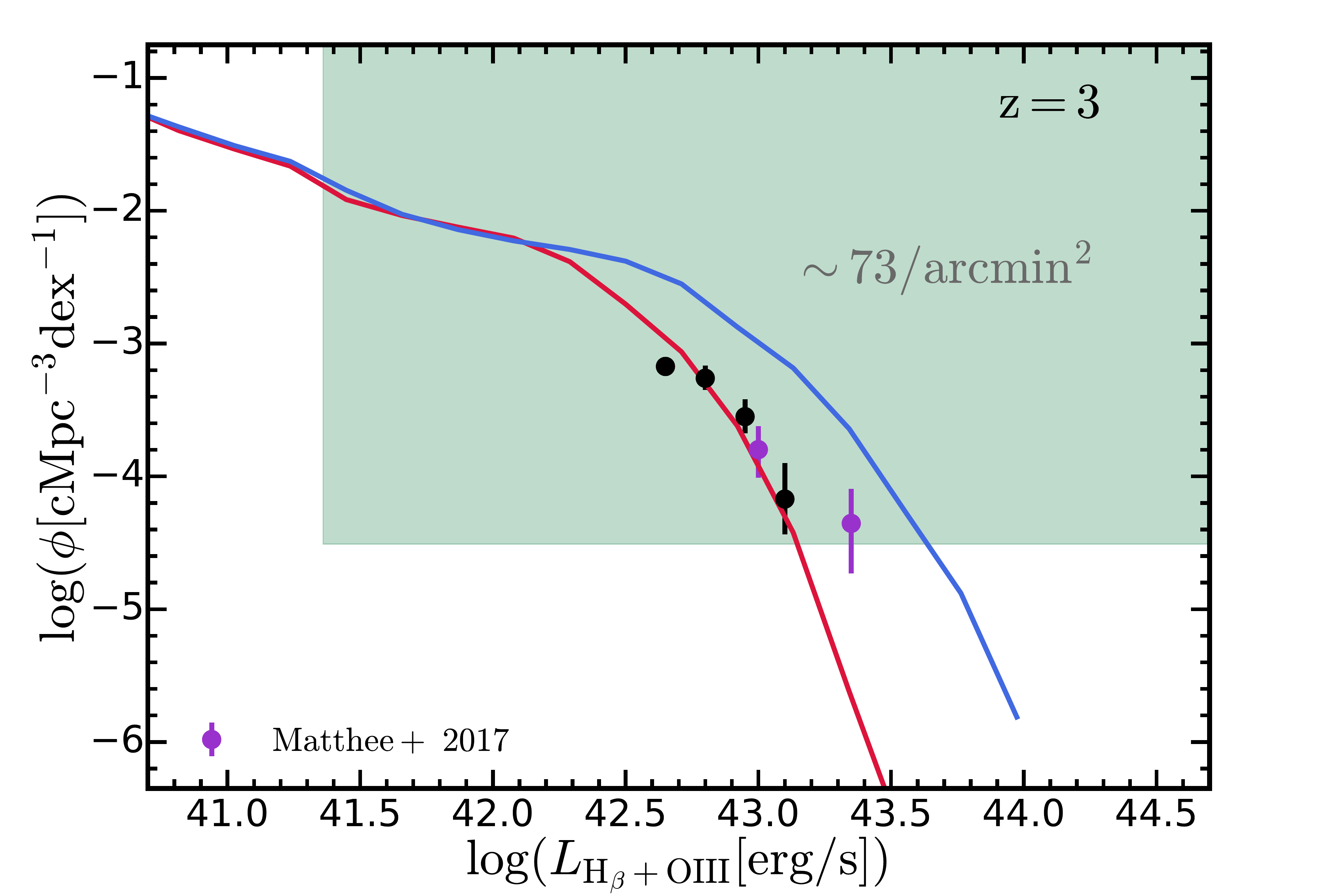}
    \includegraphics[width=0.48\textwidth]{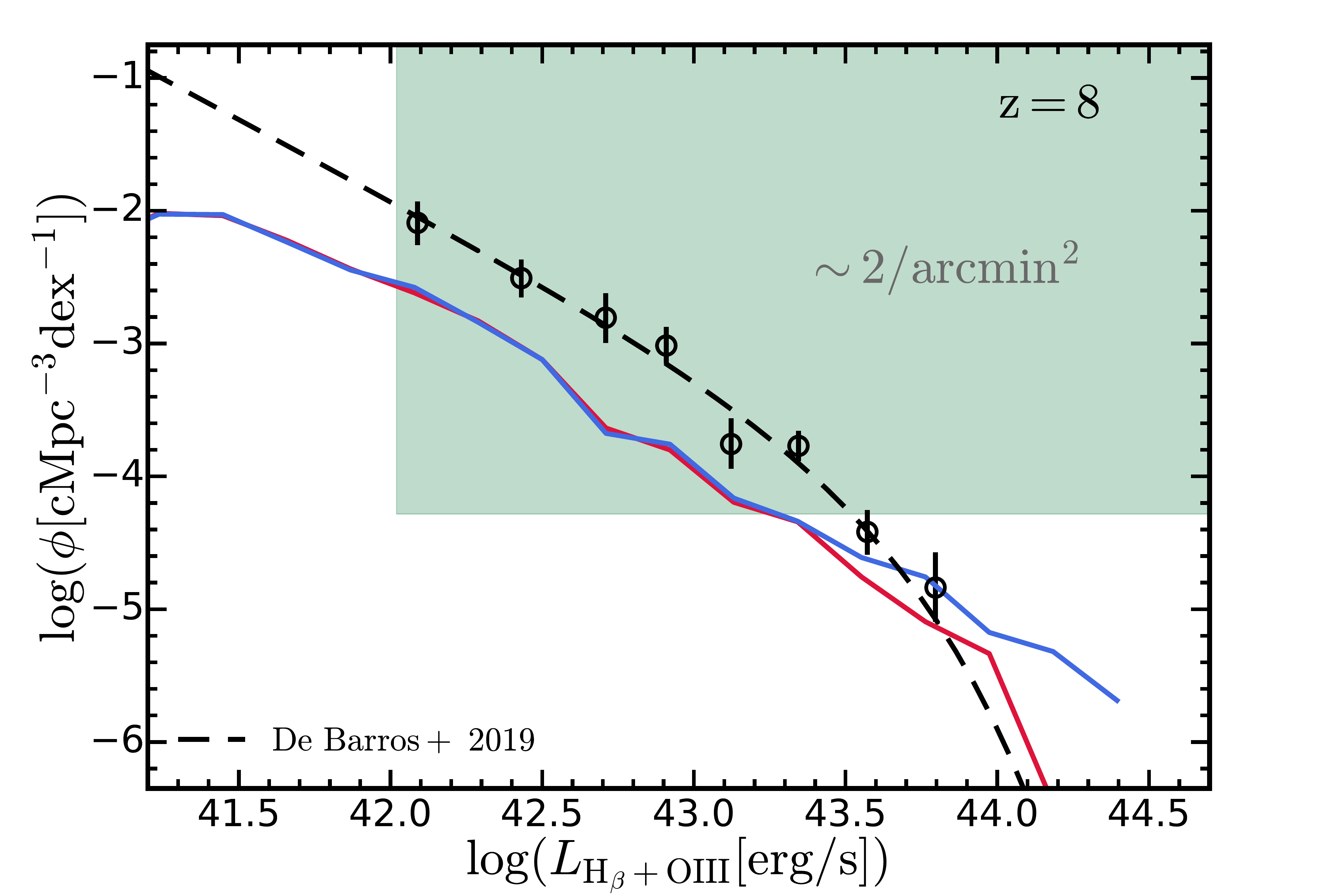}
    \includegraphics[width=0.48\textwidth]{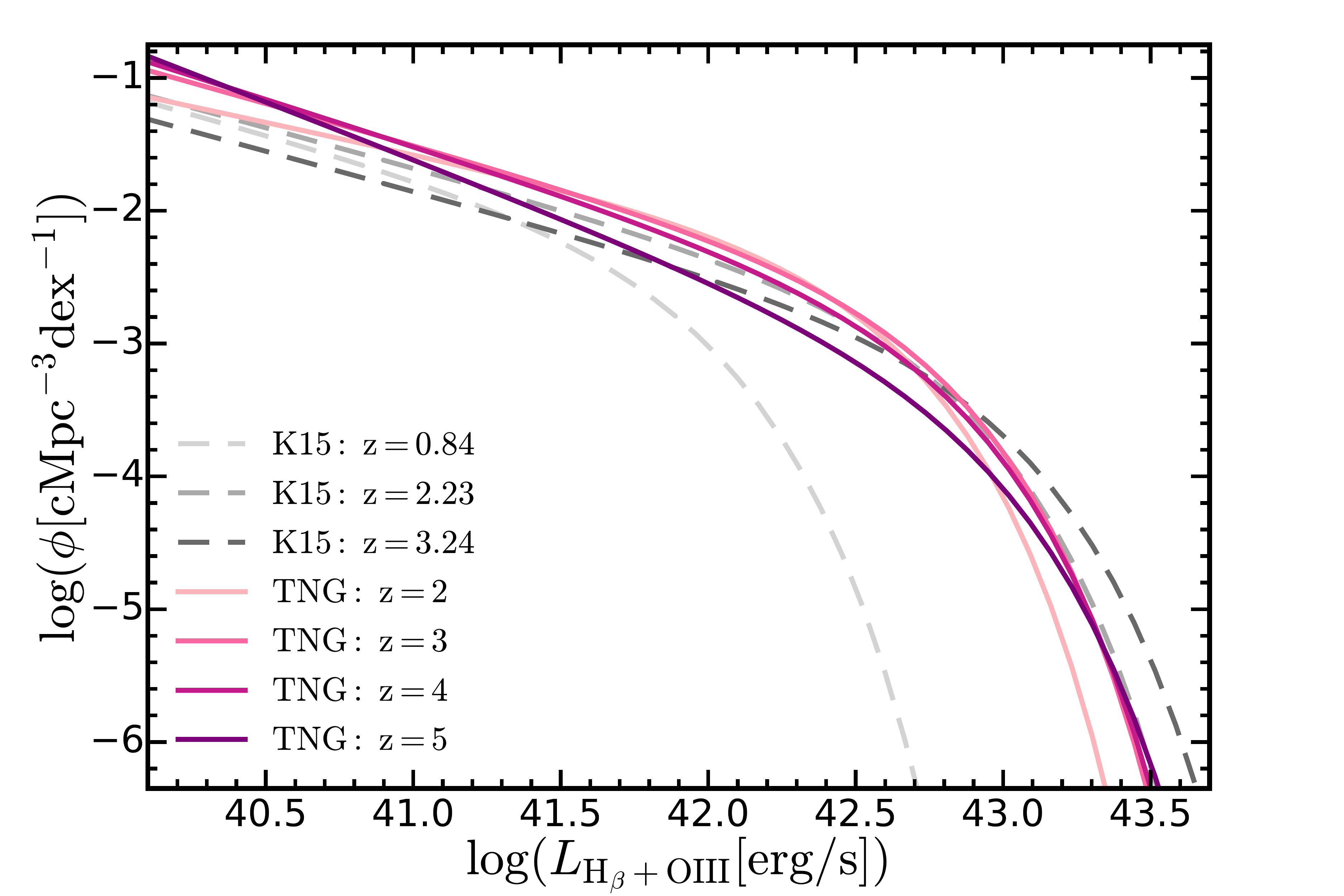}
    \caption{ \textbf{Evolution of the ${\rm H}_{\beta} + {[\rm O\,\Rmnum{3}]}$ luminosity function from the IllustrisTNG simulations.} {\it Top row and bottom left:} Predicted ${\rm H}_{\beta} + {[\rm O\,\Rmnum{3}]}$ luminosity functions at $z=2,3,8$. Blue lines show the luminosity functions when no resolved dust attenuation is taken into account. Red lines show the luminosity functions with all dust attenuation. We compare our results with observations from \citet{Khostovan2015,Matthee2017} at $z=2.2$ and $3.2$. Our predictions are in good agreement with these observations. For the $z=8$ case, we compare our luminosity function with a recent indirect measurement from \citet{DeBarros2019}. Their best-fit Schechter function is shown in the black dashed line. Their estimates based on individual UV luminosity function data points are shown in open circles. We find marginal consistency at the bright end despite our prediction being $\sim 0.5\, {\rm dex}$ lower than their results at $L\lesssim 43.5 \erg/{\rm s}$. For the green shaded regions, the vertical boundaries indicate the detection limits of a possible observational campaign, assuming a target ${\rm SNR}=10$ and an exposure time $T_{\rm exp}=10^{4}{\rm s}$, using the {\it JWST} NIRSpec instrument. The horizontal boundaries indicate a reference number density corresponding to one galaxy per dex of luminosity in an effective field of view $9.2\,{\rm arcmin}^{2}$ with a survey depth $\Delta z=1$. The numbers labeled are the numbers of galaxies expected to be observed by a survey of depth $\Delta z=1$ using the {\it JWST} NIRSpec instrument (under the assumed observation setup). {\it Bottom right:}  Evolution of the ${\rm H}_{\beta} + {[\rm O\,\Rmnum{3}]}$ luminosity function at $z=2-5$. Lines with darker colors indicate luminosity functions at higher redshifts. We compare our predicted luminosity functions with observations at lower redshift~($z\leq3.24$) from \citet{Khostovan2015}. The evolutionary pattern of the ${\rm H}_{\beta} + {[\rm O\,\Rmnum{3}]}$ luminosity function is similar to that of the ${\rm H}_{\alpha}$ luminosity function.}
    \label{fig:hbeta}
\end{figure*}

\subsubsection{${\rm H}_{\beta}$ + $[\rm O \,\Rmnum{3}]$ luminosity function}

Redshift $z=2$ is about the maximum redshift that ${\rm H}_{\alpha}$ surveys can be performed on the ground since ${\rm H}_{\alpha}$ at higher redshift falls into the mid-IR and is blocked by water vapour and carbon dioxide in the atmosphere. Measurements of galaxy SFRs at higher redshift with ground based facilities would require other emission lines as star formation indicators. For example, ${\rm H}_{\beta}$ at $4861\text{\AA}$, $[\rm O \,\Rmnum{3}]$ doublet at $4959\text{\AA}$, $5007\text{\AA}$ and $\rm O[\Rmnum{2}]$ at $3727\text{\AA}$ can be probed up to $z\sim3-5$. Here we explore the luminosity function of ${\rm H}_{\beta}+{\rm O[\Rmnum{3}]}$. Because the ${\rm H}_{\beta}$ and $[\rm O \,\Rmnum{3}]$ emission lines are close in wavelengths and are hard to distinguish from each other, their combined luminosity is usually measured in observations. Since we have accurate high resolution SEDs of galaxies and can distinguish the two lines properly, we measure their line luminosities separately and then add them to make comparisons with observations. Following the approach used to derive the ${\rm H}_{\alpha}$ luminosity, we construct broad band and narrow band top-hat filters around the ${\rm H}_{\beta}$ and $[\rm O \,\Rmnum{3}]$ emission lines respectively. The detailed properties of these filters are also listed in Table~\ref{tab:line_band}. In Figure~\ref{fig:filters}, we show all the filters constructed to derive the emission line luminosities along with the SEDs of a star-forming galaxy in TNG50.

In the top two panels of Figure~\ref{fig:hbeta}, we compare the predicted ${\rm H}_{\beta} + {[\rm O\,\Rmnum{3}]}$ luminosity functions at $z=2$ with observations at $z=2.2$~\citep{Khostovan2015} and compare the luminosity function at $z=3$ with observations at $z=3.2$~\citep{Khostovan2015,Matthee2017}. We find that the predicted ${\rm H}_{\beta} + {[\rm O\,\Rmnum{3}]}$ luminosity functions are in good agreement with observations at both redshifts. The differences in luminosities are $\lesssim 0.1\,{\rm dex}$, except for one binned value at the bright end at $z\simeq3$ which could be affected by contamination. For the green shaded regions in the figures, the vertical boundaries indicate the detection limits of a possible observational campaign, assuming a target SNR$=10$ and an exposure time $T_{\rm exp}=10^{4}{\rm s}$, using the {\it JWST} NIRSpec instrument. The detection limits are also listed in Table~\ref{tab:nirspec_sensitivity}. The horizontal boundaries indicate a reference number density that corresponds to one galaxy per dex of luminosity in an effective field of view $9.2\,{\rm arcmin}^{2}$ with a survey depth $\Delta z=1$. The calculations for the detection limits and the introduction of the field of view of a possible observational campaign carried out by the {\it JWST} NIRSpec instrument are discussed in Section~\ref{sec:jwst_fore}. The numbers labeled are the numbers of galaxies expected to be observed by a survey of depth $\Delta z=1$ using the {\it JWST} NIRSpec instrument (under the assumed observation setup). In the bottom left panel of Figure~\ref{fig:hbeta}, we present the predicted ${\rm H}_{\beta} + {[\rm O\,\Rmnum{3}]}$ luminosity function at $z=8$ which has never been reached by any direct survey. Recently, \citet{DeBarros2019} presented the first constraint on the ${\rm H}_{\beta} + {[\rm O\,\Rmnum{3}]}$ luminosity function at $z\sim8$. They measured the ${\rm H}_{\beta} + {[\rm O\,\Rmnum{3}]}$ line luminosities and the UV luminosities of $\sim100$ Lyman Break galaxies at $z\sim 8$. They fitted a $L_{{\rm H}_{\beta} + {\rm O[\Rmnum{3}]}}-M_{\rm UV}$ scaling relation and then convolved the well-measured galaxy rest-frame UV luminosity function with this relation to derive the ${\rm H}_{\beta} + {[\rm O\,\Rmnum{3}]}$ luminosity function. We compare our result with their best-fit Schechter function and the individual data points convolved from the rest-frame UV luminosity function binned estimations. We find marginal consistency at the bright end despite our prediction being $\sim 0.5\, {\rm dex}$ lower than their result at $L<43.5 \erg/{\rm s}$. This is roughly consistent with our prediction for the galaxy rest-frame UV luminosity function at $z=8$, which is also slightly lower than observations. We fit the predicted ${\rm H}_{\beta} + {[\rm O\,\Rmnum{3}]}$ luminosity functions with a Schechter function. In the bottom right panel of Figure~\ref{fig:hbeta}, we present the evolution of the predicted ${\rm H}_{\beta} + {[\rm O\,\Rmnum{3}]}$ luminosity functions at $z=2-5$. We compare our best-fit Schechter functions with observations by \citet{Khostovan2015} at $z\leq3$. We find a similar evolutionary pattern to the ${\rm H}_{\alpha}$ luminosity function in that the break luminosity is evolving rapidly at low redshift followed by a mild decrease in the number density normalization towards $z\gtrsim 2-3$. 

\begin{figure}
    \centering
    \includegraphics[width=0.48\textwidth]{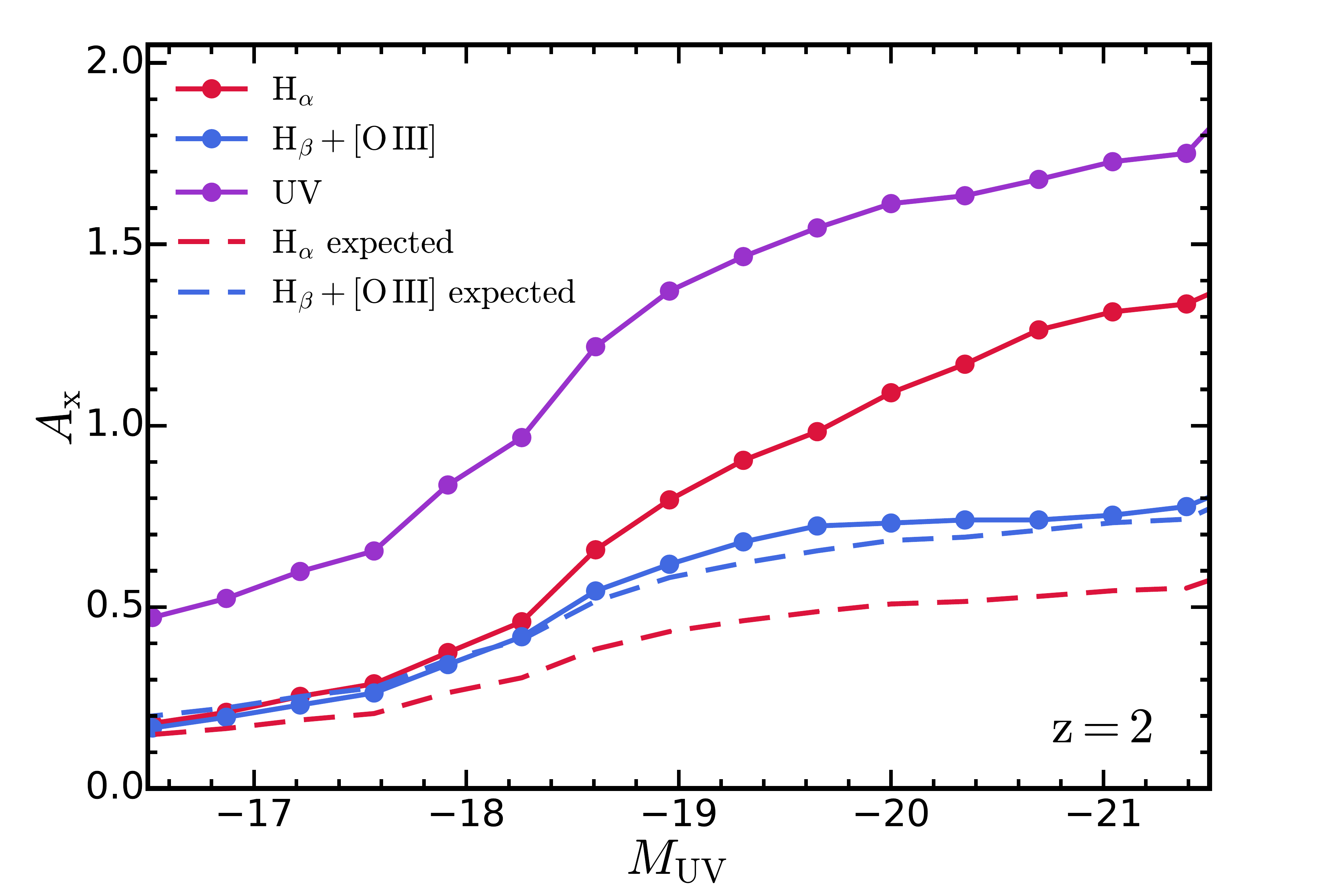}
    \caption{{\bf Resolved dust attenuation of galaxy UV and emission line luminosities versus rest-frame UV magnitude at $z=2$ from the IllustrisTNG simulations.} Dashed lines represent the expected attenuation of the lines from $A_{\rm UV}$ assuming the canonical Calzetti attenuation curve. The resolved dust attenuation of ${\rm H}_{\alpha}$, which is located at $6563\text{\AA}$, surprisingly is stronger than that of ${\rm H}_{\beta} + {[\rm O\,\Rmnum{3}]}$, which is located at around $5000\text{\AA}$. This is in contradiction with the prediction from the attenuation curve of the continuum that the attenuation at lower wavelength is stronger. The ${\rm H}_{\alpha}$ line is more heavily attenuated than expected from the attenuation curve of the continuum. These phenomena demonstrate the differential attenuation of emission lines.}
    \label{fig:att_vs_Muv}
\end{figure}

\subsubsection{Differential attenuation of emission lines}
\label{sec:diff_att}

\begin{figure*}

\includegraphics[width=0.465\textwidth]{images/faceon33_140057_nd.png}
\includegraphics[width=0.51\textwidth]{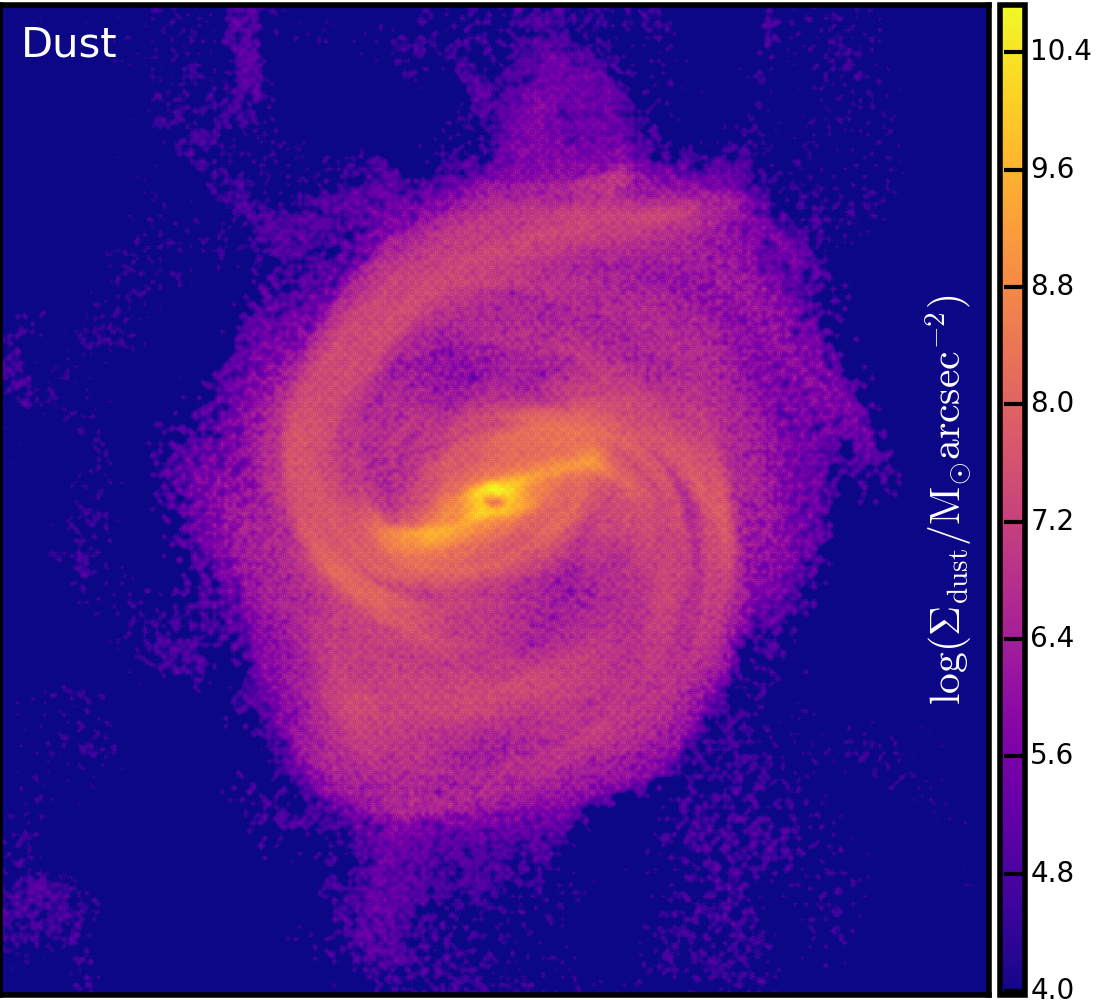}
\includegraphics[width=0.465\textwidth]{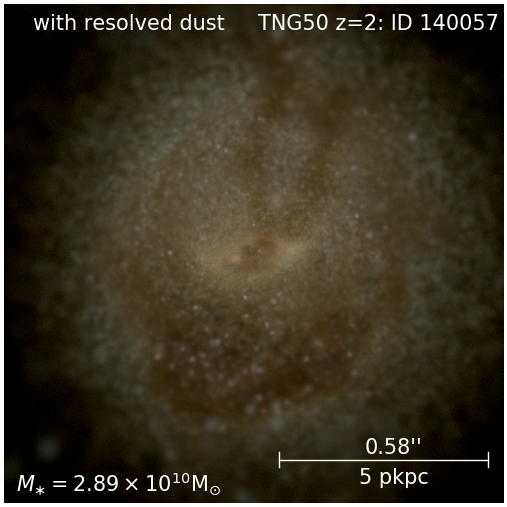}
\includegraphics[width=0.51\textwidth]{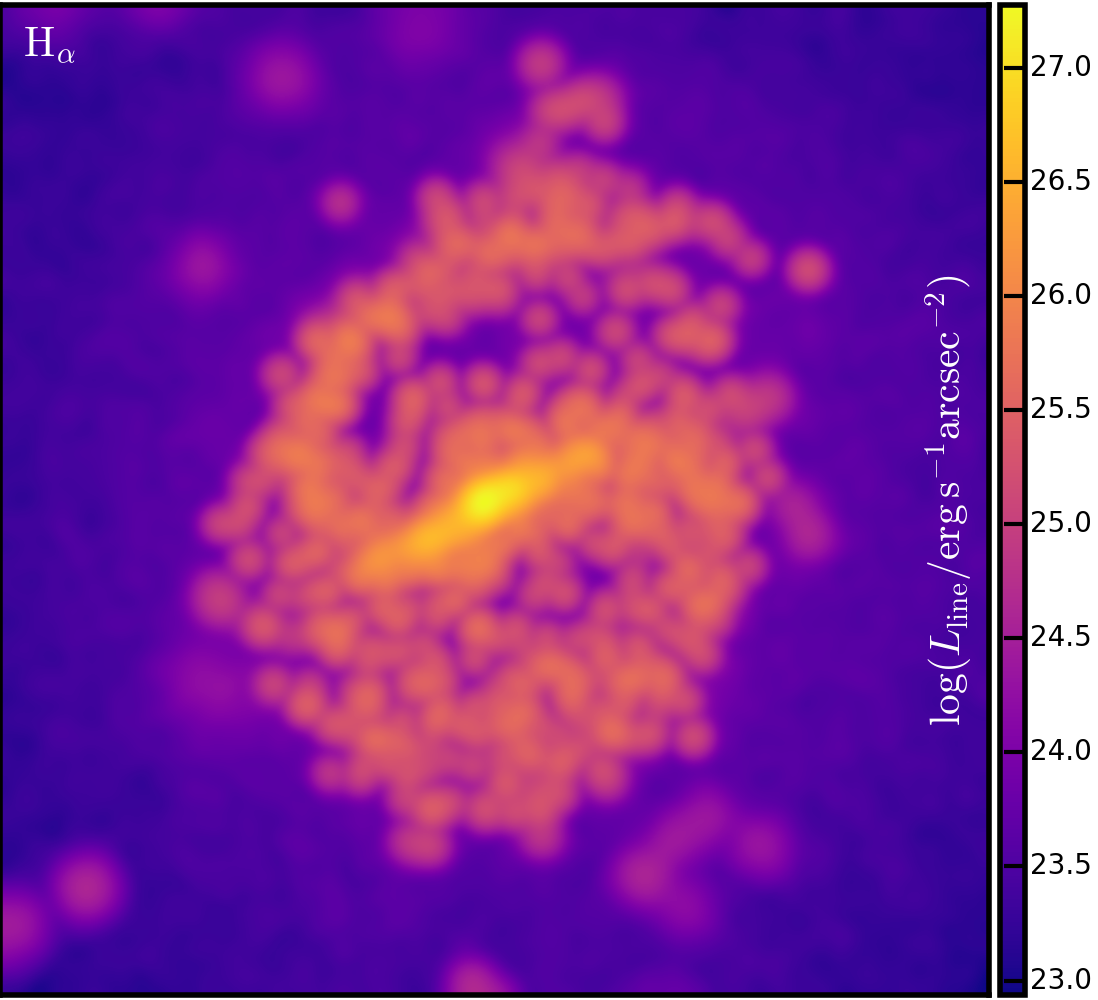}\\

\caption{ {\bf Images of a TNG50 galaxy at $\mathbf{z=2}$.} {\it Top left:} {\it JWST} NIRCam F115W, F150W, F200W bands synthetic images without the resolved dust; {\it Bottom left:} {\it JWST} NIRCam F115W, F150W, F200W bands synthetic image with all dust, assuming the dust-to-metal ratio to be $0.9$; {\it Top right:} resolved dust mass distribution with warmer color indicating higher dust mass density; {\it Bottom right:} ${\rm H}_{\alpha}$ emission line strength map. It is smoothed in the same way as the maps in Figure~\ref{fig:image1}. The images demonstrate that the origins of line emission are spatially correlated with the resolved dust distribution, which would result in the differential attenuation of emission lines.}
\label{fig:image2}
\end{figure*}

In many studies, dust attenuation of emission lines has been assumed to follow the attenuation curve of the continuum emission. Following this assumption, the attenuation of emission lines at long wavelengths is naturally thought to be limited. However, as shown in Figure~\ref{fig:att_vs_Muv}, the resolved dust attenuation of ${\rm H}_{\alpha}$, which is located at $6563\text{\AA}$, is surprisingly stronger than that of 
${\rm H}_{\beta} + {[\rm O\,\Rmnum{3}]}$, which is located at around $5000\text{\AA}$. This is in contradiction with the prediction from the attenuation curve of the continuum that the attenuation at lower wavelengths is stronger. The ${\rm H}_{\alpha}$ line is indeed less attenuated than the UV but is more heavily attenuated than expected from the attenuation curve of the continuum. This phenomenon is also illustrated in Section~\ref{sec:attenuation_curve} and Figure~\ref{fig:attenuation_curve} where spikes appear in the radiative transfer based attenuation curves where emission lines are located. We note that the dust attenuation we compare to here is the resolved dust attenuation. So the differential attenuation found here is not caused by the unresolved dust attenuation in our modelling of young stellar populations. If we also take the unresolved dust attenuation into account, the preferential attenuation of emission lines will be even stronger.

We attribute the differential attenuation of emission lines to dust geometry~(the geometrical and spatial distribution dust with respect to radiation sources). In Figure~\ref{fig:image2}, we present the {\it JWST} bands synthetic images (with and without dust attenuation), the resolved dust mass distribution and the ${\rm H}_{\alpha}$ emission strength map of a TNG50 galaxy at $z=2$. As shown in the right column of Figure~\ref{fig:image2}, the spatial distribution of the resolved dust and the emission line strength are well-correlated. Since we have assumed that dust is traced by metals in cold, star-forming gas in our post-processing procedure, dust naturally concentrates in star-forming regions where the nebular line emission originates. Therefore, the column density of dust is relatively larger for line emission than the continuum emission which is distributed more evenly in the galaxy. This results in the differential attenuation of emission lines. Furthermore, as illustrated in Figure~\ref{fig:image1}, the spatial origin of ${\rm H}_{\alpha}$ and ${\rm H}_{\beta} + {[\rm O\,\Rmnum{3}]}$ are different in galaxies at $z=2$. The ${\rm H}_{\alpha}$ emission is stronger in star-forming regions, e.g. spiral arms and disks, where metallicity is high and dust is abundant. The ${\rm H}_{\beta} + {[\rm O\,\Rmnum{3}]}$ emission is stronger in the outskirts of the galaxies where metallicity is low and the regions are dust-poor. As a consequence of these differences, the sources of ${\rm H}_{\alpha}$ are more obscured by dust than those of ${\rm H}_{\beta} + {[\rm O\,\Rmnum{3}]}$, which explains the inverse hierarchy of the resolved dust attenuation of ${\rm H}_{\beta} + {[\rm O\,\Rmnum{3}]}$ and ${\rm H}_{\alpha}$. In observations, there is evidence for this differential attenuation of emission lines. For example, in \citet{Reddy2015}, the color excess of ionized gas was found to be larger than that of stellar continuum. The difference in the attenuation of ${\rm H}_{\alpha}$ and the continuum showed a positive correlation with galaxy SFR.

\subsubsection{ {\it JWST} forecast}
\label{sec:jwst_fore}
{\it JWST} is going to provide the first spectroscopic access to optical emission lines of high redshift galaxies~\citep[e.g.,][]{Greene2017,Chevallard2019,DeBarros2019}. For example, the {\it JWST} NIRCam can perform imaging or slitless spectroscopy in either of the two $2.2\times2.2\,{\rm arcmin}^{2}$ fields of view~\citep{Gardner2006}. A wide-field high redshift survey of $\rm H_{\alpha}$ ($\rm H_{\beta}$ and $[\rm O \,\Rmnum{3}]$) can be performed at $z\sim2.7-6.6$ ($z\sim3.8-9.0$)~\citep{Greene2017}. The detection limits of a possible observational campaign using this technique are listed in Table~\ref{tab:nircam_sensitivity}. In addition, the NIRSpec instrument on {\it JWST} offers multi-object spectroscopy covering a large wavelength range from $0.6$ to $5.3\micron$ with resolving power up to $R\sim 2700$~\citep{Gardner2006}. It enables the detection of $\rm H_{\alpha}$ ($\rm H_{\beta}$ and $[\rm O \,\Rmnum{3}]$) up to $z=7$ ($z=9$). 
When operating with four quadrants, each of which covers $2.3\,{\rm arcmin}^2$, NIRSpec will cover an effective field of view of $9.2\,{\rm arcmin}^2$. 
The sensitivity of the NIRSpec technique can be calculated with the {\it JWST} Exposure Time Calculator~\footnote{\href{https://jwst.etc.stsci.edu/}{https://jwst.etc.stsci.edu/}}~\citep[ETC,][]{Pontoppidan2016}. 

Here, we provide two example calculations of the detection limits of the emission line luminosities under different assumptions. For the first case, we assume that the target emission line is well-resolved and the continuum around it is detected with a desired SNR. The online documentation~\footnote{\href{https://jwst-docs.stsci.edu/near-infrared-spectrograph/nirspec-predicted-performance/nirspec-sensitivity}{https://jwst-docs.stsci.edu/near-infrared-spectrograph/nirspec-predicted-performance/nirspec-sensitivity}}~\citep{Jdoc} of the NIRSpec expected performance provides a calculation  of the wavelength dependent limiting flux sensitivity with a target ${\rm SNR}=10$ and an exposure time $T_{\rm exp}=10^{4}{\rm s}$~(see the documentation for the detailed parameter setup). We then take the limiting flux sensitivity, $S^{\rm lim}_{\nu}$, where the emission lines are located. In this case, $S^{\rm lim}_{\nu}$ represents the detection limit of the continuum around the lines. We convert it to the detection limit of the line luminosity as:
\begin{equation}
    L^{\rm lim}_{\rm line} = 4\pi D_{\rm L}(z)^{2}\dfrac{{\rm c}}{\lambda_{\rm obs}^{2}} S^{\rm lim}_{\nu}\,{\rm EW} ,
\end{equation}
where $\lambda_{\rm obs}$ is the observed wavelength of the line, $D_{\rm L}(z)$ is the luminosity distance of an object at redshift $z$, and EW is the equivalent width of the line, assumed to be $100\text{\AA}$ here. At this line luminosity limit, the lines with EW below the assumed value can be detected. For the second case, we assume that the continuum cannot be properly detected and the emission lines are not completely resolved. The continuum flux in this case is subdominant with respect to the noise so we can ignore it. Using the {\it JWST} ETC, assuming the lines have width $\sim 150 \kms$ and no continuum contribution, we calculate the detection limits of line luminosities with a target ${\rm SNR}=10$ and an exposure time $T_{\rm exp}=10^{4}{\rm s}$. We adopt $10$ groups per integration, $24$ integrations per exposure and $1$ exposure per specification to achieve the total exposure time $10^{4}{\rm s}$. The read-out pattern is chosen to be NRS. The background configuration is assumed to be the same as in \citetalias{Vogelsberger2019} and the source is assumed to be a point source. The calculated detection limits with the two approaches as well as the details of the {\sc JWST} NIRSpec dispersers and filters are all listed in Table~\ref{tab:nirspec_sensitivity}. We note that the detection limits we use for related predictions in this paper are calculated using the second approach.

\begin{table*}
    \centering
    \begin{tabular}{p{0.13\textwidth}|p{0.06\textwidth}|p{0.06\textwidth}|p{0.22\textwidth}|p{0.36\textwidth}}
        \hline
        Filter & $R$ & $\lambda_{\rm obs}$[\micron] & Target line \& Redshift & Detection limit (${\rm SNR}=10$, $T_{\rm exp}=10^{4}{\rm s}$) \\
        & & & & $S$[$\rm nJy$] \hspace{0.04\textwidth} $L_{\rm H_\alpha}$[$\erg\,{\rm s}^{-1}$] \hspace{0.03\textwidth}  $L_{{\rm H_\beta}+{[\rm O\,\Rmnum{3}]}}$[$\erg\,{\rm s}^{-1}$] \\
        \hline
        \hline
        F322W2   & $\sim$1500 & $\sim$2.5 & $\rm H_{\alpha}$ at $z=3$; ${\rm H}_{\beta} + {[\rm O\,\Rmnum{3}]}$ at $z=4$ & $\sim$9100 \hspace{0.06\textwidth}41.90 \hspace{0.08\textwidth}42.19\\
        \hline
        F444W    & $\sim$1500 & $\sim$4   & $\rm H_{\alpha}$ at $z=5$; ${\rm H}_{\beta} + {[\rm O\,\Rmnum{3}]}$ at $z=7$ & $\sim$7800 \hspace{0.06\textwidth}42.04 \hspace{0.08\textwidth}42.38\\
        \hline
        F444W    & $\sim$1500 & $\sim$4.5 & $\rm H_{\alpha}$ at $z=6$; ${\rm H}_{\beta} + {[\rm O\,\Rmnum{3}]}$ at $z=8$ & $\sim$10000 \hspace{0.052\textwidth}42.30
        \hspace{0.08\textwidth}42.59\\
        \hline
    \end{tabular}
    \caption{{\bf Detection limits of a possible observational campaign (under the assumed setup) using the {\it JWST} NIRCam slitless spectroscopy.} $R$ is the nominal resolving power and $S$ is the continuum sensitivity. Here we take the sensitivity of the Module A long-wavelength grism. Detection limits of the line luminosites were calculated by \citet{Greene2017} assuming a target ${\rm SNR}=10$ and an exposure time $T_{\rm exp}=10^{4}{\rm s}$.}
    \label{tab:nircam_sensitivity}
\end{table*}

\begin{table*}
    \centering
    \begin{tabular}{p{0.13\textwidth}|p{0.06\textwidth}|p{0.06\textwidth}|p{0.22\textwidth}|p{0.36\textwidth}}
        \hline
        Disperser \& filter & $R$ & $\lambda_{\rm obs}$[\micron] & Target line \& Redshift & Detection limit (${\rm SNR}=10$, $T_{\rm exp}=10^{4}{\rm s}$) \\
        ($\lambda$ coverage[\micron]) & & & & $S$[$\rm nJy$] \hspace{0.04\textwidth} $L_{\rm H_\alpha}$[$\erg\,{\rm s}^{-1}$] \hspace{0.02\textwidth}  $L_{{\rm H_\beta}+{[\rm O\,\Rmnum{3}]}}$[$\erg\,{\rm s}^{-1}$] \\
        \hline
        \hline
        G235M/F170LP    & $\sim$1,000 & $\sim$2   & $\rm H_{\alpha}$ at $z=2$; ${\rm H}_{\beta} + {[\rm O\,\Rmnum{3}]}$ at $z=3$ & $\sim$1100 \hspace{0.04\textwidth}\circled{1}41.41 \circled{2}40.93 \hspace{0.02\textwidth}\circled{1}41.82 \circled{2}41.36\\
        (1.66-3.07)     &             & $\sim$2.5 & $\rm H_{\alpha}$ at $z=3$; ${\rm H}_{\beta} + {[\rm O\,\Rmnum{3}]}$ at $z=4$ & $\sim$1200 \hspace{0.04\textwidth}\circled{1}41.63 \circled{2}41.23 \hspace{0.02\textwidth}\circled{1}41.97 \circled{2}41.53\\
        \hline
        G395M/F290LP    & $\sim$1,000 & $\sim$4   & $\rm H_{\alpha}$ at $z=5$; ${\rm H}_{\beta} + {[\rm O\,\Rmnum{3}]}$ at $z=7$ & $\sim$1100 \hspace{0.04\textwidth}\circled{1}41.76 \circled{2}41.50 \hspace{0.02\textwidth}\circled{1}42.09 \circled{2}41.84\\
        (2.87-5.10)     &             & $\sim$4.5 & $\rm H_{\alpha}$ at $z=6$; ${\rm H}_{\beta} + {[\rm O\,\Rmnum{3}]}$ at $z=8$ & $\sim$1500 \hspace{0.04\textwidth}\circled{1}41.95 \circled{2}41.73 \hspace{0.02\textwidth}\circled{1}42.25 \circled{2}42.02\\
        \hline
        G235H/F170LP    & $\sim$2,700 & $\sim$2   & $\rm H_{\alpha}$ at $z=2$; ${\rm H}_{\beta} + {[\rm O\,\Rmnum{3}]}$ at $z=3$ & $\sim$3000 \hspace{0.04\textwidth}\circled{1}41.85 \circled{2}41.11 \hspace{0.02\textwidth}\circled{1}42.26 \circled{2}41.54\\
        (1.66-3.05)     &             & $\sim$2.5 & $\rm H_{\alpha}$ at $z=3$; ${\rm H}_{\beta} + {[\rm O\,\Rmnum{3}]}$ at $z=4$ & $\sim$3300 \hspace{0.04\textwidth}\circled{1}42.07 \circled{2}41.45 \hspace{0.02\textwidth}\circled{1}42.41 \circled{2}41.75\\
        \hline
        G395H/F290LP    & $\sim$2,700 & $\sim$4   & $\rm H_{\alpha}$ at $z=5$; ${\rm H}_{\beta} + {[\rm O\,\Rmnum{3}]}$ at $z=7$ & $\sim$3000 \hspace{0.04\textwidth}\circled{1}42.20 \circled{2}41.74 \hspace{0.02\textwidth}\circled{1}42.52 \circled{2}42.08\\
        (2.87-5.14)     &             & $\sim$4.5 & $\rm H_{\alpha}$ at $z=6$; ${\rm H}_{\beta} + {[\rm O\,\Rmnum{3}]}$ at $z=8$ & $\sim$4000 \hspace{0.04\textwidth}\circled{1}42.38 \circled{2}42.00  \hspace{0.02\textwidth}\circled{1}42.68 \circled{2}42.28\\
        \hline
    \end{tabular}
    \caption{{\bf Detection limits of a possible observational campaign (under the assumed setup) using the {\it JWST} NIRSpec instrument.}  $R$ is the nominal resolving power and $S$ is the continuum sensitivity. Detection limits of the line luminosites are calculated assuming a target ${\rm SNR}=10$ and an exposure time $T_{\rm exp}=10^{4}{\rm s}$. We present the calculated results with the two approaches~(see Section~\ref{sec:jwst_fore}) labeled by numbers. The first approach assumes the continuum around the line is detected while the second approach does not. We note that the detection limits we use for related predictions in this paper are calculated using the second approach. Characteristics of the dispersers and filters are also presented.}
    \label{tab:nirspec_sensitivity}
\end{table*}

\begin{figure}
    \centering
    \includegraphics[width=0.48\textwidth]{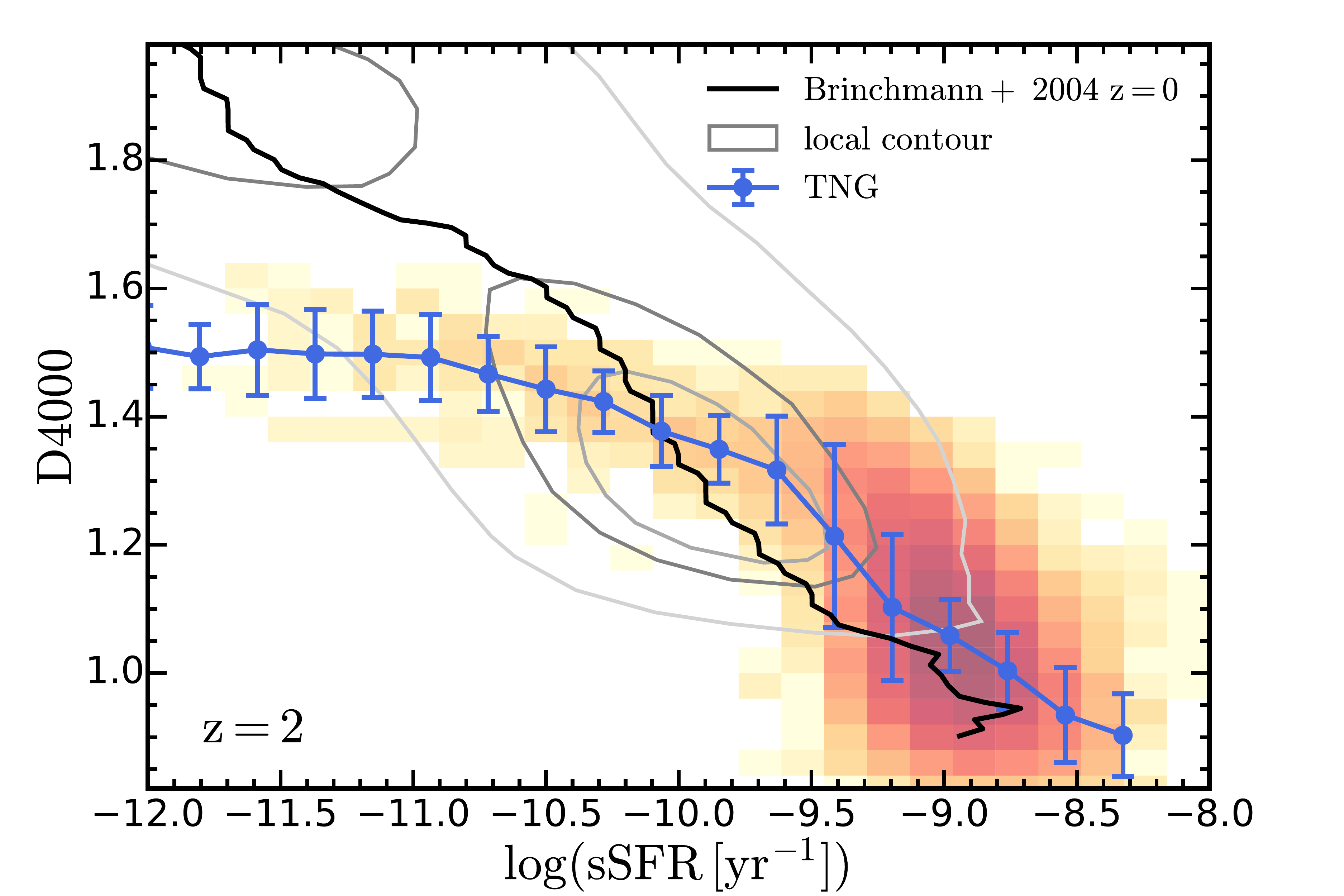}
    \caption{{\bf D4000 versus sSFR relation at $z=2$ from the IllustrisTNG simulations.} The median relation along with 1$\sigma$ dispersion are shown in the blue line and circles with error bars. The median relation and its scatter are combined results from TNG50, TNG100 and TNG300. We also present the distribution of galaxies from TNG100 with a color mesh. Darker color indicates a larger number of galaxies in the grid. We compare our results with the calibration in the local Universe~\citep{Brinchmann2004} shown in the black solid line. The distribution of the observed galaxies in the local Universe is shown with the gray contours.}
    \label{fig:d4000}
\end{figure}

\subsection{Balmer break at 4000\AA}

In addition to the emission line luminosities, another important spectral feature as a star formation indicator is the Balmer break at $\sim 4000$\AA, often referred to as D4000. In old stellar populations, opacity due to several ions in the stellar atmosphere induces a discontinuity in the flux around $4000\text{\AA}$. This spectral feature is missing in young stellar populations. Therefore, the strength of the discontinuity around $4000\text{\AA}$ can be used to measure the contribution of young stellar populations to the galaxy's total emission and equivalently the specific star formation rate~(sSFR). D4000 is usually defined as the ratio between the averaged flux in $4050-4250\text{\AA}$ and that in $3750-3950\text{\AA}$~\citep{Bruzual1983}. Unlike the emission lines we studied above, measurements of D4000 do not require very high quality spectra and are less affected by dust attenuation. It is therefore suitable for the study of star formation in galaxies at high redshift. D4000 has been used in the study of the star formation history of galaxies as well as the selection of star-forming galaxies~\citep[e.g.,][]{Kauffmann2003,Brinchmann2004,Kriek2011,Johnston2015,Haines2017}. However, a direct survey of the D4000s of galaxies at high redshift ($z\geq 2$) is still lacking.

To calculate D4000, we construct two top-hat filters to measure the averaged flux (luminosity per unit wavelength) in the wavelength range $4050-4250\text{\AA}$ and $3750-3950\text{\AA}$ respectively. These two filters are also listed in Table~\ref{tab:line_band}. The filters are also visualized in Figure~\ref{fig:filters} along with the SEDs of a star-forming galaxy in TNG50. D4000 is then calculated as the ratio between the flux measured in $4050-4250\text{\AA}$ and that measured in $3750-3950\text{\AA}$.

In Figure~\ref{fig:d4000}, we present the predicted D4000 versus sSFR relation at $z=2$. The relation combined from TNG50, TNG100 and TNG300 and its dispersion are presented with the blue line and circles. We also show the distribution of galaxies from TNG100 with a color mesh. Darker color indicates more galaxies in the grid. The dust attenuation has only a limited impact on the D4000 values~($\lesssim 0.05$), so we only show the D4000s of the dust attenuated SEDs here. The distribution pattern is similar for galaxies in TNG50/100/300, so we choose TNG100 galaxies as examples here. In the simulations, D4000 shows a tight correlation with sSFR at $z=2$. The predicted relation almost overlaps with the local calibration~\citep{Brinchmann2004} at the high sSFR end. A cluster of star-forming galaxies clearly shows up at the high sSFR end followed by a tail of passive galaxies. The pattern is quite different from the bimodal distribution found in the local Universe~\citep[e.g.,][]{Kauffmann2003,Brinchmann2004}. The D4000s of galaxies at $z=2$ are systematically lower than those in the local Universe while the sSFRs are systematically higher. This is consistent with the trend found at $z\lesssim 1$~\citep{Haines2017} that the sSFRs of massive blue cloud galaxies decline steadily with time and their D4000s rise correspondingly. Another interesting fact is that, at the low sSFR end, the tail of passive galaxies at $z=2$ does not sit on the local calibration. At fixed sSFR, the D4000s of the passive galaxies at $z=2$ are smaller than those of the passive galaxies in the local Universe. D4000 traces the age of the stellar population.  The passive galaxies at $z=2$ were quenched relatively recently and are quite young by the time. The passive galaxies at $z=0$ are overall older than those early-quenched galaxies and therefore exhibit higher values of D4000s.

\begin{figure}
    \centering
    \includegraphics[width=0.48\textwidth]{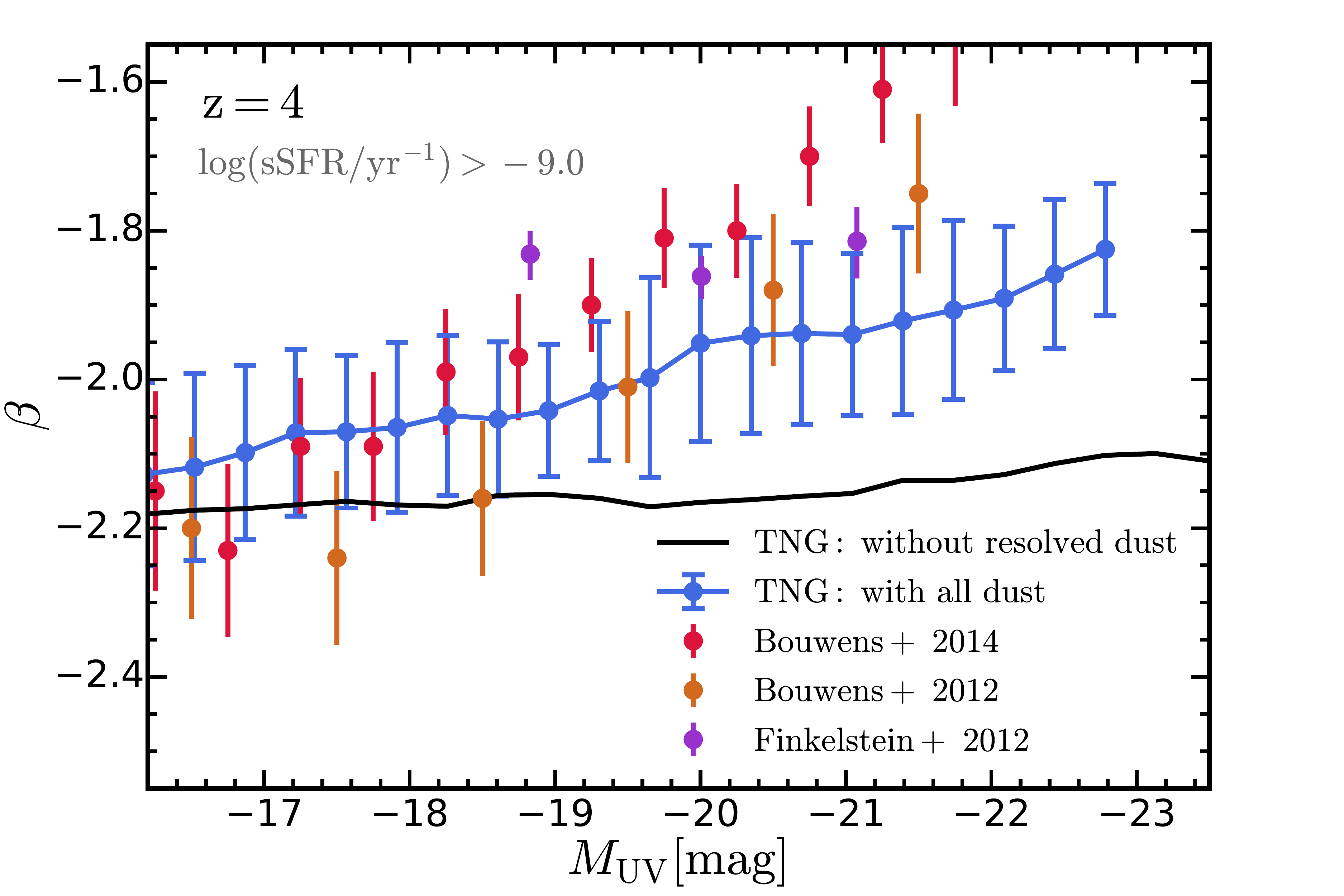}
    \includegraphics[width=0.48\textwidth]{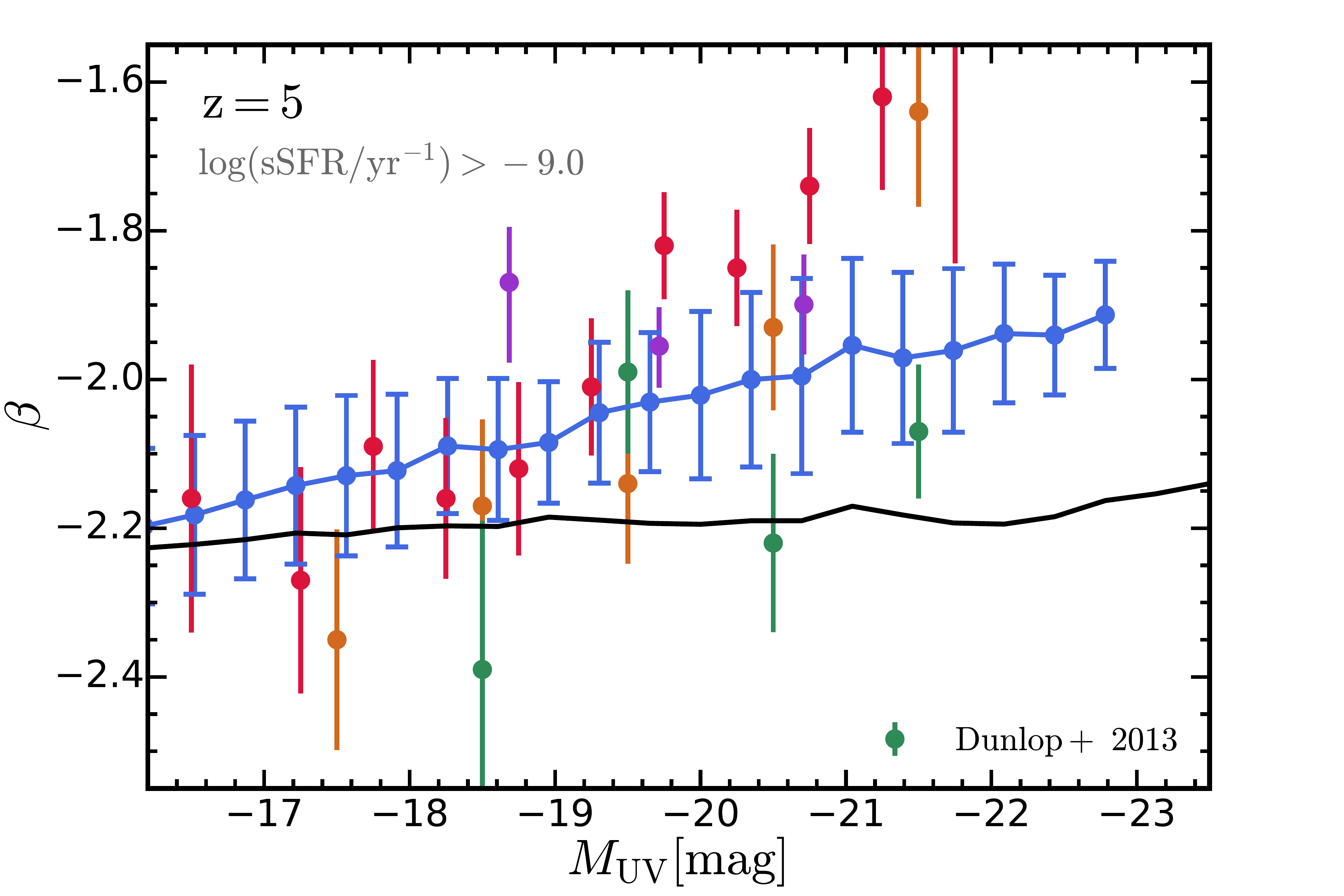}
    \includegraphics[width=0.48\textwidth]{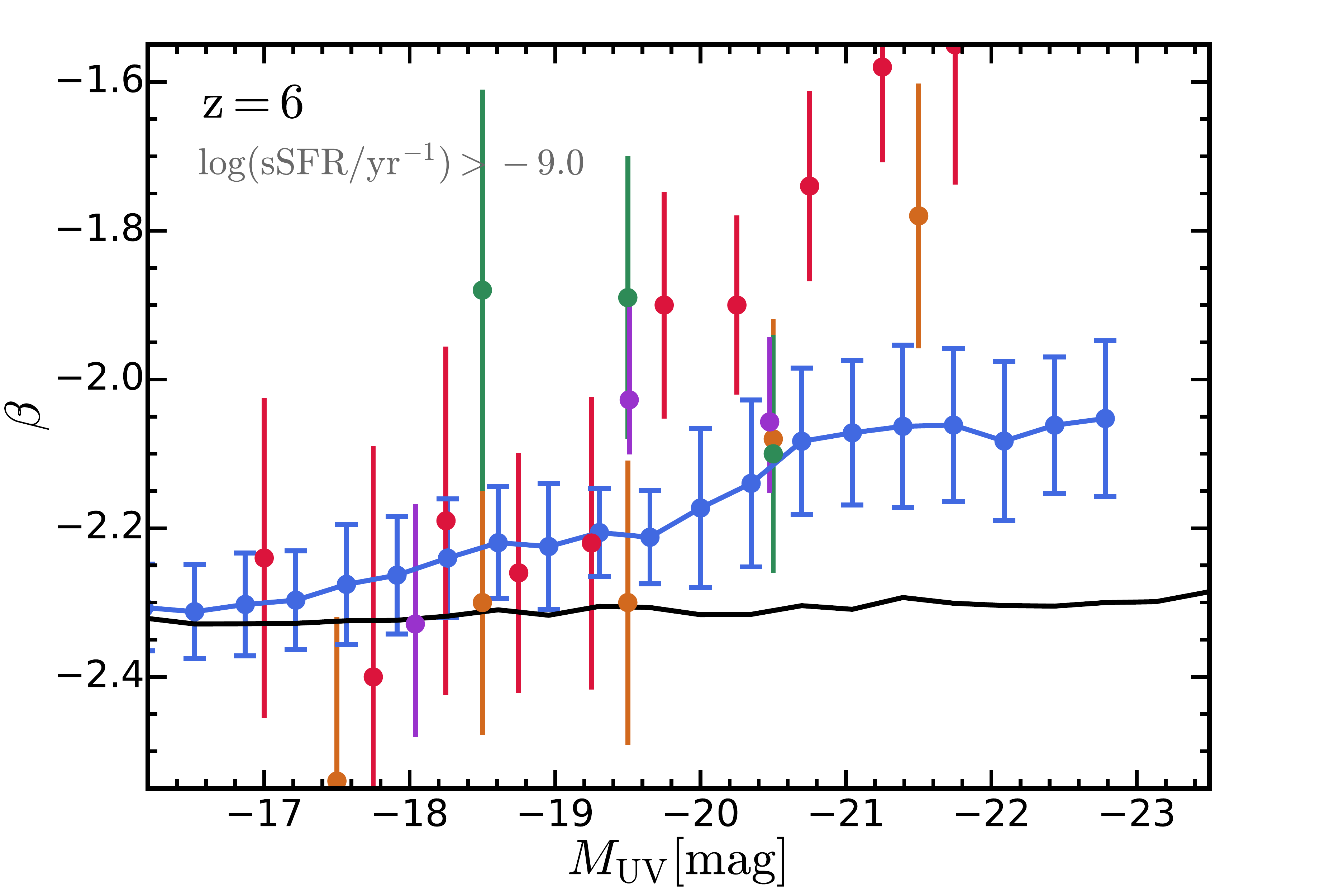}
    \caption{{\bf Predicted UV continuum slope $\beta$ versus $M_{\rm UV}$ relations at $z=4-6$ from the IllustrisTNG simulations.} Here we select only star-forming galaxies with $\log{({\rm sSFR}/{\rm yr}^{-1})}>-9.0$. Blue circles with error bars and blue lines represent relations with all dust attenuation and 1$\sigma$ dispersion. Black lines show relations when the resolved dust attenuation is not taken into account. We compare them with observations from \citet{Bouwens2012,Finkelstein2012,Dunlop2013,Bouwens2014}. The predictions are consistent with observations at the faint end. At the bright end, the predicted relation is lower than observations and the slope of the predicted relation is apparently flatter. The under-prediction of $\beta$ is related with the deficiency in heavily attenuated galaxies at the massive end as shown in Figure~\ref{fig:Auv-Mstar}. } 
    \label{fig:slope-Muv}
\end{figure}

\subsection{UV continuum slope $\beta$}
\label{sec:beta}

\citet{Calzetti1994} parameterized the UV color of galaxies through the UV continuum slope $\beta$ defined as $f_{\lambda}\sim \lambda^{\beta}$, which can be measured from galaxy spectra or multi-band photometry. \citet{Meurer1997,Meurer1999} found a correlation between galaxy far-IR dust emission excess and $\beta$, the IRX-$\beta$ relation, and thus established $\beta$ as a dust attenuation indicator~\citep[see][a study of this relation based on the IllustrisTNG]{Schulz2020}. Such a correlation was found to exist up to $z\sim 5$~\citep[e.g.,][]{Seibert2002,Overzier2011,Reddy2012,Koprowski2018}. Due to accessibility, the UV continuum slope $\beta$ acts as one of the best and widely used dust attenuation indicators in the study of high redshift galaxies. As a result, a large amount of work has been devoted to determining the distribution of $\beta$ and its dependence on galaxy luminosity at different redshifts~\citep[e.g.,][]{Meurer1999,Adelberger2000,Ouchi2004,Papovich2004,Hathi2008}. In particular, a relation between $\beta$ and galaxy rest-frame UV luminosity, or equivalently $M_{\rm UV}$, has been revealed and investigated in many studies~\citep[e.g.,][]{Bouwens2009,Finkelstein2012,Dunlop2013,Bouwens2012,Bouwens2014}. The relation has enabled an empirical estimation of dust attenuation solely based on the galaxy rest-frame UV luminosity.

\begin{figure}
    \centering
    \includegraphics[width=0.48\textwidth]{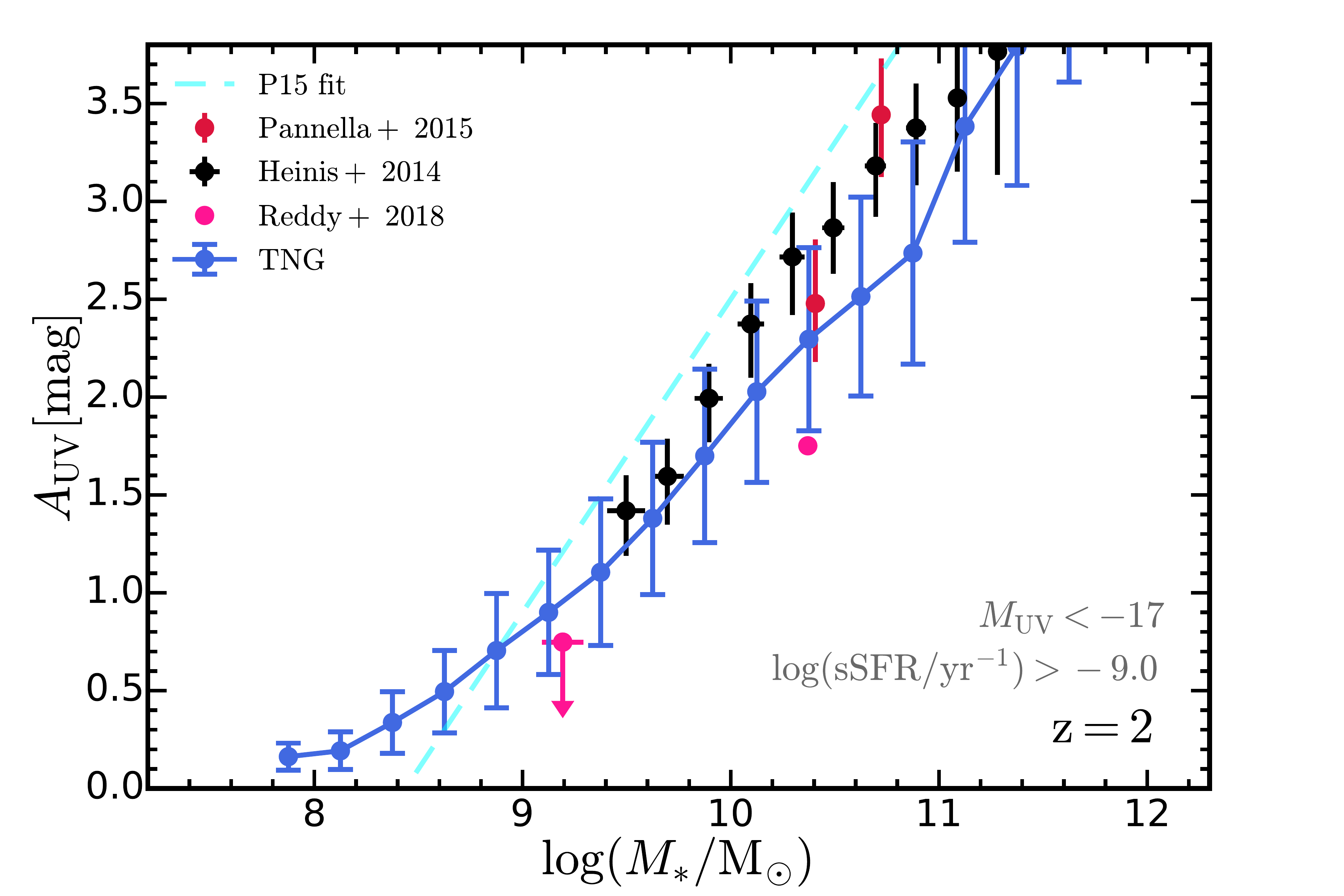}
    \includegraphics[width=0.48\textwidth]{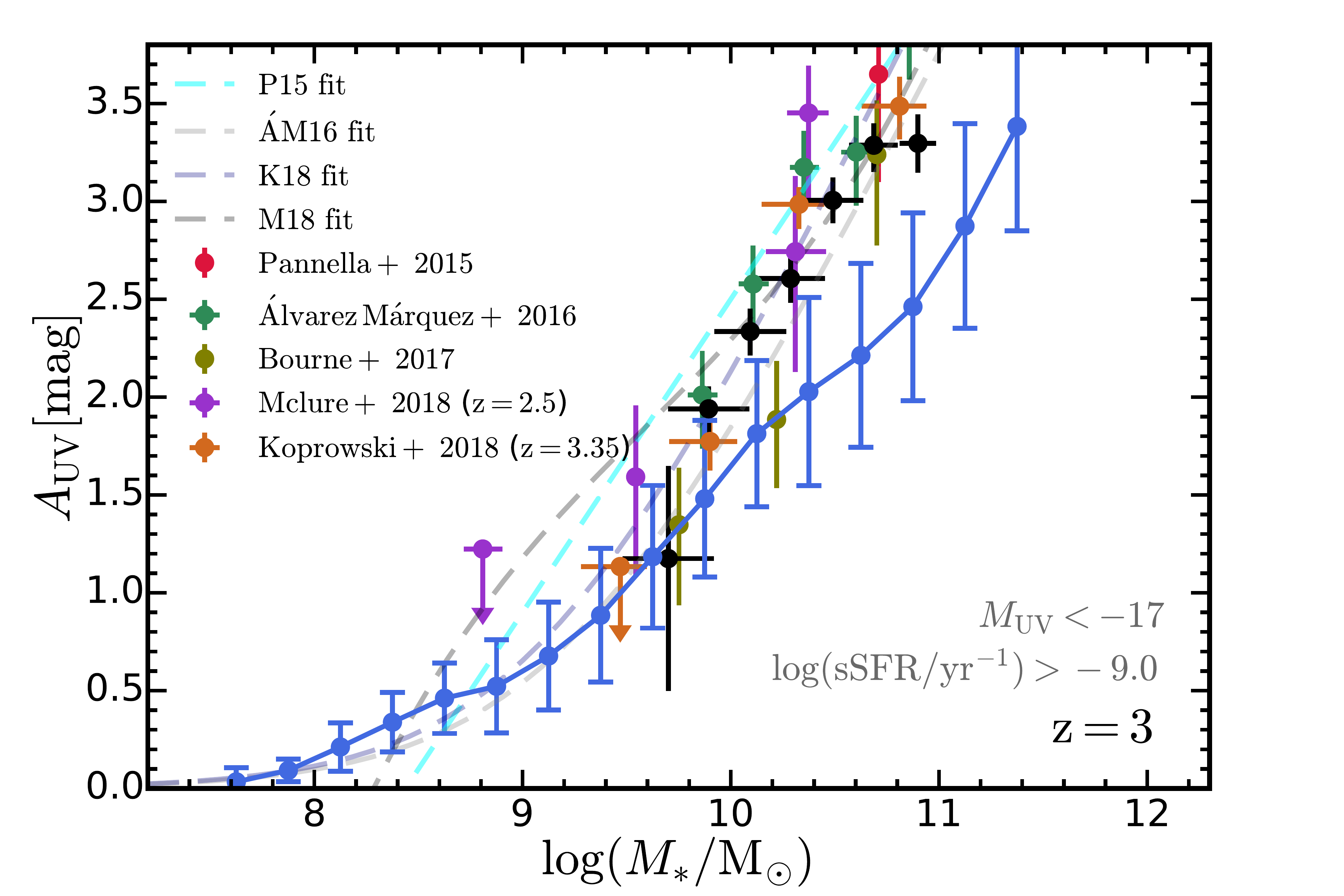}
    \includegraphics[width=0.48\textwidth]{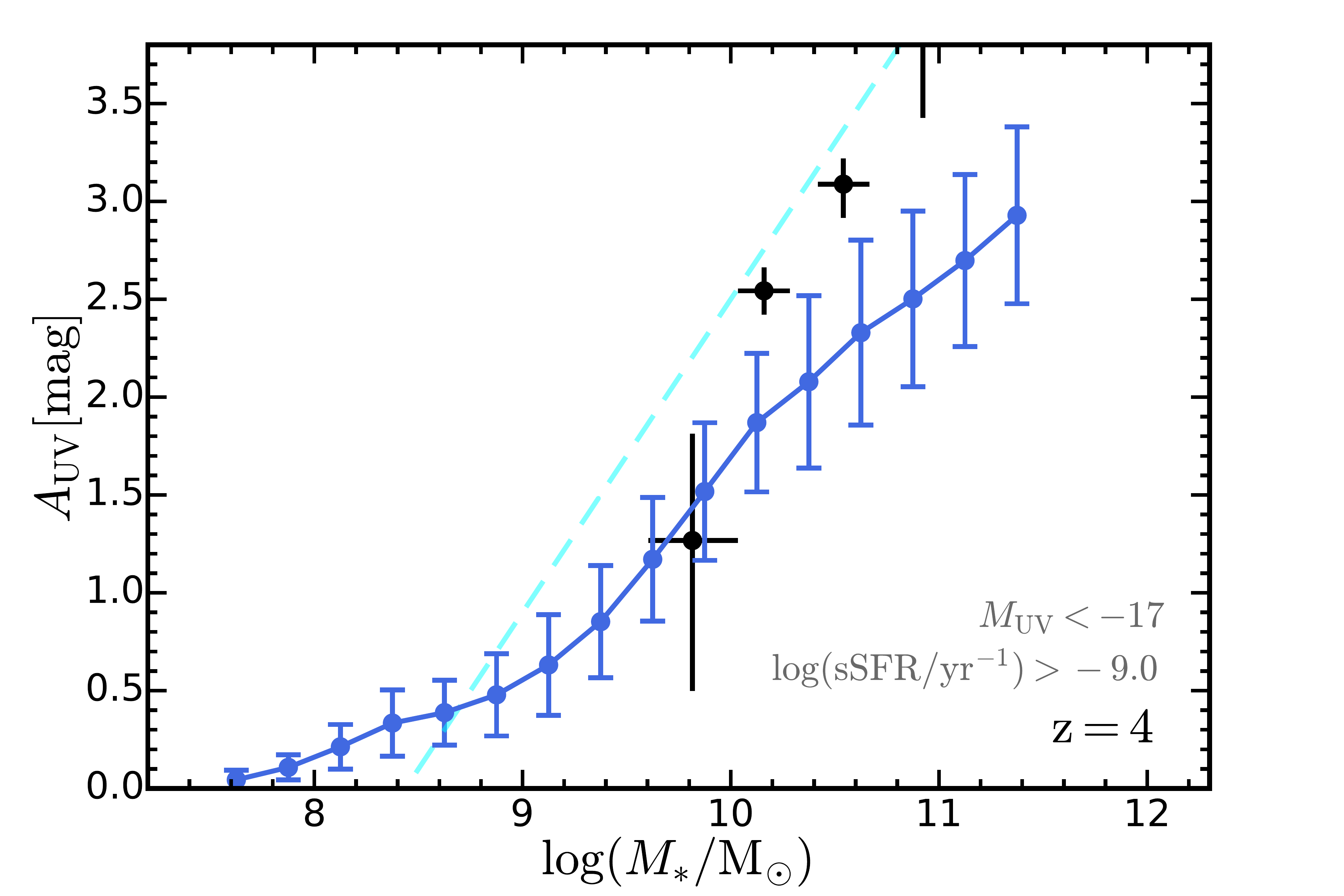}
    \caption{{\bf Predicted resolved dust attenuation in the rest-frame UV, $A_{\rm UV}$, versus stellar mass relations at $z=2-4$ from the IllustrisTNG simulations.} We use the term ``resolved dust attenuation'' because the unresolved dust component in the {\sc Mappings-\Rmnum{3}} SED library is always present and cannot be subtracted. However, we note that the impact of the unresolved dust attenuation on galaxy continuum emission is limited. Here we select only star-forming galaxies with $\log{({\rm sSFR}/{\rm yr}^{-1})}>-9.0$ and rest-frame UV magnitude $M_{\rm UV}<-17$. We compare our results with observations from \citet{Heinis2014,Pannella2015,JAM2016,Bourne2017,Koprowski2018,Mclure2018,Reddy2018}. The fitted relations in observations are shown as dashed lines. The predictions are consistent with observations at the low mass end despite the under-prediction of $A_{\rm UV}$ at the massive end.} 
    \label{fig:Auv-Mstar}
\end{figure}

In practice, the UV continuum slope $\beta$ is measured by performing a $f_{\lambda}\sim \lambda^{\beta}$ power-law fit on either photometric data~\citep[e.g.,][]{Bouwens2009,Bouwens2012,Bouwens2014} or SED synthesized based on photometric data~\citep[e.g.,][]{Finkelstein2012}. Since we have an accurate galaxy UV SED and do not rely on photometric measurements to obtain the flux, as our fiducial method, we perform a power-law fit on the galaxy UV SED from $1450\text{\AA}$ to $3000\text{\AA}$ to derive $\beta$. The wavelength range is selected to cover $1450\text{\AA}-1550\text{\AA}$ where we measured the rest-frame UV luminosity and to avoid the influence of the UV bump feature of the dust attenuation curve at $\sim 2175\text{\AA}$ on the measurement of $\beta$. To test the robustness of this method, we try alternative ways to measure the UV continuum slope: varying the maximum wavelength to $2500\text{\AA}$ and $2000\text{\AA}$ or calculating the slope purely with the fluxes at the head and the end of the wavelength range. We present a comparison between the results derived with different methods at the end of this section.

In Figure~\ref{fig:slope-Muv}, we compare the predicted $\beta$ versus $M_{\rm UV}$ relations with recent observations~\citep{Bouwens2012,Finkelstein2012,Dunlop2013,Bouwens2014}. We note that here we select only star-forming galaxies with $\log{({\rm sSFR}/{\rm yr}^{-1})}>-9.0$. This criterion is roughly $0.5\,{\rm dex}$ below the star formation main sequence measured at $z\simeq 2-6$~\citep[e.g.,][]{Salmon2015,Tomczak2016,Santini2017}. The $\Delta \log{\rm sSFR} > -0.5$ criterion has been demonstrated to be consistent with the UVJ-diagram selection adopted in observations~\citep{Fang2018,Donnari2019,Pillepich2019}. At $z\gtrsim 2$, the main sequence sSFR is slightly lower at the massive end~\citep[e.g.,][]{Tomczak2016}. So we also explore varying this selection criterion to $-9.5$, but we find that our results are not significantly affected. When the resolved dust attenuation is not taken into account (referred to as ``intrinsic''), $\beta$ remains a constant value $\sim -2.1-2.3$ that is independent of galaxy rest-frame UV luminosities and becomes lower at higher redshift where stellar populations are younger. At the same redshift, the stellar populations of star-forming galaxies have similar ages~(the sSFRs and mass doubling time scales of galaxies are similar), which results in ``intrinsic'' $\beta$s almost independent of galaxy luminosities. However, the dust attenuated $\beta$ shows a clear dependence on galaxy UV luminosity, with shallower slope at higher UV luminosity. This indicates that dust attenuation is stronger in galaxies with higher UV luminosity and thus makes the UV continuum shallower. Compared with observations, 
the $\beta-M_{\rm UV}$ relation in the simulations is not as steep as that found by \citet{Bouwens2014} at the bright end. The discrepancy is more apparent towards higher redshifts indicating a deficiency of galaxies with high $\beta$ in the simulations.

\begin{figure*}
    \centering
    \includegraphics[width=0.49\textwidth]{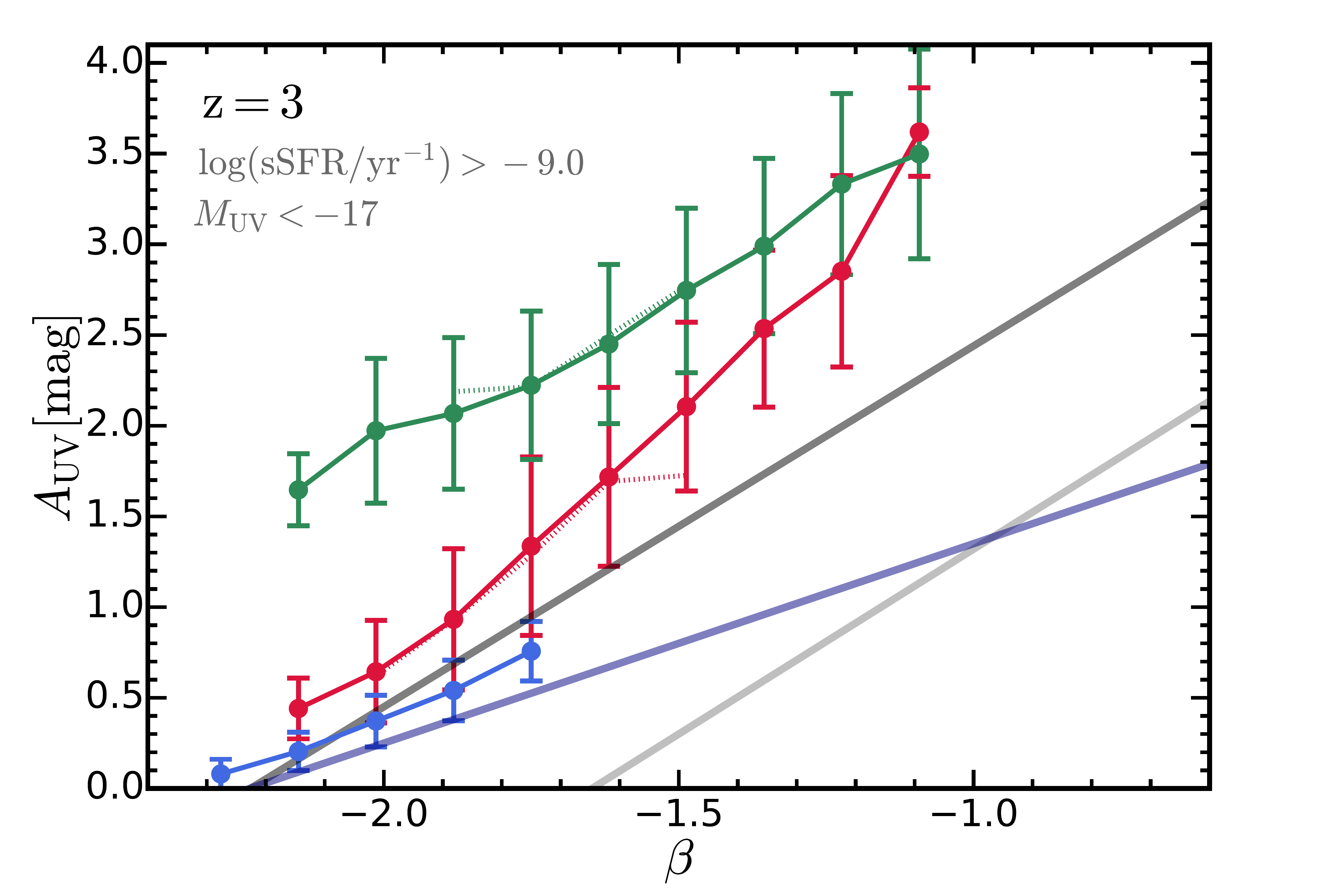}
    \includegraphics[width=0.49\textwidth]{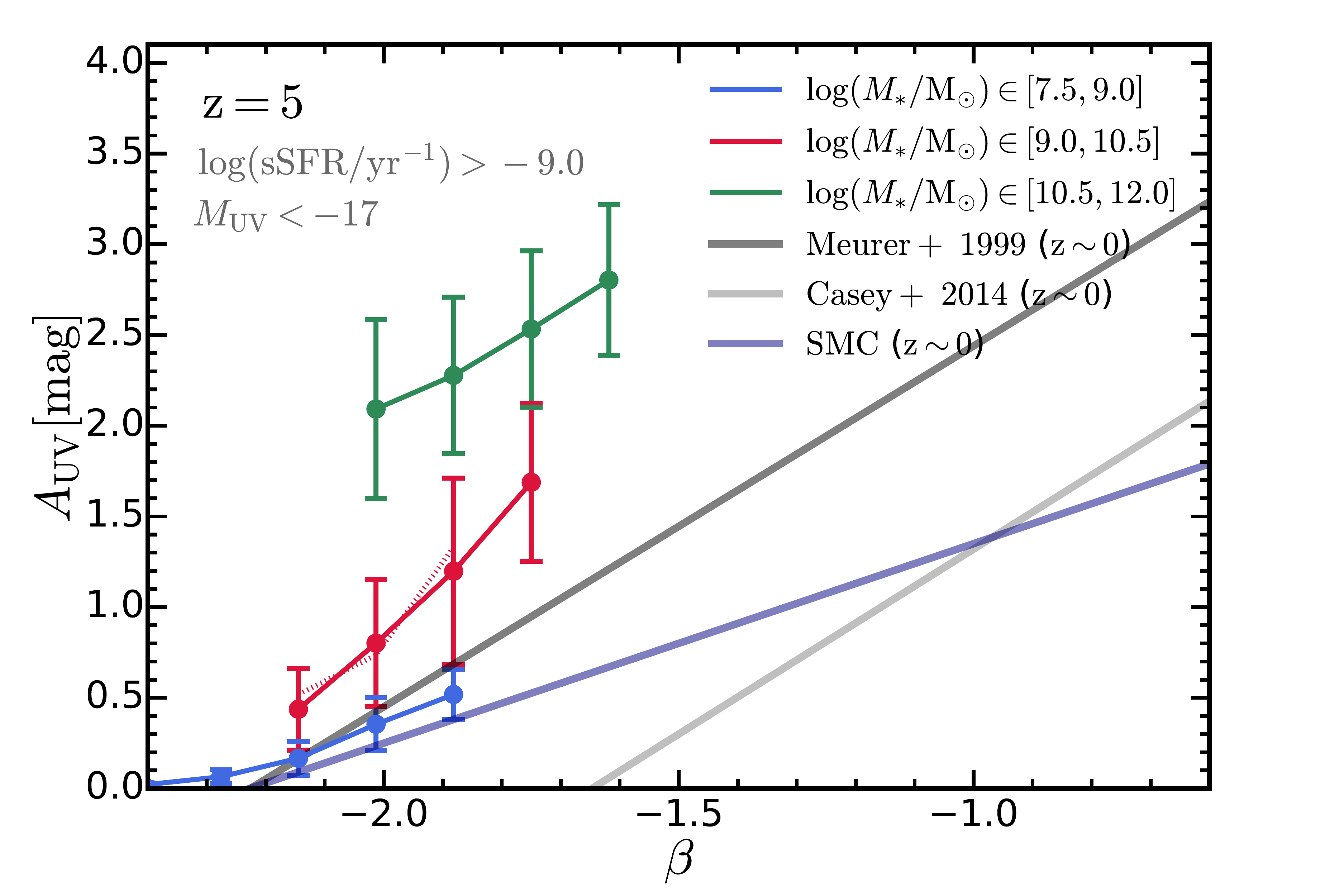}
    \caption{{\bf Predicted $A_{\rm UV}$ versus $\beta$ relations at $z=3, 5$ from the IllustrisTNG simulations.} Here we select only star-forming galaxies with $\log{({\rm sSFR}/{\rm yr}^{-1})}>-9.0$ and rest-frame UV magnitude $M_{\rm UV}<-17$. We divide galaxies into three stellar mass bins: $\log{(M_{\ast}/\msun)}\in[7.5,9.0],\,[9.0,10.5],\,[10.5,12.0]$, and evaluate the $A_{\rm UV}-\beta$ relation in each stellar mass bin. Blue circles with error bars and blue solid lines show the median relation with 1$\sigma$ dispersion in the $\log{(M_{\ast}/\msun)}\in[7.5,9.0]$ bin from TNG50. Red solid lines show the relation in the $\log{(M_{\ast}/\msun)}\in[9.0,10.5]$ bin from TNG100. We also present the relation in the same stellar mass bin from TNG50 with red dotted lines, which almost overlaps with the one from TNG100. This indicates that the stellar mass does control the $A_{\rm UV}-\beta$ relation. Similarly, we show the relation in the $\log{(M_{\ast}/\msun)}\in[10.5,12.0]$ bin from TNG300 with green solid lines and the relation in the same stellar mass bin from TNG100 with green dotted lines. These results are compared with the canonical relations~\citep{Meurer1999,Casey2014}. We find that the $A_{\rm UV}-\beta$ relations predicted at high redshifts generally lie at higher positions~(stronger attenuation or bluer continuum) on the $A_{\rm UV}-\beta$ plane than the canonical relations. The $A_{\rm UV}-\beta$ relations of more massive galaxies tend to lie at higher positions.} 
    \label{fig:slope-Auv}
\end{figure*}

As discussed, the relationship between galaxy FIR excess (IRX) and $\beta$ established $\beta$ as a dust attenuation indicator. By definition, we have:
\begin{equation}
    {\rm IRX}=\dfrac{L_{\rm IR}}{L_{\rm UV}}=B\dfrac{L^{\rm absorbed}_{\rm UV}}{L^{\rm dust}_{\rm UV}}=B\dfrac{L^{\rm nodust}_{\rm UV}-L^{\rm dust}_{\rm UV}}{L^{\rm dust}_{\rm UV}}=B(10^{0.4A_{\rm UV}}-1),
\end{equation}
where $B$ is a constant correction factor describing how much the far-IR luminosity can represent the total amount of energy emitted by dust and how much the dust absorption in the UV can represent the total amount of energy absorbed by dust. The typical value for this constant is $1.66-2.08$~\citep[e.g.,][]{Meurer1999,Mclure2018}. Therefore, IRX has a direct connection with the attenuation in the UV. The IRX versus $\beta$ relation can be translated to the $A_{\rm UV}$ versus $\beta$ relation and vice versa. The $A_{\rm UV}-\beta$ relation is often parameterized linearly as: $A_{\rm UV}=k\beta+C$. However, this observational relation has a large scatter, $\sim 1$ dex, and no consensus has been reached on parameter choices for this relation at high redshift. In the simulations, we find that the resolution corrected relations from TNG100, TNG300 do not overlap either with each other or with the relation derived from TNG50. This indicates that there are other galaxy properties that affect the $A_{\rm UV}-\beta$ relation and such galaxy properties are different in galaxies selected from TNG50/100/300. The phenomenon is consistent with the findings in observations that galaxies of different types lie differently on the IRX-$\beta$ or $A_{\rm UV}-\beta$ plane. A simple and straightforward candidate for this controlling property would be the stellar mass, since there is a well-established relation between $A_{\rm UV}$ and stellar mass~\citep[e.g.,][]{Pannella2009,Reddy2010,Buat2012,Heinis2014,Pannella2015,JAM2019} which is even tighter than the $A_{\rm UV}-\beta$ relation. Recently, \citet{JAM2019} even proposed a new relation combining $\beta$ and stellar mass as the dust attenuation proxy. In Figure~\ref{fig:Auv-Mstar}, we predict the $A_{\rm UV}-M_{\ast}$ relations at $z=2-4$ compared with observations from \citet{Heinis2014,Pannella2015,JAM2016,Bourne2017,Koprowski2018,Mclure2018,Reddy2018}. Here we have also restricted our analysis to star-forming galaxies with $\log{({\rm sSFR}/{\rm yr}^{-1})}>-9.0$ and rest-frame UV magnitude $M_{\rm UV}<-17$. Again, we find the results here not affected by varying the selection criterion for sSFR. We note that the $A_{\rm UV}$ presented is the resolved dust attenuation and the unresolved dust attenuation is subdominant in affecting broadband photometry. In the simulations, there is a tight relation between $A_{\rm UV}$ and stellar mass. The relation is approximately linear on the $A_{\rm UV}-\log{M_{\ast}}$ plane when $\log{(M_{\ast}/\msun)}\in [9,11]$. The relation becomes flatter towards the low mass end, which is consistent with the findings in \citet{Mclure2018}. We predict consistent results with existing observations at the low mass end, but under-predict the attenuation at the massive end. The discrepancy is more apparent at higher redshift and could reach $\sim 1\mmag$ at the massive end at $z=4$. This is consistent with what we have found in the $\beta$ versus $M_{\rm UV}$ relation that there is a deficiency of heavily attenuated, UV red galaxies in the simulations at the massive (bright) end. 

Acknowledging the influence of stellar mass on the UV attenuation, we next study the $A_{\rm UV}-\beta$ relation at fixed stellar masses. For each simulation, we divide galaxies into three stellar mass bins, $\log{(M_{\ast}/\msun)}\in[7.5,9.0],\,[9.0,10.5],\,[10.5,12]$, and calculate the $A_{\rm UV}-\beta$ relation in each stellar mass bin respectively. Here we have also restricted our analysis to star-forming galaxies with $\log{({\rm sSFR}/{\rm yr}^{-1})}>-9.0$ and rest-frame UV magnitude $M_{\rm UV}<-17$. In Figure~\ref{fig:slope-Auv}, we present the median $A_{\rm UV}-\beta$ relation with 1$\sigma$ dispersion for each stellar mass bin. For the $\log{(M_{\ast}/\msun)}\in[7.5,9.0]$ bin, we show the relation of TNG50 galaxies as blue solid lines. For the $\log{(M_{\ast}/\msun)}\in[9.0,10.5]$ bin, we show the relation of TNG100 galaxies as red solid lines and the relation of TNG50 galaxies as red dotted lines. For the $\log{(M_{\ast}/\msun)}\in[10.5,12.0]$ bin, we show the relation of TNG300 galaxies as green solid lines and the relation of TNG100 galaxies as green dotted lines. We compare our results with the canonical relations measured in the local Universe: the \citet{Meurer1999} relation, the \citet{Casey2014} relation and the SMC relation derived in \citet{Bouwens2016}. The median relations for the same stellar mass bin from different simulations overlap, which indicates that the stellar mass does control the $A_{\rm UV}-\beta$ relation. The slope of the $A_{\rm UV}-\beta$ relation is shallower in the high and low stellar mass bins. The relation in the intermediate stellar mass bin lies above the canonical \citet{Meurer1999} relation while its slope is close to the \citet{Meurer1999} relation. The relation in the low stellar mass bin lies below the relation derived in the intemediate mass bin and its slope is close to the SMC relation. Galaxies in the high stellar mass bin that encounter the strongest dust attenuation lie significantly above all the canonical relations in the local Universe. In observations, a consistent picture has emerged in the local Universe. Metal and dust poor galaxies tend to have the SMC-like $A_{\rm UV}-\beta$ relation with redder UV continuum and less dust, which corresponds to the simulated galaxies in the low stellar mass bin; normal star-forming galaxies as those in the intermediate mass bin exhibit the canonical relations~\citep[e.g.,][]{Buat2005,Buat2010,Seibert2005,Cortese2006,Boquien2012,Mateos2009,Takeuchi2010,Hao2011,Overzier2011}. Infrared luminous galaxies were found to lie significantly above the canonical relations~\citep[e.g.,][]{Goldader2002,Burgarella2005,Buat2005,Howell2010,Takeuchi2010}. A systematic study at $z=0-3.5$ by \citet{Casey2014} showed that galaxies will lie at higher position on the $A_{\rm UV}-\beta$ plane if they have higher rest-frame IR luminosity (more dusty) or higher rest-frame UV luminosity, which is consistent with what we find in the high stellar mass bin. Due to the complexity in the $A_{\rm UV}-\beta$ relation, one should be cautious in using $\beta$ as a dust attenuation indicator. The inferred $A_{\rm UV}$ from the same $\beta$ can differ up to $\sim 1\,{\rm dex}$ for galaxies with different stellar masses. This can result in large systematic discrepancies in predicting the dust attenuated galaxy luminosities. For example, in Model A of \citetalias{Vogelsberger2019}, we have adopted an empirical dust attenuation model that utilizes the $A_{\rm UV}-\beta$ relation and the $\beta-M_{\rm UV}$ relation measured in observations. We found that the canonical $A_{\rm UV}-\beta$ relation under-predicted dust attenuation at the bright end while over-predicted the dust attenuation at the faint end. There, we had to manually increase the slope of the $\beta-M_{\rm UV}$ relation to make the predicted galaxy rest-frame UV luminosity functions consistent with observations. 

For the redshift dependence of the $A_{\rm UV}-\beta$ relation, most of the predicted relations at high redshift lie significantly above the local canonical relations measured in the Local Universe. Similar phenomena have been found in observations. Galaxies at high redshift tend to lie above or on the bluer side of the canonical relations measured in the local Universe~\citep[e.g.,][]{Howell2010,Reddy2012,Penner2012,Casey2014,Schulz2020}. However, we do not find apparent redshift evolution of the $A_{\rm UV}-\beta$ relation at $z=3,5$. Both the normalizations and the slopes of the relations in different stellar mass bins are stable at $z=3,5$, despite that the population of UV red galaxies (with high $\beta$s) gradually diminishes at higher redshift.

The location of galaxies on the $A_{\rm UV}-\beta$ plane is conjectured to be affected by the star formation history and the stellar metallicities of galaxies~\citep[e.g.,][]{Kong2004,Mateos2009,Narayanan2018,Schulz2020}. Older stellar populations or higher stellar metallicities can drive the relation downward due to the reddening of galaxy intrinsic SEDs. This would explain why high redshift galaxies generally lie above the local relations since they are more dominated by young stellar populations. But the systematic dependence of the $A_{\rm UV}-\beta$ on galaxy stellar mass would not be simply explained by the differences in galaxy intrinsic SEDs. Supported by the fact that the intrinsic $\beta$ has almost no dependence on galaxy rest-frame UV luminosity as shown in Figure~\ref{fig:slope-Muv}, the mass dependence of the $A_{\rm UV}-\beta$ relation of the simulated galaxies is not mainly driven by the intrinsic properties of the stellar populations. We note that this argument may not be true for selected subsamples of galaxies, the $A_{\rm UV}-\beta$ relation of which can still strongly depend on the intrinsic properties of the stellar populations. Alternatively, some studies attribute the complexity in the $A_{\rm UV}-\beta$ relation to the differences in dust attenuation curves~\citep[e.g.,][]{Gordon2000,Burgarella2005,Boquien2009,Narayanan2018,JAM2019,Ma2019} and dust geometry~\citep[e.g.,][]{Seibert2005,Cortese2006,Boquien2009,Narayanan2018}. 
For example, optically thin UV sightlines in galaxies with complex geometries would result in a bluer UV continuum. On the other hand, steep SMC-like attenuation curves would result in a redder UV continuum. We defer the discussion of the dust attenuation curve and dust geometry to the next section.

\begin{figure}
    \centering
    \includegraphics[width=0.48\textwidth]{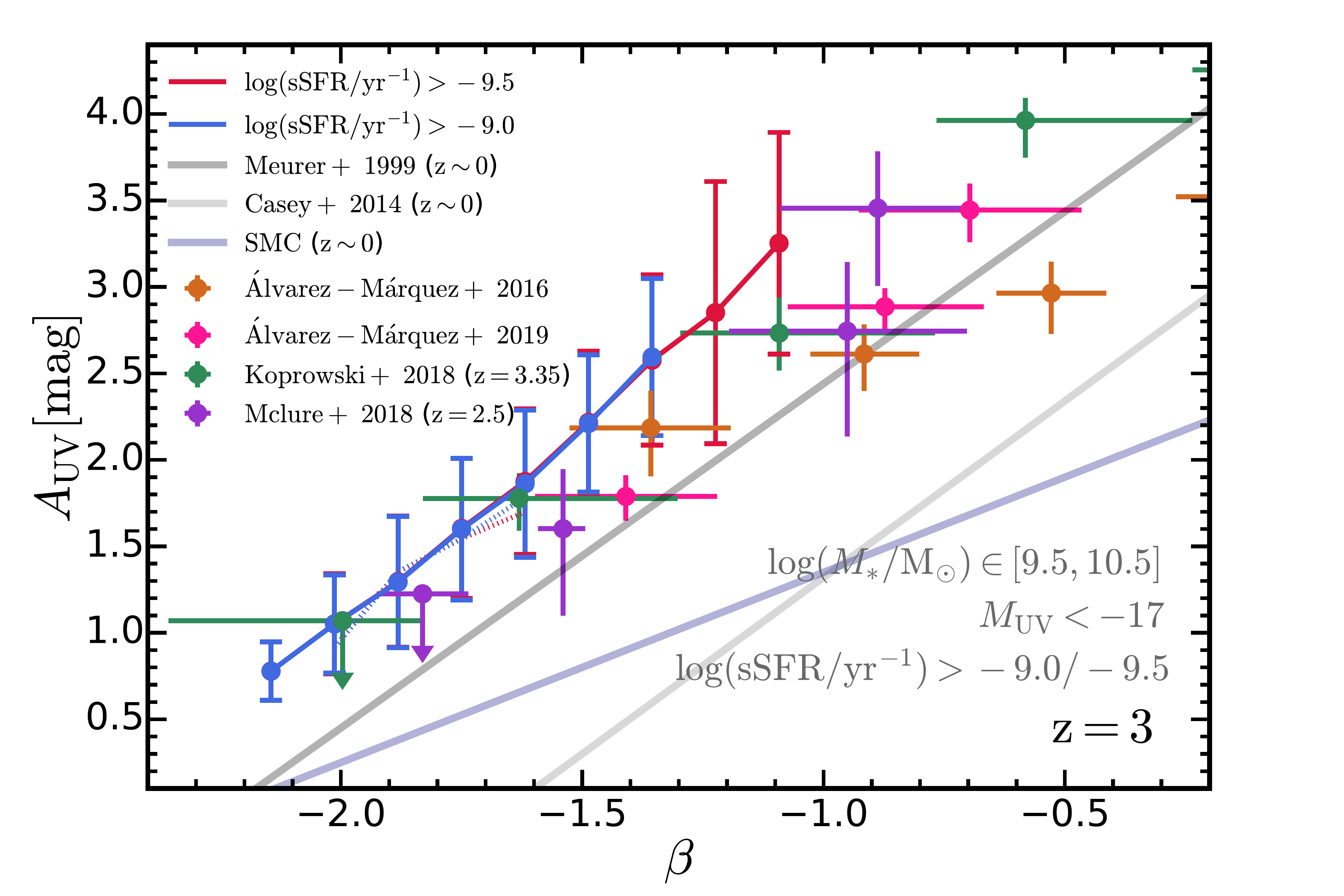}
    \includegraphics[width=0.48\textwidth]{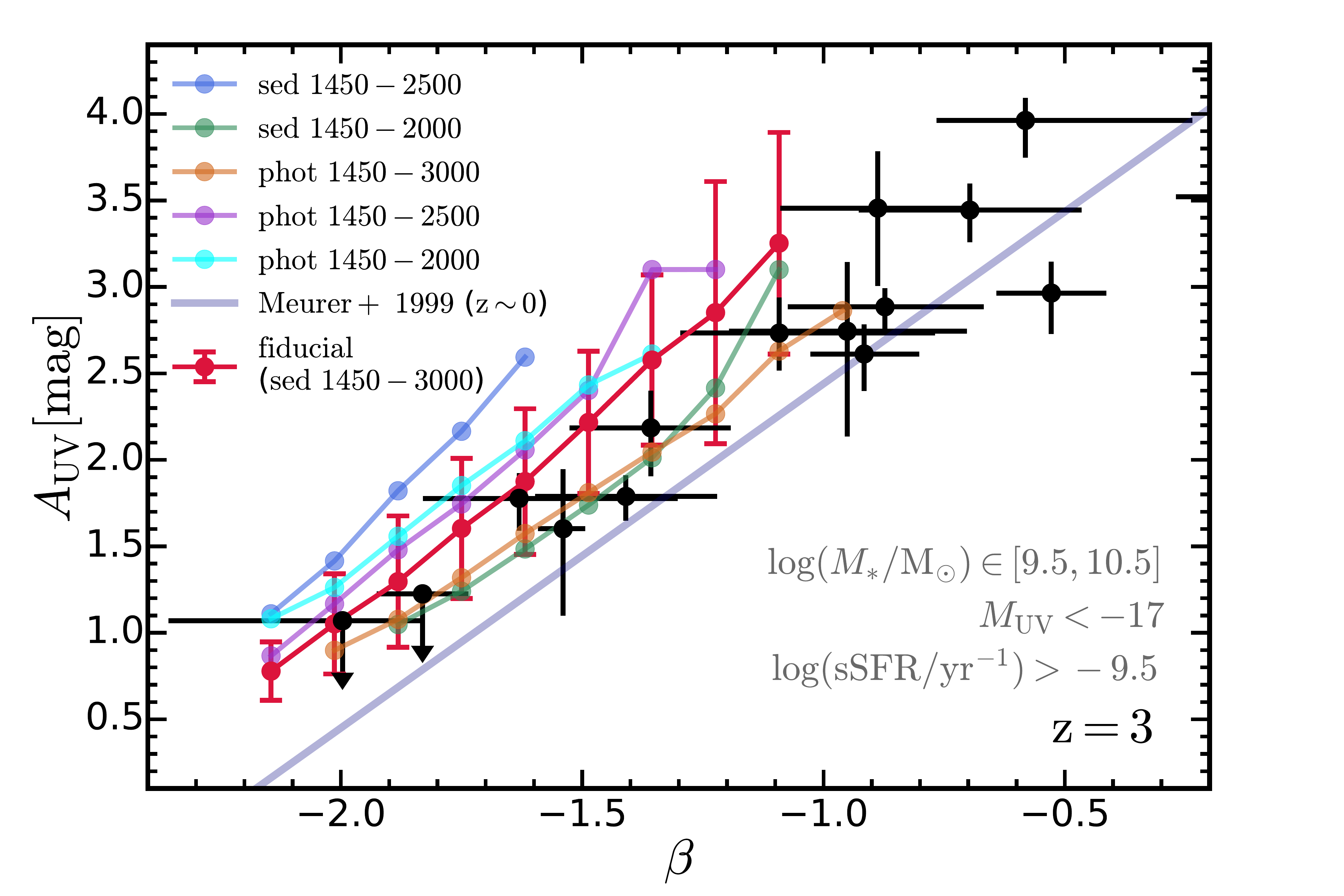}
    \caption{{\bf Predicted $A_{\rm UV}$ versus $\beta$ relations at $z=3$ from the IllustrisTNG simulations.} {\it Top:} Predicted $A_{\rm UV}$ versus $\beta$ relation in stellar mass bin $\log{(M_{\ast}/\msun)}\in [9.5,10.5]$. Here we select only star-forming galaxies with rest-frame UV magnitude $M_{\rm UV}<-17$. We explore the selection criterion on sSFR between $\log{({\rm sSFR}/{\rm yr}^{-1})}>-9.0$ and $\log{({\rm sSFR}/{\rm yr}^{-1})}>-9.5$ and show the results with the two criteria explicitly. We compare the results with selected observations~\citep{JAM2016,Koprowski2018,Mclure2018,JAM2019} at $z=3$ that have similar stellar mass binning. We also show canonical relations~\citep{Meurer1999,Casey2014} with transparent lines. The predicted relation is in reasonable agreement with observations in this stellar mass restricted comparison. However, there is still a lack of galaxies with $\beta>-1$ in the simulations in this stellar mass bin. The relations derived with different selection criteria overlap. The selection with the higher sSFR threshold $-9.5$ tends to include galaxies with high $\beta$s, but choosing a sSFR threshold lower than $-9.5$ would not change the relation further. So the lack of galaxies with $\beta>-1$ is not affected by the selection criterion. {\it Bottom:} $A_{\rm UV}$ versus $\beta$ relations derived with different methods. The label ``sed'' indicates SED fitting approach while ``phot'' indicates photometric approach. The prediction from our fiducial approach is shown in the red line. Observational data points are the same as shown in the top panel and are all shown in black circles here for clarity.}
    \label{fig:Auv-beta-detail}
\end{figure}

In the top panel of Figure~\ref{fig:Auv-beta-detail}, we provide a detailed comparison between the predicted $A_{\rm UV}-\beta$ relation with observations at $z\simeq 3$~\citep{JAM2016,Koprowski2018,Mclure2018} and the canonical relations~\citep{Meurer1999,Casey2014}. In matching the stellar masses of the selected galaxies in observations ($\sim 10^{10}\msun$), we choose a stellar mass bin $\log{(M_{\ast}/\msun)}\in[9.5,10.5]$ and correspondingly choose TNG100 to make the comparison. Here we have also restricted our analysis to star-forming galaxies with rest-frame UV magnitude $M_{\rm UV}<-17$. We switch the selection criterion on sSFR between $\log{({\rm sSFR}/{\rm yr}^{-1})}>-9.0$ and $\log{({\rm sSFR}/{\rm yr}^{-1})}>-9.5$ and show the results with the two criteria explicitly. The prediction is in reasonable agreement with these observations at $1\sigma$ level and lies above the local canonical relation~\citep{Meurer1999}. The relations derived with different selection criteria overlap. The selection with the lower sSFR threshold $-9.5$ tends to include galaxies with high $\beta$s, but choosing a sSFR threshold lower than $-9.5$ would not change the relation further. In the bottom panel, we compare the relations derived with different methods for calculating $\beta$ and with different wavelength range choices. We use the $\log{({\rm sSFR}/{\rm yr}^{-1})}>-9.5$ selection criterion here to make the visual comparison clearer. Our fiducial method is fitting the SED from $1450\text{\AA}$ to $3000\text{\AA}$ with a power-law. Alternatively, we experiment with determining $\beta$ purely by the flux measured by two top-hat filters at the head and the end of the wavelength range, referred to as the photometric approach and denoted with ``phot''. We also experiment with two other wavelength range choices: $1450\text{\AA}$ to $2000\text{\AA}$ and $1450\text{\AA}$ to $2500\text{\AA}$. Our fiducial approach and wavelength choice produce the result that lies at the center among all approaches. The photometric approach with $1450\text{\AA}$ and $3000\text{\AA}$ as the head and the end produces a slightly more consistent result with observations. This is likely because the simulated UV continuum SED does not have a perfect power-law shape but has a dip around $2175\text{\AA}$ caused by the UV bump feature of the dust attenuation curve. Affected by this dip, the SED fitting approach can give relatively lower $\beta$ compared with the photometric approach. 

We note that, across all our analysis on $\beta$, there is a deficiency of UV red (with high $\beta$) and heavily attenuated galaxies at high redshift in the simulations compared with observations. The deficiency appears in the $A_{\rm UV}-\beta$ relation shown in Figure\ref{fig:slope-Auv} and Figure~\ref{fig:Auv-beta-detail} where galaxies with $\beta\sim -0.5$ in observations are missing in the simulations. This deficiency is not affected by the selection criteria we choose and it also appears when no selection on sSFR is made. We note that the stellar masses of galaxies selected for Figure~\ref{fig:Auv-beta-detail} have already been chosen to match the masses of observed galaxies, so the deficiency is also not affected by biases in stellar masses of galaxies. Such a deficiency also manifests in the $A_{\rm UV}-M_{\ast}$ relation shown in Figure~\ref{fig:Auv-Mstar} where the simulated galaxies at the massive end encounter lower dust attenuation compared with the observed ones. The deficiency could explain the discrepancy we found at the bright end of the $\beta-M_{\rm UV}$ relation shown in Figure~\ref{fig:slope-Muv}. A straight explanation for the deficiency is that we under-estimate the dust abundance in massive galaxies. We have simply assumed a constant dust-to-metal ratio among all galaxies at a certain redshift to convert metal abundance to dust abundance. However, some previous studies have revealed that the dust-to-metal ratios increase with the metallicities of galaxies~\citep[e.g.,][]{RemyRuyer2014,Wiseman2017,Popping2017,DeVis2019}. Given that more massive galaxies have higher metalicities, we may under-estimated the abundance of dust in massive galaxies if we calibrate the model assuming a constant dust-to-metal ratio. 

\subsection{Dust attenuation curve}
\label{sec:attenuation_curve}
The dust attenuation curves describe the wavelength dependence of dust attenuation on galaxy intrinsic emission. As discussed in Model B of \citetalias{Vogelsberger2019}, attenuation is very different from the term extinction~\citep[e.g.,][]{Calzetti1994}. Extinction only considers the removal of photons from the line of sight. However, attenuation refers to the situation where radiation sources and dust have an extended and complex co-spatial distribution. Photons will not only be removed from the line of sight but can also be scattered into it from other points in the extended source. Therefore, the attenuation curve is seriously affected by the geometrical properties of galaxies and its shape can be very different from the extinction curve.

Modelling the stellar population from the observed galaxy SED requires a robust dust attenuation correction. The resulting physical properties of galaxies, e.g. stellar mass, are greatly influenced by the shape of the assumed dust attenuation curve. The translation between reddening and UV colors also depends on the shape of the dust attenuation curve, which affects the color selection criteria of galaxies. Moreover, the dust attenuation curves encode physical information on dust grains and are important for studying the cosmic evolution of dust.
In many observational and theoretical studies of high redshift galaxies, a universal attenuation curve has been assumed and the canonical curve measured in local starburst galaxies~\citep{Calzetti2000} has been widely used. However, observations have shown that the dust attenuation curves are far from universal~\citep[e.g.,][]{Kriek2013,Salmon2016,Salim2018,JAM2019}. For example: in \citet{Kriek2013}, the attenuation curve varies with the spectral shapes of galaxies; in \citet{Salmon2016}, galaxies at $z=1.5-3$ with higher color excess have shallower Calzetti-like attenuation curves and those with lower color excess have steeper SMC-like attenuation curves, indicating the non-universality of the attenuation curves.

In this paper and \citetalias{Vogelsberger2019}, we have assumed the dust model proposed by \citet{Draine2007}, which can reproduce the extinction properties of dust in the Milky Way. So, the potential influence of physical properties of dust on the attenuation curve will not be covered by our analysis. That is to say, if dust is simply placed as a foreground screen to the source, our radiative transfer calculated attenuation curve will be no different from the the Milky Way extinction curve. However, the attenuation curve of real galaxies is also influenced by geometrical properties of galaxies. We will focus on this aspect in the subsequent analysis and check to what extent the observed variety of the attenuation curves can be explained purely by dust geometry. 

The attenuation curve is usually defined as:
\begin{equation}
    A_\lambda = k(\lambda)\dfrac{A_{\rm V}}{k(\lambda_{\rm V})},
\end{equation}
where $A_{\rm V}$ is the attenuation in the V band, which controls the normalization of the curve, and $k(\lambda)$ is a function that controls the shape of the attenuation curve. To compare with the canonical attenuation curve, it is convenient to parameterize $k(\lambda)$ as~\citep[e.g.,][]{Noll2009,Kriek2013}:
\begin{equation}
    k(\lambda) = k^{\rm C}(\lambda) (\lambda/\lambda_{\rm V})^{\delta},
    \label{eq:curve}
\end{equation}
where $k^{\rm C}(\lambda)$ is the Calzetti attenuation curve~\citep{Calzetti2000}, $\delta$ is a correction factor on the slope of the attenuation curve, and $\lambda_{\rm V}$ is the wavelength of the V band. A higher $\delta$ indicates a shallower attenuation curve and vice versa. Under this parameterization, we have $k(\lambda_{\rm V})=k^{\rm C}(\lambda_{\rm V})=4.05$.  

\begin{figure}
    \centering
    \includegraphics[width=0.48\textwidth]{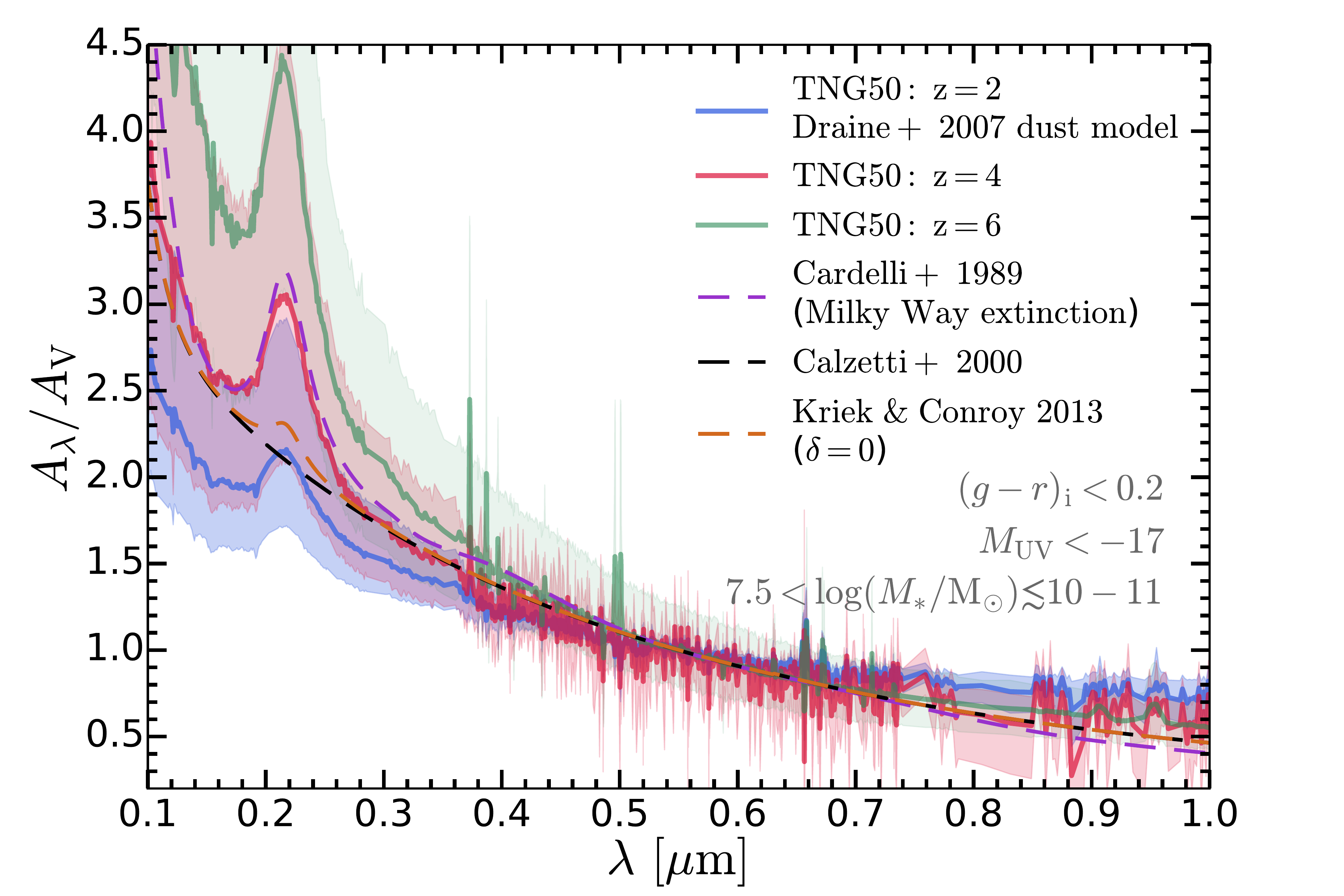}
    \includegraphics[width=0.48\textwidth]{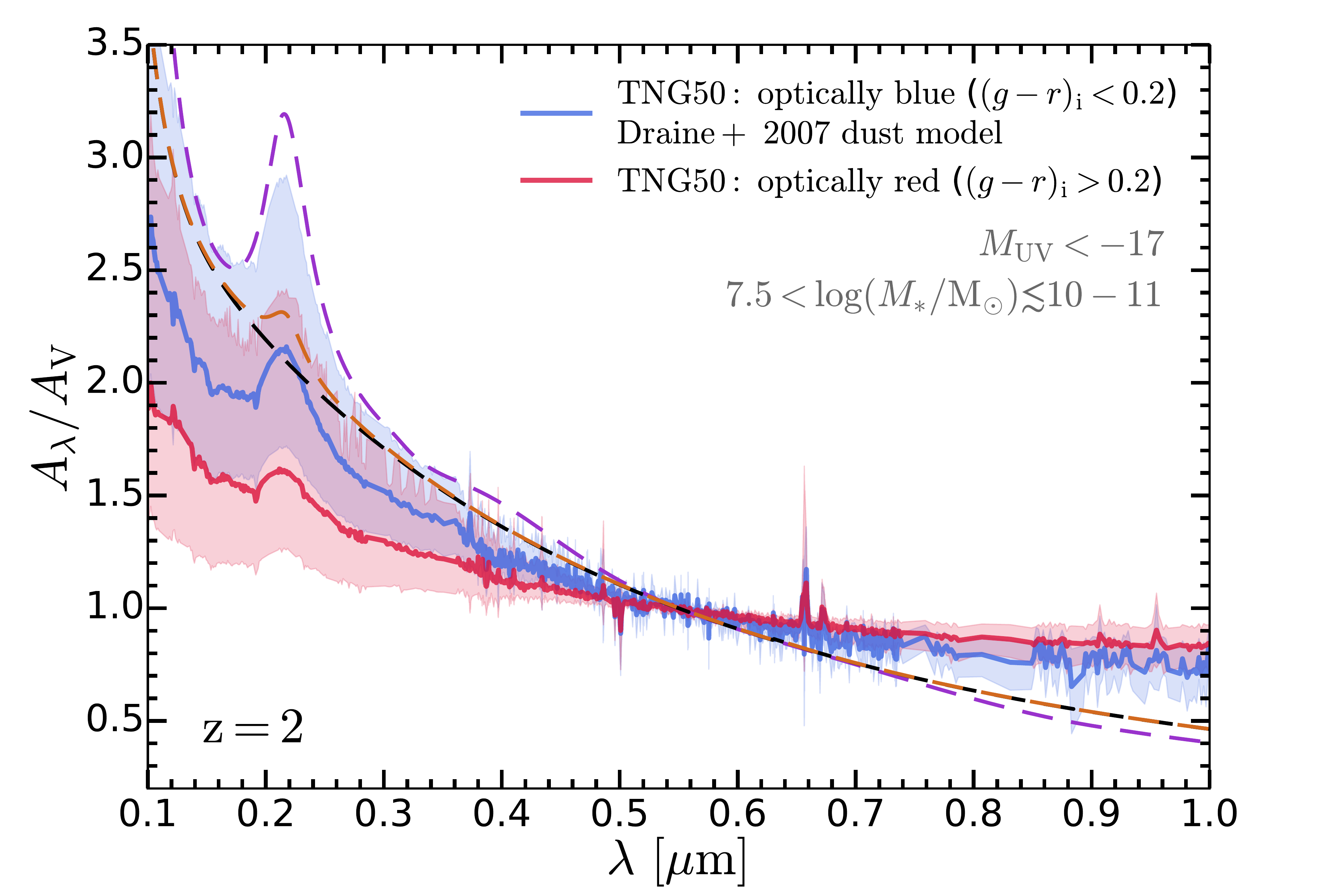}
    \caption{{\bf Predictions for the dust attenuation curves from the TNG50 simulation.} {\it Top:} Dust attenuation curves of galaxies in TNG50 at $z=2,4,6$. Only galaxies with ``intrinsic'' (without the resolved dust attenuation) optical color $g-r<0.2$, rest-frame UV magnitude $M_{\rm UV}<-17$, and stellar mass $\log{(M_{\ast}/\msun)}>7.5$ are selected. The maximum stellar mass that TNG50 can sample at these redshifts is $\log{(M_{\ast}/\msun)}\simeq 10-11$. Solid lines indicate the median dust attenuation at each wavelength while the shaded regions show 1$\sigma$ dispersion. Dashed lines show canonical attenuation curves from \citet{Cardelli1989,Calzetti2000,Kriek2013}. The dust attenuation curve of galaxies at higher redshift is steeper. {\it Bottom:} Dust attenuation curves of blue and red galaxies in TNG50 at $z=2$. Galaxies with ``intrinsic'' optical color $g-r<0.2$ are classified as blue, otherwise red. The dust attenuation curve of blue galaxies is steeper than that of red galaxies.} 
    \label{fig:attenuation_curve}
\end{figure}

\begin{figure}
    \centering
    \includegraphics[width=0.48\textwidth]{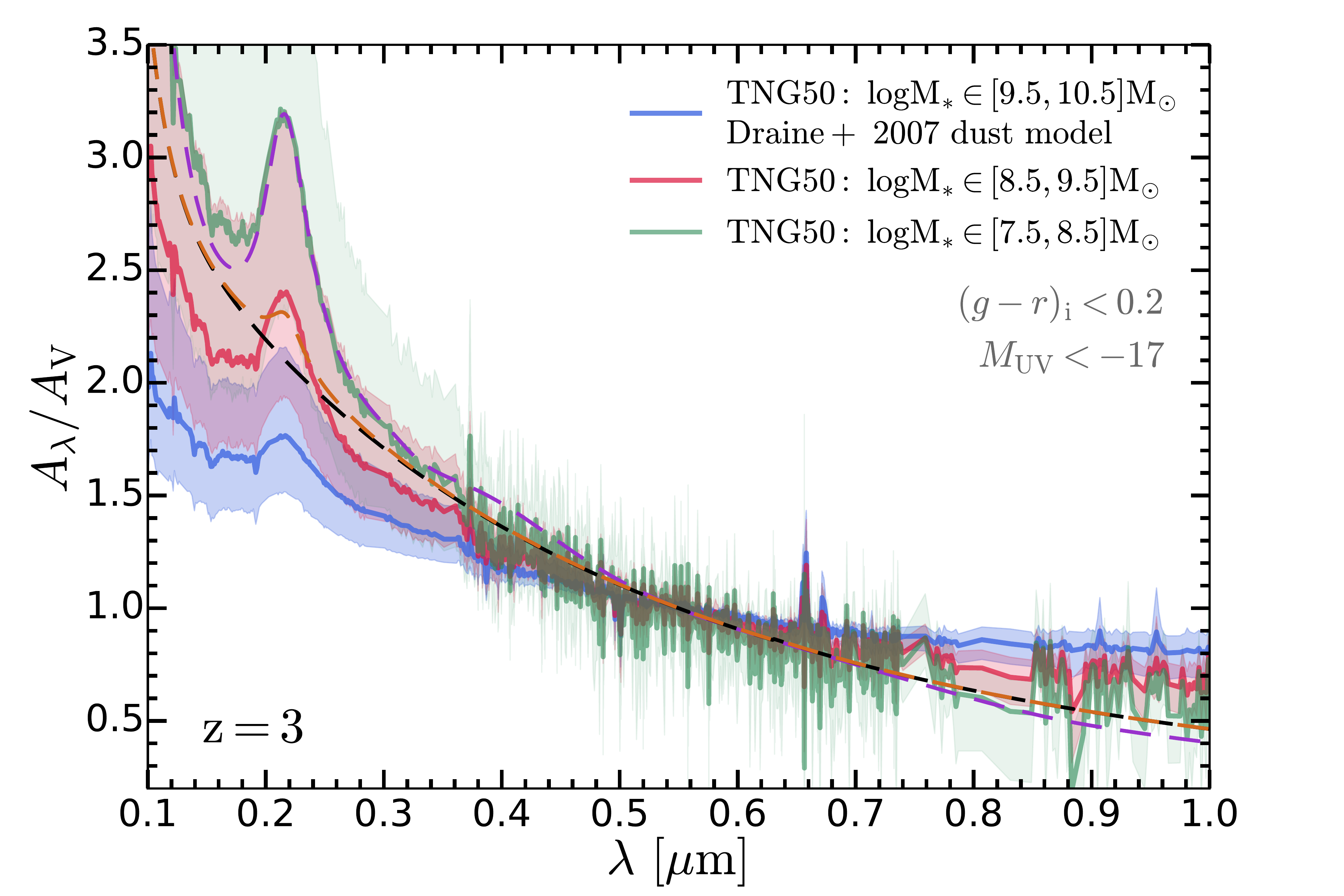}
    \includegraphics[width=0.48\textwidth]{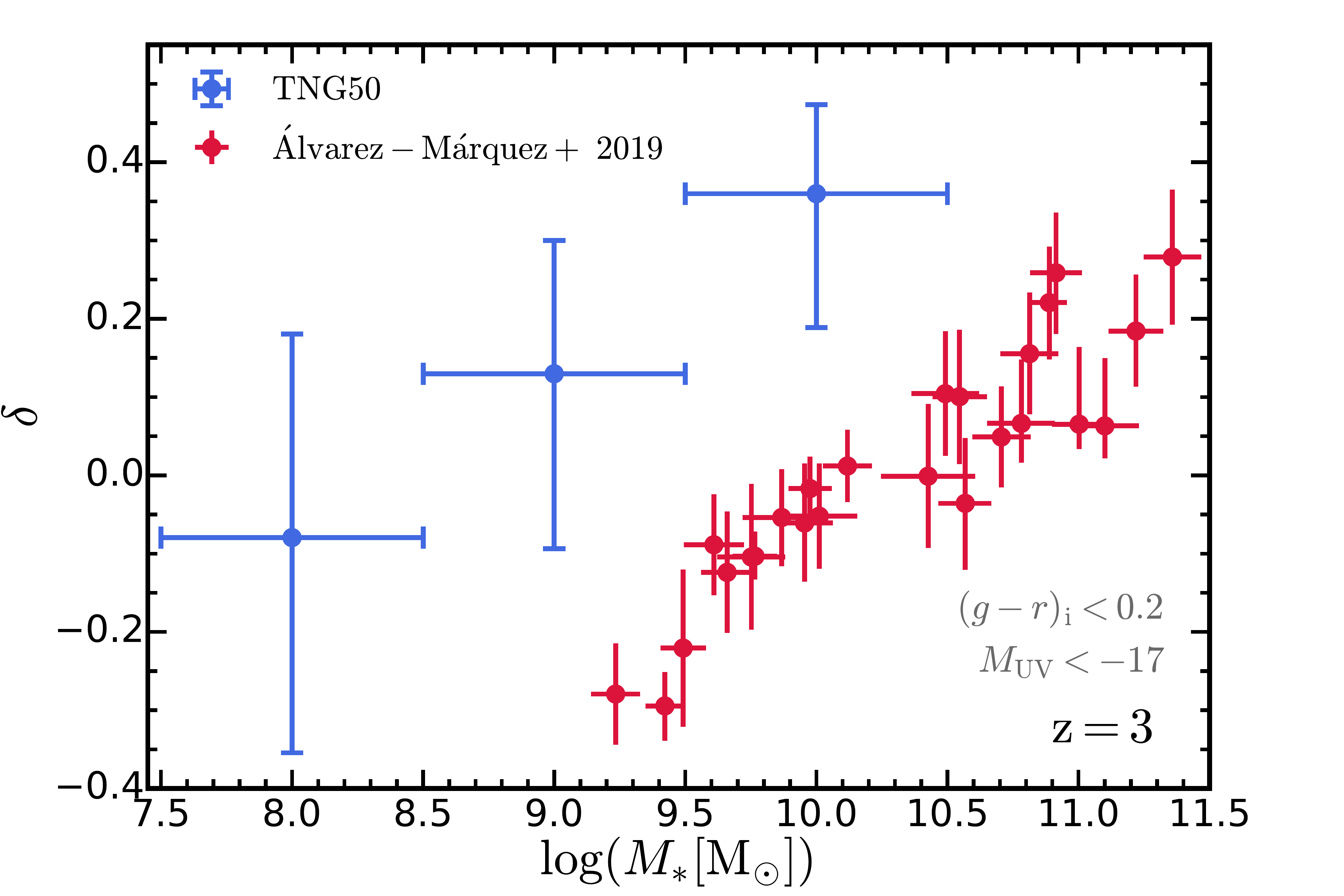}
    \caption{{\bf Predictions for the dust attenuation curves from the TNG50 simulation.} {\it Top:} Dust attenuation curves of galaxies with different stellar masses in TNG50 at $z=3$. Only galaxies with ``intrinsic'' (without the resolved dust attenuation) optical color $g-r<0.2$ and rest-frame UV magnitude $M_{\rm UV}<-17$ are selected. We divide galaxies into three stellar mass bins: $\log{(M_{\ast}/\msun)}\in[7.5,8.5]$,$[8.5,9.5]$ and $[9.5,10.5]$. Solid lines indicate the median attenuation at each wavelength while the shaded regions show 1$\sigma$ dispersion. Dashed lines show canonical attenuation curves from \citet{Cardelli1989,Calzetti2000,Kriek2013}. The dust attenuation curve of more massive galaxies is shallower. {\it Bottom:} Slope correction factor $\delta$~(defined in Equation~\ref{eq:curve}) versus galaxy stellar mass relation at $z=3$ compared with the observation from \citet{JAM2019}. We predict a qualitatively consistent trend with the observations. However, the predicted attenuation curves are less steep than the observed ones.}
    \label{fig:attenuation_curve_mass}
\end{figure}

First, we study the dust attenuation curves of star-forming galaxies at different redshifts. We calculate the median attenuation curve for galaxies in TNG50 at $z=2,4,6$ with ``intrinsic'' (without the resolved dust attenuation) optical color $g-r<0.2$, rest-frame UV magnitude $M_{\rm UV}<-17$ and stellar mass $\log{(M_{\ast}/\msun)}>7.5$. We choose these selection criteria because the calculated attenuation curve is valid only when the geometry of the galaxy is properly resolved. We also note that the attenuation calculated here is the resolved dust attenuation. In the top panel of Figure~\ref{fig:attenuation_curve}, we show the median attenuation curves of galaxies at $z=2,4,6$ along with a 1$\sigma$ dispersion. The attenuation curve at $z=2$ is shallower than the Milky Way extinction curve. This is not surprising considering the complex dust geometry in galaxies. For instance, in the case that stars and dust are uniformly co-spatially mixed, dust attenuation would depend on optical depth as~\citep{Calzetti1994}:
$A_\lambda=-\dfrac{2.5}{\ln{10}}\ln{\Big[\dfrac{1-e^{-\tau(\lambda)}}{\tau(\lambda)}\Big]}.$ Given the same optical depth $\tau(\lambda)$, the attenuation curve here would be shallower than the extinction curve which behaves as $A_\lambda = 2.5/\ln{10}\,\tau(\lambda)$. At higher redshift, we find that the attenuation curve becomes steeper and is accompanied by sharp spikes around emission lines. Both of the phenomena can be attributed to dust geometry. As shown in Figure~\ref{fig:image2}, the origin of the line emission clearly has a spatial correlation with the dust mass distribution in the galaxy. Such a correlation with dust is less apparent in the broadband galaxy image. Young and massive stars that are responsible for the continuum emission in the UV and the nebular line emission are usually embedded in environments that are rich in cold gas and dust. Therefore, their emission suffers preferentially stronger dust attenuation. Then, in terms of the attenuation curve for the entire galaxy, the attenuation in the UV and the attenuation of emission lines will become relatively higher, resulting in a steeper attenuation curve as well as spikes around emission lines in the curve. Similar phenomena have also been revealed in observations~\citep[e.g.,][]{Reddy2015} that the dust attenuation of emission lines was found to be larger than that of the continuum in star-forming galaxies, and the difference can reach as much as a factor $\sim 2$ in magnitude. The effect is stronger when the galaxy is richer in dense clumps of cold gas where star formation takes place and is stronger when the stellar population is more dominated by young stars. The steepening of the attenuation curve towards higher redshift can be explained by the younger stellar populations and the stronger star formation in high redshift galaxies. To make it more clear, we compare the attenuation curves of red and blue galaxies, shown in the bottom panel of Figure~\ref{fig:attenuation_curve}. The attenuation curve of red galaxies is clearly shallower than that of blue galaxies. 

In observations, the slope correction $\delta$ of the attenuation curve was found to have a positive correlation with both stellar mass and color excess $E(B-V)$, with shallower attenuation curves found in more massive or dusty galaxies~\citep[e.g.,][]{Salmon2016,Salim2018,JAM2019}. These findings can be explained to some extent by dust geometry. More massive galaxies are usually lower in their sSFR, less dominated by young stellar populations and thus less affected by the differential attenuation of blue emission discussed above. Moreover, the complex dust geometry in massive galaxies gives rise to more optically thin UV sightlines. Emission through these sightlines would suffer much less attenuation in the UV. Both of the two factors drive the attenuation curve shallower in more massive galaxies.

Furthermore, as revealed by the $A_{\rm UV}-M_{\ast}$ relation we presented above, more massive galaxies are more heavily attenuated and thus have higher color excess. So, dust geometry can also qualitatively explain why galaxies with higher color excess have shallower attenuation curves. We demonstrate these explanations in the top panel of Figure~\ref{fig:attenuation_curve_mass} where the attenuation curves of galaxies in the three different stellar mass bins are presented. The attenuation curves in low mass galaxies are shallower than those in massive galaxies. We fit the attenuation curves with the formula in Equation~\ref{eq:curve}. We neglect the wavelength range from $1500\text{\AA}$ to $3000\text{\AA}$ in our fit to avoid the influence of the UV bump feature. The fitted slope correction factor $\delta$ as a function of stellar mass is presented in the bottom panel of Figure~\ref{fig:attenuation_curve_mass}. Compared with the observational measurements from \citet{JAM2019}, we qualitatively produce a consistent dependence of $\delta$ on galaxy stellar mass, while finding a systematic offset $\Delta \delta \simeq 0.4$. In all three stellar mass bins, the attenuation curves are not as steep as the ones found in the observations. This indicates missing factors other than the geometrical properties of galaxies that affect the shape of the attenuation curve. The physical properties of dust grains, e.g., the grain size distribution, the chemical composition, etc. in galaxies at high redshift can be very different from those in the local Universe.

The variety in the dust attenuation curves also plays a role in the complexity of the $A_{\rm UV}-\beta$ relation. A steeper attenuation curve in the UV will result in a redder observed UV continuum with higher $\beta$. Galaxies with lower (higher) stellar masses exhibiting steeper (shallower) attenuation curves have higher (lower) UV continuum slope $\beta$. This offers an explanation for the dependence of the $A_{\rm UV}-\beta$ relation on galaxy stellar mass explored in Section~\ref{sec:beta}. In observations, the shape of the attenuation curves has also been favored as the main driver of the complexity of the $A_{\rm UV}-\beta$ relation~\citep[e.g.,][]{Salmon2016,LoFaro2017,JAM2019}.

\section{Conclusions}
\label{sec:conclusions}
In this paper, we have studied properties of galaxy populations at high redshift from the IllustrisTNG simulation suite consisting of TNG50, TNG100 and TNG300. Most importantly, the IllustrisTNG simulations employ a galaxy formation model that produces a realistic low redshift Universe that is consistent with a wide range of observational data. As introduced in \citetalias{Vogelsberger2019} of this series, galaxies from the simulations were post-processed with full Monte Carlo radiative transfer calculations to model dust attenuation. We have calibrated the dust attenuation model~(assuming the \citet{Draine2007} dust mix) based on the galaxy rest-frame UV luminosity functions measured at $z=2-10$ and have made predictions for the {\it JWST} apparent bands' luminosity functions in \citetalias{Vogelsberger2019}. In this paper, we focus on providing further predictions for high redshift galaxies that involve a variety of photometric and spectroscopic features of galaxies. We make comparisons with existing observations and provide predictions that can be tested by future {\it JWST} observations. Our findings can be summarized as:
\begin{itemize}
    \item {\bf $M_{\rm halo}-M_{\rm UV}$ relation:} The predicted $M_{\rm halo}-M_{\rm UV}$ relation at $z=4-8$ is consistent with most of the previous theoretical works with $\lesssim 0.3\,{\rm dex}$ differences in halo masses at all luminosities and redshifts. Dust attenuation plays a crucial role in shaping this relation, producing a steepening at the bright end. We confirm the success of previous empirical models that link galaxy star formation with halo assembly history. However, we find larger scatter in the predicted $M_{\rm halo}-M_{\rm UV}$ relation compared to that predicted by an empirical model~\citep{Tacchella2018}. The discrepancy indicates that the halo assembly time is not the only factor that causes the scatter in galaxy-halo scaling relations. The star formation history of a galaxy does not perfectly follow the mass accretion history of its host halo.
    
    \item {\bf Emission line luminosity functions:} We predict the ${\rm H}_{\alpha}$ luminosity functions at $z=2-5$ and derive the best-fit Schechter function parameters. The evolution of the predicted ${\rm H}_{\alpha}$ luminosity function at $z\geq 2$ is consistent with the evolutionary trend found in observations at $z\leq 2.33$. The predicted ${\rm H}_{\alpha}$ luminosity function is $\sim 0.3\,{\rm dex}$ dimmer than the observational binned estimations at $z\simeq 2$. In addition, we make predictions for the ${\rm H}_{\beta}$ + $[\rm O \,\Rmnum{3}]$ luminosity function at $z=2,3,8$ and find good agreement with observational binned estimations at $z\leq 3.2$ with $\lesssim 0.1\,{\rm dex}$ differences in luminosities. The evolutionary pattern is similar to that of the ${\rm H}_{\alpha}$ luminosity function. We calculate the detection limits of a possible future {\it JWST} observational campaign, assuming a target SNR$=10$ and an exposure time $T_{\rm exp}=10^{4}{\rm s}$. We find that, with the NIRSpec instrument, the ${\rm H}_{\alpha}$ line can be detected down to $\sim$ $10^{41.7}\, \erg/{\rm s}$ at $z=6$ and the ${\rm H}_{\beta}$ + $[\rm O \,\Rmnum{3}]$ line can be detected down to $\sim$ $10^{42.0}\,\erg/{\rm s}$ at $z=8$. With the {\it JWST} NIRSpec instrument, $\sim$ $2/{\rm arcmin}^{2}$ ${\rm H}_{\beta}$ + $[\rm O \,\Rmnum{3}]$ emitters at $z=8$ are expected to be observed assuming a survey depth $\Delta z = 1$.
    
    \item {\bf D4000 versus sSFR relation:} The predicted D4000 versus sSFR relation at $z=2$ lies on the local calibrations at the high sSFR end. The galaxies at $z=2$ distribute systematically in the high sSFR, low D4000 regime compared with local galaxies. This is consistent with the trend found at $z\lesssim1$ in observations. The distribution pattern of galaxies at $z=2$ on the D4000-sSFR plane is very different from that of local galaxies. A cluster of star-forming galaxies followed by a tail of passive galaxies shows up instead of the bimodal distribution found in the local Universe. The tail of passive galaxies has lower D4000 at fixed sSFR compared with the local calibration.
    
    \item {\bf UV continuum slope $\beta$:} Both the predicted $\beta-M_{\rm UV}$ relation and the $A_{\rm UV}$ versus stellar mass relation are only consistent with observations at the faint (low mass) end. $\beta$ ($A_{\rm UV}$) is under-predicted at the bright (massive) end. The differences are more apparent at higher redshift and could reach $\sim 1\mmag$ at $\log{(M_{\ast}/\msun)}\simeq 11$ at $z=4$. These discrepancies indicate a deficiency of galaxies with high $\beta$ or strong attenuation in the simulations. We study the $A_{\rm UV}-\beta$ relation in different stellar mass bins at $z=3,5$ and find that galaxies with higher (lower) stellar mass lie higher (lower) on the $A_{\rm UV}-\beta$ plane. Such a dependence can systematically explain observational findings for this relation. At the stellar mass similar to that of observed galaxy samples, the predicted $A_{\rm UV}-\beta$ relation is consistent with observations at $z\simeq 3$ at $1\sigma$ level. 
    
    \item {\bf Dust attenuation curves:} We focus on the influence of dust geometry on the shape of the dust attenuation curves of galaxies. We study the dependence of the attenuation curve on redshift, galaxy optical color, and stellar mass. We find that the attenuation curves are steeper in galaxies at higher redshifts, with lower stellar masses, or bluer optical colors. The dependence can be related to a positive correlation between curve steepness and domination of young stellar populations. The continuum emission at short wavelengths from young stellar populations suffers preferentially stronger dust attenuation in the birth environment of stars, which results in a steeper attenuation curve. Our findings are qualitatively consistent with the observed trend despite that the predicted curves are not as steep as the observed ones. This indicates that the geometrical property of galaxies is not the only factor that affects the shape of the attenuation curve at high redshift.
\end{itemize}

In this paper, based on the IllustrisTNG simulation suites, using our radiative post-processing technique, we make detailed predictions for galaxy populations at high redshift. These predictions are in the regime that will be studied by future {\it JWST} observations.  In general, our predictions produce ``qualitative'' trends that are consistent with observations in all the aspects explored in this paper. Quantitatively, our results are consistent with previous theoretical models on galaxy halo scaling relations and are consistent with existing observations on emission line luminosity functions within $\sim 0.3\,{\rm dex}$ (a factor of $\sim 2$) differences in masses/luminosities. However, at the bright/massive end, our results significantly under-predict the amount of dust attenuation and results in a deficiency of heavily attenuated, UV red galaxies. We also find that dust attenuation curves measured in observations are steeper than predicted. We note that the tensions with observations we found in this paper could be caused by our simplified treatment of dust in the post-processing procedure, e.g. our assumption of the dust model~\citep{Draine2007}, and do not necessarily imply that the IllustrisTNG galaxy formation model is flawed. In order to break the degeneracy, a self-consistent treatment of dust as well as the nebular emission and the unresolved dust attenuation will be needed. These aspects will be explored in follow-up works. 


\section*{Acknowledgements}
FM acknowledges support through the Program ``Rita Levi Montalcini'' of 
the Italian MIUR. The primary TNG simulations were realized with compute time granted by
the Gauss Centre for Supercomputing (GCS): TNG50 under GCS
Large-Scale Project GCS-DWAR (2016; PIs Nelson/Pillepich), and
TNG100 and TNG300 under GCS-ILLU (2014; PI Springel) on
the GCS share of the supercomputer Hazel Hen at the High Performance Computing Center Stuttgart (HLRS). MV acknowledges support through an MIT RSC award, a Kavli Research Investment Fund, NASA ATP grant NNX17AG29G, and NSF grants AST-1814053, AST-1814259 and AST-1909831.






\appendix


\bsp	
\label{lastpage}
\end{document}